\documentclass[aps,prd,10pt,twocolumn,floatfix,superscriptaddress,nofootinbib,amsmath,amssymb]{revtex4-2}

\usepackage[T1]{fontenc}
\usepackage{graphicx}
\usepackage{bm}
\usepackage{microtype}

\usepackage{natbib}
\usepackage[dvipsnames]{xcolor}
\usepackage[colorlinks=true,linkcolor=blue,citecolor=blue,urlcolor=blue]{hyperref}
\usepackage[capitalize]{cleveref}

\usepackage{etoolbox}
\usepackage{orcidlink}
\usepackage{enumitem}

\usepackage{physics}
\usepackage{booktabs}

\begin{document}
\title{Gravitating sources with zero energy density in cosmology}

\author{\"{O}zg\"{u}r Akarsu\,\orcidlink{0000-0001-6917-6176}}
\email{akarsuo@itu.edu.tr}
\affiliation{Department of Physics, Istanbul Technical University, Maslak 34469 Istanbul, T\"{u}rkiye}

\author{Cihad K{\i}br{\i}s\,\orcidlink{0000-0002-4129-2199}}
\email{kibrisc@itu.edu.tr}
\affiliation{Department of Physics, Istanbul Technical University, Maslak 34469 Istanbul, T\"{u}rkiye}

\author{Burcu \"{O}zt\"{u}rk\,\orcidlink{0000-0003-2845-5106}}
\email{ozturkburc@itu.edu.tr}
\affiliation{Department of Physics, Istanbul Technical University, Maslak 34469 Istanbul, T\"{u}rkiye}

\author{N. Merve Uzun\,\orcidlink{0000-0001-5555-4500}}
\email{uzunmer@itu.edu.tr}
\affiliation{Department of Physics, Istanbul Technical University, Maslak 34469 Istanbul, T\"{u}rkiye}

\begin{abstract}
A cosmological source can have identically zero comoving energy density and still gravitate through its pressure. We investigate such sources, whose inertial and active gravitational mass densities are $\mathcal I_{\rm ze}=p_{\rm ze}$ and $\mathcal M_{\rm ze}=3p_{\rm ze}$. Unlike an isolated zero-crossing of the energy density, maintaining $\rho_{\rm ze}\equiv0$ with $p_{\rm ze}\neq0$ in a nonstatic Friedmann--Lema\^{\i}tre--Robertson--Walker (FLRW) background requires energy exchange, with $\mathcal Q=-3Hp_{\rm ze}$ in our two-component convention. We construct proportional-pressure and constant-pressure solutions, including steady-state, bouncing, recollapsing, and $\Lambda$CDM expansion histories, and establish their background correspondences with creation pressure, bulk viscosity, and time-dependent vacuum descriptions. Canonical, phantom, and power-law scalar fields realize zero density along reconstructed trajectories. Imposing zero density as an identity on a timelike $P(X,\phi)$ domain instead selects $P=A(\phi)\sqrt X$, locally equivalent to a signed potential-free cuscuton wherever $A\neq0$. We formulate linear perturbations without dividing by the vanishing background density. For a nondegenerate $P(X,\phi)$ scalar with a timelike field gradient, negative pressure at zero density is incompatible with positive kinetic and gradient coefficients when the interaction leaves its principal kinetic structure unchanged. Two explicit models couple the scalar to number-conserving dust through a field-dependent particle mass. In the proportional-pressure model, the reconstructed trajectory is a saddle in homogeneous phase space, and its accelerating branch admits no growing power-law dust mode in the matter-sourced quasistatic subhorizon approximation. A distinct constant-pressure model reproduces the $\Lambda$CDM expansion history, but in the same approximation the perturbation mode that initially grows during matter domination reaches a maximum at low redshift and subsequently decays. These restrictions concern the stated scalar class and models, not all zero-density sources. Beyond FLRW, exact Bianchi Type~I solutions show how anisotropic stress supports separately conserved zero-density sources with persistent relative anisotropy. An inverse-pressure construction distinguishes the density of an interacting constituent from an inferred dark-energy density that can cross zero.
\end{abstract}

\maketitle

\section{Introduction}
\label{sec:intro}

The gravitational role of a cosmological source is determined by its energy--momentum tensor (EMT), not by its energy density alone. For a source of perfect fluid form, the comoving energy density $\rho$ and isotropic pressure $p$ enter the field equations of general relativity (GR) through combinations with distinct physical roles. In spatially flat Friedmann--Lema\^{\i}tre--Robertson--Walker (FLRW) cosmology, the total density enters the Hamiltonian constraint, the sum of the inertial mass densities $\mathcal I\equiv\rho+p$ governs the evolution of the Hubble parameter, and the sum of the active gravitational mass densities $\mathcal M\equiv\rho+3p$ supplies the source in the timelike Raychaudhuri equation. A negative contribution to $\mathcal M$ favors acceleration, whereas a positive contribution favors deceleration. The quantity $\mathcal I$ also multiplies the acceleration in the relativistic Euler equation and governs the evolution of the signed density of a separately conserved component. For such a component in an expanding universe, positive $\mathcal I$ decreases the signed energy density, whereas negative $\mathcal I$ increases it. These roles motivate organizing the description around the combinations selected by the field equations, while retaining the density information required by the constraint~\cite{Akarsu:2026pia}.

For a perfect fluid EMT, the null energy condition (NEC) requires $\mathcal I\geq0$, and the strong energy condition (SEC) additionally requires $\mathcal M\geq0$~\cite{Hawking:1973uf}. We use \emph{quintessence-like} and \emph{phantom-like} as shorthand for the sector-level signs $\mathcal I>0$ and $\mathcal I<0$, respectively, with $\mathcal I=0$ defining the NEC boundary (NECB)~\cite{Akarsu:2026anp,Akarsu:2026pia,Gokcen:2026pkq}. These labels characterize an individual source; they do not by themselves determine the energy conditions of a multicomponent system or the sign of the cosmic acceleration. For components sharing the FLRW comoving congruence, it is the sums $\mathcal I_{\rm tot}$ and $\mathcal M_{\rm tot}$ that enter the gravitational evolution equations~\cite{Akarsu:2026pia}.

From the perspective of relativistic inertia, a zero-density source is a natural limiting stress--energy structure. Dust has $p_{\rm m}=0$ and $\mathcal I_{\rm m}=\mathcal M_{\rm m}=\rho_{\rm m}$; vacuum energy $p_{\rm vac}=-\rho_{\rm vac}$ has $\mathcal I_{\rm vac}=0$ and $\mathcal M_{\rm vac}=2p_{\rm vac}$; the complementary case $\rho_{\rm ze}=0$, $p_{\rm ze}\neq0$ has $\mathcal I_{\rm ze}=p_{\rm ze}$ and $\mathcal M_{\rm ze}=3p_{\rm ze}$. Its relativistic inertia and Raychaudhuri contribution are therefore carried entirely by pressure. In particular, it does not lie on the NEC boundary since its nonzero pressure fixes its NEC character. Throughout the FLRW analysis, \emph{perfect fluid form} refers to the algebraic form of the EMT. It does not assume a microscopic thermodynamic fluid with a specified particle number, entropy, or constitutive relation.

The equation of state (EoS) parameter $w\equiv p/\rho$ remains useful where the density is nonzero, but its interpretation depends on the density sign. For either sign of $\rho\neq0$, $w=-1$ is equivalent to $\mathcal I=0$. What reverses for negative density is the assignment of the two sides of $w=-1$ to the two NEC characters. At $\rho=0$, the ratio itself is undefined, even when the EMT and the combinations $\mathcal I$ and $\mathcal M$ remain finite. Negative-density contributions and sign-changing dark energy histories, including those motivated by graduated dark energy (gDE), make this distinction particularly relevant~\cite{Sahni:2002dx,Sahni:2014ooa,Dutta:2018vmq,Visinelli:2019qqu,Akarsu:2019hmw,Calderon:2020hoc,Acquaviva:2021jov,Sen:2021wld,Adil:2023exv,Malekjani:2023ple,Tiwari:2023jle,Akarsu:2024qsi,Akarsu:2024eoo,Dwivedi:2024okk,Akarsu:2024nas,Akarsu:2025gwi,Akarsu:2026pia,Akarsu:2026lva,Bouhmadi-Lopez:2025ggl,Bouhmadi-Lopez:2025spo,Adil:2026gjl}. A divergence of $w$ at a regular density zero is a failure of the ratio parametrization, not necessarily a physical singularity~\cite{Ozulker:2022slu,Akarsu:2025gwi,Akarsu:2026pia}.

The issue is timely in precision cosmology. DESI DR2 BAO combinations with CMB and supernova data have renewed interest in time-dependent dark energy, with conclusions depending on the dataset combination and parametrization~\cite{DESI:2025DR2}. The subsequent DR2 Lyman-$\alpha$ full-shape Alcock--Paczy\'nski analysis adds a complementary intermediate-redshift test~\cite{DESI:2026Lya}. Neither a preference within an EoS family nor a regular expansion history reconstruction uniquely specifies the dark sector constituents. The relevance of zero-density sources is therefore not a claim that they already explain these data, but a precise way to examine which assumptions enter the interpretation of a measured expansion history. Simple inertial mass density laws, such as graduated dark energy, illustrate why this enlargement of the description is useful~\cite{Akarsu:2019hmw,Acquaviva:2021jov}. For broader reviews of cosmological tensions and consistency tests, see Refs.~\cite{Perivolaropoulos:2021jda,Abdalla:2022yfr,Vagnozzi:2023nrq,Akarsu:2024qiq,CosmoVerseNetwork:2025alb}.

The present work investigates a stronger condition than an isolated density zero-crossing: a source whose comoving density \emph{vanishes throughout a cosmological interval}, while its pressure remains nonzero. For a sign-changing density, the conventional EoS parameter is defined on the nonzero-density intervals but becomes ill-defined at the zero-crossing. For the persistently zero-density source considered here, it is instead undefined throughout the interval, so the source must be characterized directly by its pressure and stress--energy dynamics. The distinction is also dynamical. For a separately conserved component in an expanding FLRW background, $\dot\rho=-3H(\rho+p)$ permits an isolated zero at $t=t_*$ with $p_*=-\dot\rho_*/(3H_*)$. By contrast, maintaining $\rho\equiv0$ requires $3Hp=0$ unless the component exchanges energy with another source. We shall also distinguish a zero-density background trajectory from an identity-level condition on a scalar field Lagrangian. These alternatives have different implications for perturbations and for the number and character of the dynamical degrees of freedom.

The distinction between a background correspondence and a field realization is central to the analysis. We first determine which conservation conditions and pressure laws admit a source with zero density, and then ask what additional restrictions follow when that source is represented by a scalar field. For an ordinary, nondegenerate $P(X,\phi)$ scalar with a timelike field gradient, zero density and negative pressure are incompatible with simultaneously positive kinetic and gradient coefficients if the interaction leaves the principal scalar structure unchanged. This result specializes the familiar relation between the NEC and scalar stability; it does not apply to the positive-pressure canonical construction or replace the constraint analysis needed for a degenerate scalar theory.

To examine dynamics beyond this intrinsic kinetic test, we specify two models in which a scalar field controls the mass of number-conserving dust particles. The first realizes the proportional-pressure background. Its reconstructed zero-density trajectory is a saddle in the homogeneous phase space, and the accelerating branch admits no growing power-law dust mode in the matter-sourced quasistatic subhorizon approximation. The second uses a different mass function and a constant potential to reproduce the $\Lambda$CDM expansion history. In this case, the perturbation mode selected by growing-mode initial conditions during matter domination reaches a maximum at low redshift and subsequently decays. Thus, even an exact match of the expansion history need not imply the same clustering. These examples make the role of the interaction explicit while leaving room for constrained scalar sectors, more general interactions, and the anisotropic realizations developed below.

Pressure-only stress--energy structures already occur in several physically distinct settings. In the fluid/gravity correspondence, a relativistic fluid dual to vacuum Einstein gravity has a zero-density equilibrium state with nonzero pressure and nontrivial hydrodynamics~\cite{Compere:2011dx,Compere:2012mt,Eling:2012ni}. There the fluid is an effective boundary description of a higher-dimensional gravitational solution. Gravitational-aether models provide another closely related structure. In their incompressible limit, an additional pressure contribution participates in a modification of the gravitational response to matter and vacuum energy~\cite{Afshordi:2008xu,Aslanbeigi:2011si}. These examples differ from an independently introduced additional matter sector in GR, but demonstrate that vanishing comoving density need not eliminate gravitationally relevant stresses.

There are also direct examples within GR. Static, spherically symmetric perfect fluid solutions with zero density and nonzero pressure were obtained by Kuchowicz~\cite{Kuchowicz1968} and discussed subsequently in Ref.~\cite{Semiz:2022iyh}. A zero-density limit of the Schwarzschild constant-density solution has recently been examined in isotropic coordinates~\cite{Simmonds:2026jln}. At the field theory level, scalar field Lagrangian densities $P(X,\phi)$ admit a perfect fluid representation for timelike field gradients~\cite{Garriga:1999vw}. The potential-free cuscuton is particularly relevant. Its square-root kinetic structure gives an identically vanishing scalar rest-frame density, with the field equation becoming a constraint rather than the evolution equation of an ordinary propagating scalar~\cite{Afshordi:2006ad,Afshordi:2007yx}.

A related cosmological construction dates back to Hoyle's steady-state model. In its homogeneous and isotropic formulation, the creation tensor has $C_{00}=0$ and nonzero spatial components, so that it can be represented as an effective pressure-only contribution~\cite{Hoyle:1948zz}. Particle production models likewise introduce a creation pressure into the cosmological equations~\cite{Prigogine1988,Calvao:1991wg}, while bulk viscosity supplies an additional isotropic pressure at the homogeneous level~\cite{Zimdahl:1996fj}. Their background equations can coincide even though their thermodynamic interpretations differ~\cite{Lima:1992np}. We retain these distinctions when establishing the corresponding relations below.

Our aim is a systematic cosmological analysis of this stress--energy structure: the conservation conditions that sustain it, representative background and scalar field realizations, its linear perturbative formulation, and how the FLRW conservation restriction changes when anisotropic stresses are admitted. An algebraic decomposition of an EMT into interacting contributions does not, on its own, identify independent physical constituents. Their interpretation requires constitutive information or an action that fixes the split and its exchange. Accordingly, the existence of a background solution, the regularity of stress--energy variables, and the stability of the underlying model are treated as distinct questions. This extends the density-zero diagnostic perspective to the dynamical problem of sustaining a nontrivial zero-density source.

We use the notation of Ref.~\cite{Akarsu:2026pia} for the shared diagnostics: $\mathcal I_i=\rho_i+p_i$, $\mathcal M_i=\rho_i+3p_i$, $\rho_{\rm cr0}=3H_0^2/(8\pi G)$, and $z=a^{-1}-1$. Dots denote cosmic-time derivatives in homogeneous models and derivatives along the chosen congruence in covariant equations; a prime on a background function denotes ${\rm d}/{\rm d}z$, and scale factor derivatives are written explicitly. Conformal-time derivatives in Sec.~\ref{sec:perturbation} are denoted by ${\rm d}/{\rm d}\tau$. There $\Psi$ is the lapse perturbation, $\Phi$ the spatial-curvature perturbation, $Q_i$ the momentum-density divergence, and $\Pi_i$ the unnormalized scalar anisotropic stress. We use $\mathcal Q_i$ for energy transfer, so it is not confused with $Q_i$.

The paper is organized as follows: Section~\ref{sec:density} distinguishes zero comoving energy density from a vanishing EMT and discusses its observer interpretation. Section~\ref{sec:flrw} derives the FLRW conservation restriction and the required energy exchange, and relates the framework to creation and bulk viscous pressures. Section~\ref{sec:particular} presents the steady-state, proportional-pressure, and constant-pressure cosmological solutions. Section~\ref{sec:scalar-realizations} constructs their scalar field counterparts and distinguishes trajectory-level from identity-level zero density. Section~\ref{sec:perturbation} formulates linear perturbations without density normalization and examines intrinsic scalar kinetic properties. Section~\ref{sec:anisotropic} treats imperfect sources and exact Bianchi Type~I geometries. Section~\ref{sec:inference} develops the connection to cosmological inference, and Sec.~\ref{sec:Conclusion} concludes. Appendix~\ref{app:action} supplies covariant dust--scalar realizations, the homogeneous stability test of the proportional-pressure model, and subhorizon dust-growth calculations for both pressure laws.

\section{Relativistic Description of Energy Density}
\label{sec:density}

Before introducing a source with zero energy density, it is useful to clarify the observer dependence of this quantity in the theory of relativity. The energy density measured by an observer with four-velocity $v^\mu={\rm d}x^\mu/{\rm d}\tau_{\rm obs}$ in the observer's local rest frame is given by
\begin{align}
    \label{eq:rho-comv}
    \rho_{\rm obs}=T_{\mu\nu}v^\mu v^\nu,
\end{align}
where $\tau_{\rm obs}$ is the proper time measured by the observer. Hence, the decomposition of the EMT $T_{\mu\nu}$ into energy density, isotropic pressure, momentum density (energy flux), and anisotropic stresses is not unique and depends on the observer's four-velocity. We consider the EMT in the perfect fluid form
\begin{align}
    \label{eq:emt-pf}
    T^{\mu\nu}=(\rho+p)u^\mu u^\nu+pg^{\mu\nu},
\end{align}
where $u^{\mu}={\rm d}x^{\mu}/{\rm d}\tau_{\rm fl}$ is the fluid four-velocity, with $\tau_{\rm fl}$ being the proper time along its worldlines. The quantities $\rho$ and $p$ are the energy density and isotropic pressure measured in this rest frame. We adopt metric signature $(-,+,+,+)$ and normalize all observer four-velocities to minus unity.

Equation~\eqref{eq:emt-pf} implies
\begin{align}
T^{\mu}{}_{\nu}u^{\nu}=-\rho u^{\mu},
\end{align}
showing that $u^\mu$ is a timelike eigenvector of $T^{\mu}{}_{\nu}$ with the corresponding eigenvalue $-\rho$. For a vector $w^{\mu}$ orthogonal to the fluid's four-velocity satisfying $w_{\mu}u^{\mu}=0$, Eq.~\eqref{eq:emt-pf} yields
\begin{align}
T^{\mu}{}_{\nu} w^{\nu}=p w^{\mu}.
\end{align}
Hence, every spatial vector orthogonal to $u^{\mu}$ is an eigenvector of $T^{\mu}{}_{\nu}$ with the degenerate eigenvalue $p$. The vacuum energy, $p_{\rm vac}=-\rho_{\rm vac}$, case is exceptional: since $T^{\mu}{}_{\nu}=-\rho_{\rm vac}\delta^{\mu}{}_{\nu}$ every vector is an eigenvector and the EMT does not select a unique timelike eigenvector or fluid congruence. This intrinsic eigenvector structure should be distinguished from the observer-dependent decomposition of the EMT. At a given spacetime event, there are infinitely many future-directed unit timelike vectors $v^\mu$, corresponding to the possible observer four-velocities at that event. If the observer is comoving with the fluid, so that $v^\mu=u^\mu$, Eqs.~\eqref{eq:rho-comv} and~\eqref{eq:emt-pf} give $\rho_{\rm obs}=\rho$.

For a noncomoving observer, with $v^\mu\neq u^\mu$, the measured energy density is instead
\begin{align}
    \label{eq:rho-gen}
    \rho_{\mathrm{obs}}=(\rho+p)\gamma^2-p,
\end{align}
where the relative Lorentz factor between the observer and the fluid is defined as $\gamma=-u_\mu v^\mu$. Since $u^\mu$ and $v^\mu$ are both future-directed unit timelike four-velocities, $\gamma\geq1$, with equality only in the comoving case, viz., when $v^\mu=u^\mu$. In the flat spacetime (special relativistic) limit, this takes the familiar form $\gamma=(1-v_{\rm rel}^2)^{-1/2}$, where $v_{\rm rel}$ denotes the relative three-speed between the observer and the fluid. Similarly, the isotropic pressure, momentum density and spatial stresses measured by the observer are given by
\begin{align} \label{eq:momentum}
p_{\rm obs}
&= p+\frac{1}{3}(\rho+p)(\gamma^2-1),\\
q_{\mu}^{\rm obs}&=(\rho+p)\gamma(u_{\mu}-\gamma v_{\mu}), \\
\pi_{\mu\nu}^{\rm obs}
&=(\rho+p)
\left[
(u_\mu-\gamma v_\mu)(u_\nu-\gamma v_\nu)
-\frac{\gamma^2-1}{3}
h_{\mu\nu}^{\rm obs}
\right], \label{eq:stress}
\end{align}
where the spatial projector associated with the observer congruence is defined by $h_{\mu\nu}^{\rm obs}\equiv g_{\mu\nu}+v_{\mu}v_{\nu}$. The explicit appearance of $v^\mu$ in these expressions makes the observer dependence of the EMT decomposition manifest; in particular, a noncomoving observer generally measures a nonzero momentum density and direction-dependent spatial stresses, so that the EMT decomposition in that observer’s frame is no longer of perfect fluid form. The vacuum energy has a distinctive EoS, $p_{\rm vac}=-\rho_{\rm vac}$, so one obtains $\rho_{\rm obs}=-p_{\rm obs}=\rho_{\rm vac}$ and $q_{\mu}^{\rm obs}=\pi_{\mu\nu}^{\rm obs}=0$, consistently with $T_{\mu\nu}^{\rm vac}=-\rho_{\rm vac} g_{\mu\nu}$. Thus, its EMT has the same form for every observer, as expected for a Lorentz-invariant source. Equations~\eqref{eq:rho-gen}--\eqref{eq:stress} show that setting both $\rho$ and $p$ to zero makes every observer-dependent EMT component vanish. This is the covariant condition $T_{\mu\nu}=0$, not merely the vanishing of one temporal projection. It means that the sector supplies no stress--energy contribution; however, it need not imply that an underlying field configuration is trivial. In contrast, $\rho=0$ with $\gamma=1$ permits $p\neq0$. A genuinely noncomoving observer can also measure $\rho_{\rm obs}=0$ for nonzero $\rho$ and $p$ when $\rho p<0$ and $|\rho|<|p|$, since Eq.~\eqref{eq:rho-gen} then admits $\gamma^2=p/(\rho+p)>1$. A vanishing density measured by one observer therefore does not imply a vanishing EMT.

\begin{figure}[htbp]
\centering
\includegraphics[width=\columnwidth]{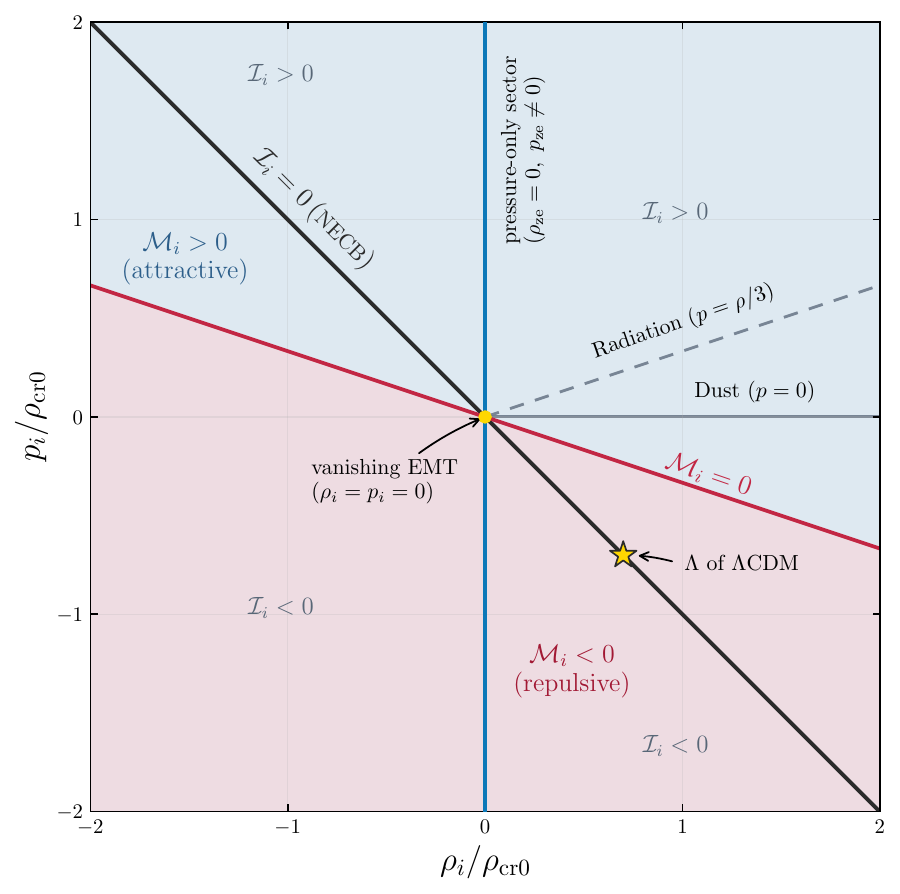}
\caption{Algebraic map of candidate-source stress states in the signed
$(\rho_i/\rho_{\rm cr0},p_i/\rho_{\rm cr0})$ plane. The black line is the vacuum relation $p_i=-\rho_i$, corresponding to
$\mathcal I=0$, i.e., the null energy condition boundary (NECB), and to the cosmological-constant form of the EMT. The red line $\mathcal M_i=0$ separates attractive (blue, $\mathcal M_i>0$) from repulsive (pink, $\mathcal M_i<0$) contributions to the Raychaudhuri equation; it does not by itself mark the onset of total cosmic acceleration. The highlighted vertical line, excluding the origin, denotes the pressure-only sector studied here, with $\rho_{\rm ze}=0$ and $p_{\rm ze}\neq0$. The filled marker at the origin represents $(\rho_i,p_i)=(0,0)$ and hence $T_{\mu\nu}=0$ within the perfect-fluid class. The star denotes the cosmological-constant component of $\Lambda$CDM, with the illustrative choice $\Omega_{\Lambda}=0.7$, located at $(\rho_\Lambda,p_\Lambda)/\rho_{\rm cr0}=(0.7,-0.7)$ on the NECB. Dust ($p=0$) and radiation ($p=\rho/3$) are shown only as positive-density reference rays. The normalization is $\rho_{\rm cr0}=3H_0^2/\kappa>0$. No cosmological trajectory is implied.}
\label{fig:stress-plane}
\end{figure}

We focus on $\rho=0$ and $p\neq0$ in the source rest frame, which is the comoving frame in the homogeneous models below. From Eq.~\eqref{eq:rho-gen}, the energy density measured by a noncomoving observer is then
\begin{align}
\rho_{\rm{obs}}=(\gamma^2-1)p \qquad   {\rm when}\qquad \rho=0.
\end{align}
Therefore, for $p\neq0$, any noncomoving observer ($\gamma\neq1$) measures a nonzero energy density, whose sign is determined by the rest-frame pressure:
\begin{align}
{\rm sgn}(\rho_{\rm{obs}})={\rm sgn}(p).
\end{align}
The condition $\rho=0$ sets the timelike eigenvalue of $T^{\mu}{}_{\nu}$ to zero, while its three spatial eigenvalues remain equal to $p$. For $p\neq0$, the timelike eigendirection is distinct from the spatial eigenspace. Thus the zero-density condition is a covariant property of this EMT, although arbitrary observers generally measure a different density.

Introducing the spatial projection tensor $h_{\mu\nu}$ associated with the comoving congruence, the spacetime metric $g_{\mu\nu}$ can be decomposed into components orthogonal and parallel to the fluid's four-velocity $u^\mu$ as $g_{\mu\nu}=h_{\mu\nu}-u_{\mu}u_{\nu}$, where $u^{\mu}$ is the unique unit timelike field that is tangent to the worldlines of comoving observers and satisfies the relation $u_\mu u^{\mu}=-1$. Then, the perfect fluid EMT~\eqref{eq:emt-pf} can equivalently be written as
\begin{align}
T_{\mu\nu}=-\rho g_{\mu\nu}+\mathcal I h_{\mu\nu}.
\label{eq:emt-rho-I}
\end{align}
For zero density this becomes $T_{\mu\nu}=p h_{\mu\nu}$, with invariant contractions
\begin{align}
T^\mu{}_{\mu}=3p,\qquad T_{\mu\nu}T^{\mu\nu}=3p^2.
\label{eq:zero-invariants}
\end{align}
The source is therefore nonvacuum whenever $p\neq0$. Equation~\eqref{eq:emt-rho-I} also shows that a vacuum-form term and a pressure-only term can be separated algebraically in any perfect fluid EMT. Such a rearrangement does not by itself define two independent material constituents or determine their interaction. In particular, $\rho\mapsto\rho+C$ and $p\mapsto p-C$ leave $\mathcal I$ fixed but shift $\mathcal M\mapsto\mathcal M-2C$: an inertial mass density history does not determine the vacuum offset in the Hamiltonian constraint. A physical interpretation of the split thus requires additional constitutive or action-level information.

For this zero-density perfect fluid, the NEC, weak energy condition, and SEC hold for $p>0$ and fail for $p<0$. The dominant energy condition would require $\rho\geq|p|$ and hence fails for every nonzero $p$ when $\rho=0$. These results apply to the individual source, not necessarily the total EMT of an interacting system, and are distinct from the stability properties of any particular field realization.

Figure~\ref{fig:stress-plane} shows where this limiting stress-energy structure lies in the signed $(\rho_i,p_i)$ plane. The diagram uses the index $i$ for a generic source; the dust and radiation lines are familiar positive-density reference stress relations. The vertical axis is the zero-density limit, denoted by $\rho_{\rm ze}=0$ in the dynamical models below. Both coordinates are normalized by the same positive reference, namely, the present-day critical density $\rho_{\rm cr0}=3H_0^2/\kappa$. The blue and pink regions distinguish the sign of the sector's Raychaudhuri contribution, not the sign of the total cosmic acceleration. The starred point denotes $\Lambda$ itself, the dark-energy component of $\Lambda$CDM; $p_\Lambda=-\rho_\Lambda$ places it on the $\mathcal I_i=0$ line, while $T_{\mu\nu}^{(\Lambda)}=-\rho_\Lambda g_{\mu\nu}\neq0$ distinguishes it from the origin. The map classifies local stress states; it does not establish whether a cosmological solution can remain on the $\rho=0$ axis with nonzero pressure. That dynamical question is governed by conservation and energy exchange, addressed in Sec.~\ref{sec:flrw}.

We now proceed with Robertson--Walker (RW) spacetime described by the metric
\begin{align}
\label{eq:rw}
{\rm d} s^2 = -{\rm d}t^2 + a^2(t)\,^{(3)}g_{ij}{\rm d}\xi^{i} {\rm d} \xi^{j},
\end{align}
where $t$ is the timelike coordinate, $\xi^i$ are coordinates on the maximally symmetric spatial three-manifold $\Sigma$, and $^{(3)}g_{ij}$ is the metric on $\Sigma$.
The spacetime~\eqref{eq:rw} admits a preferred timelike unit vector field $n^\mu$, given by the future-directed normal to the homogeneous and isotropic spatial hypersurfaces. For the metric~\eqref{eq:rw}, $n_\mu=-\nabla_\mu t$ and hence $n^\mu=\delta^\mu_0$. This congruence is associated with the foliation of spacetime by spacelike hypersurfaces of constant cosmic time $t$. We now consider a single perfect fluid source described by the EMT~\eqref{eq:emt-pf}, without assuming \emph{a priori} that the fluid is comoving with this congruence. In this case, one has
\begin{align}
    T_{0i}=(\rho+p)u_0 u_i.
\end{align}
However, for the metric given in Eq.~\eqref{eq:rw}, the temporal-spatial components of the Einstein tensor vanish identically, viz., $G_{0i}=0$, which implies $(\rho+p)u_0u_i=0$ through the field equations of GR. Since the timelike character of $u^\mu$ ensures $u_0\neq0$, this leads to $u_i=0$ for any perfect fluid with $\rho+p\neq0$. Only the special vacuum energy case, $\rho_{\rm vac}+p_{\rm vac}=0$, for which $T_{\mu\nu}^{\rm vac}=-\rho_{\rm vac} g_{\mu\nu}$, does not determine a unique four-velocity. For any perfect fluid source with $\rho+p\neq0$, however, the fluid four-velocity must align with the geometrically preferred congruence,
\begin{align}
 u^{\mu}=n^{\mu}=\delta^{\mu}_0.
\end{align}
Thus, for a perfect fluid source with $\rho+p\neq0$, its comoving nature is not an independent assumption but follows from compatibility with the symmetries of the RW metric in GR. As for the observer, we restrict our attention to a congruence with vanishing vorticity, $\omega_{\mu\nu}^{\rm obs}=0$, which is therefore orthogonal to a family of spacelike hypersurfaces. For the metric~\eqref{eq:rw}, the spatial components of the spatial projector associated with the observer congruence are
\begin{align}
    h_{ij}^{\rm obs}=a^2(t)\,^{(3)}g_{ij}+v_iv_j.
\end{align}
The first term is the induced metric on the homogeneous and isotropic spatial sections associated with the RW foliation, whereas the second term arises from the observer's tilt relative to that foliation. If $v^i\neq0$, the additional tensor $v_iv_j$ introduces a preferred spatial direction, so that homogeneity and isotropy are no longer manifest in the spatial decomposition in the frame of that observer. This does not imply a loss of the underlying RW symmetry but means that the RW symmetry is not manifest in the observer-adapted spatial geometry. The comoving congruence, characterized by $v^i=0$, or equivalently $v^\mu=n^\mu$, is the unique future-directed unit timelike congruence orthogonal to the maximally symmetric spatial hypersurfaces. For this congruence, the observer-dependent spatial projector reduces to
\begin{align}
    h_{ij}^{\rm obs}=a^2(t)\,^{(3)}g_{ij} \qquad \text{for}\qquad v^i=0,
\end{align}
so that the spatial geometry has the canonical RW form. This normal congruence is distinguished by the chosen RW foliation, without excluding other admissible observers. The latter can measure nonzero energy flux and direction-dependent stresses even for an EMT of perfect fluid form in the fluid rest frame. Neither this observer dependence nor the vanishing of the comoving energy density projection eliminates the invariant pressure contribution in Eq.~\eqref{eq:zero-invariants}.

\section{Zero Energy Density Source within FLRW Framework}
\label{sec:flrw}

We now examine a persistently zero-density source in spatially flat FLRW cosmology. Each background component is assumed to respect the RW symmetries and to share the comoving congruence. In Cartesian comoving coordinates, Eq.~\eqref{eq:rw} becomes
\begin{align}
\label{rwmetric}
{\rm d} s^2 = -{\rm d}t^2 + a^2(t) ({\rm d}x^2+{\rm d} y^2+{\rm d} z^2),
\end{align}
where $a(t)$ is the scale factor. The perfect fluid form given in Eq.~\eqref{eq:emt-pf} is the most general EMT that respects the symmetries of \cref{rwmetric}.
We work in standard GR in which the relation between the geometry of spacetime and its material content is described by the Einstein field equations~(EFE) $G_{\mu\nu} =  \kappa T_{\mu\nu}$. Here $\kappa = 8\pi G$ with $G$ being the Newtonian gravitational constant and we take the speed of light $c=1$. The twice-contracted Bianchi identity, together with EFE, guarantees that EMT is divergence-free, $\nabla_{\mu}T^{\mu\nu} = 0$, which, when projected parallel $u_{\nu}\nabla_{\mu}T^{\mu\nu}=0$ and orthogonal $h_{\nu}^{\;\;\sigma}\,\nabla_{\mu}T^{\mu\nu}=0$ to $u_\mu$, yields the generic energy and momentum conservation equations, namely, the continuity and Euler equations,
\begin{align}
\dot{\rho}+\Theta(\rho+p) &= 0,
\label{cont}\\[6pt]
{\rm D}^\mu p+(\rho+p)A^\mu &= 0.
\label{Euler}
\end{align}
Here $\Theta=\nabla_\mu u^\mu$ is the volume expansion scalar, $A^\mu\equiv\dot u^\mu=u^\nu\nabla_\nu u^\mu$ is the four-acceleration, and ${\rm D}_\mu$ denotes the covariant derivative projected orthogonally to $u^\mu$ on all free indices. In particular, ${\rm D}_\mu p=h_\mu{}^\nu\nabla_\nu p$ and ${\rm D}_\nu u_\mu=\nabla_\nu u_\mu+u_\nu A_\mu$. The definition of $\Theta$ does not require shear or vorticity to vanish; in FLRW it reduces to $3H$ where $H\equiv\dot{a}/a$ is the Hubble parameter.

The combination $\rho+p$, which appears in these two equations, is naturally identified as \textit{inertial mass density} $\mathcal{I}$ since, in the relativistic Euler equation~\eqref{Euler}, it plays the role of the inertial mass in Newton's second law. At the exact FLRW background level, though, Eq.~\eqref{Euler} is identically satisfied as the homogeneity condition requires ${\rm D}^\mu p=0$ and the fluid worldlines are geodesics, $A^\mu=0$. It therefore provides no independent background constraint, whereas the continuity equation~\eqref{cont} remains nontrivial. Treating the universal material content as composed of multiple components that share the same four-velocity $u^\mu$, the total EMT can be written as the summation of each contributing source as $T_{\mu\nu}^{\rm tot}=\sum_i T_{\mu\nu}^{(i)}$, where $T_{\mu\nu}^{(i)}=(\rho_i+p_i)u_\mu u_\nu+p_i g_{\mu\nu}$.

Using the metric~\eqref{rwmetric} along with the total EMT in the EFE, we obtain the corresponding Friedmann equations
\begin{align}  \label{fe:00}
     3H^2 &= \kappa \sum_i\rho_i , \\
   -2\dot{H} - 3H^2 &= \kappa  \sum_i p_i, \label{fe:11}
\end{align}
 Accordingly, from Eq.~\eqref{cont}, the total continuity equation reads
\begin{align} \label{cont-total}
\sum_i\left[\dot{\rho}_i+3H(\rho_i+p_i)\right]=0.
\end{align}
The Friedmann equations~\eqref{fe:00} and ~\eqref{fe:11} further imply that the time evolution of the Hubble parameter is determined by the total inertial mass density,
\begin{align}
\dot{H}=-\frac{\kappa}{2}\sum_i(\rho_i+p_i)
\equiv-\frac{\kappa}{2}\mathcal{I}_{\rm tot},
\end{align}
whereas the acceleration equation is determined by the total active gravitational mass density,
\begin{align}
\frac{\ddot{a}}{a}
=-\frac{\kappa}{6}\sum_i(\rho_i+3p_i)
\equiv-\frac{\kappa}{6}\mathcal{M}_{\rm tot}.
\end{align}
These are the GR counterparts of the curvature contractions used in Ref.~\cite{Akarsu:2026pia}. For the FLRW congruence and a null vector $k^\mu$ with $\mathcal E=-u_\mu k^\mu$, the Ricci contractions entering the timelike and null Raychaudhuri equations are~\cite{Hawking:1973uf}
\begin{align}
R_{\mu\nu}u^\mu u^\nu&=\frac{\kappa}{2}\mathcal M_{\rm tot},\qquad
R_{\mu\nu}k^\mu k^\nu=\kappa\mathcal E^2\mathcal I_{\rm tot}.
\label{eq:curvature-diagnostics}
\end{align}
Here timelike defocusing refers to a negative comoving Ricci contraction, equivalently $\ddot a>0$ in FLRW, and does not require $\dot H>0$. A sector's pressure-only contribution may be repulsive while the other sources keep the total Ricci contraction positive.

The sign of an individual $\mathcal M_i$ determines the sign of its contribution to the acceleration equation, whereas the sign of $\mathcal M_{\rm tot}$ determines whether the Universe accelerates. Similarly, the sign of $\mathcal{I}_i$ indicates the NEC character of the individual source while the sign of $\mathcal{I}_{\rm tot}$ indicates whether the total NEC is violated. We now introduce a component labeled ``$\mathrm{ze}$'' whose comoving energy density is zero throughout the interval considered.
Accordingly, we impose $\rho_{\rm ze}\equiv T_{\mu\nu}^{({\rm ze})}u^\mu u^\nu=0$ while maintaining the pressure $p_{\rm ze}(t)\equiv \frac{1}{3}h^{\mu\nu}T_{\mu\nu}^{({\rm ze})}\neq0$. For this source, the inertial and active gravitational mass densities reduce to
\begin{align}
\label{zero:I-M}
\mathcal{I}_{\rm ze}\equiv \rho_{\rm ze}+p_{\rm ze}=p_{\rm ze},
\qquad
\mathcal{M}_{\rm ze}\equiv \rho_{\rm ze}+3p_{\rm ze}=3p_{\rm ze}.
\end{align}
Thus $p_{\rm ze}<0$ gives a phantom-like, Raychaudhuri-repulsive component, while $p_{\rm ze}>0$ gives a quintessence-like, attractive component. Since $p_{\rm ze}\neq0$, neither case lies on the NECB. For example, with pressureless matter of density $\rho_{\rm m}>0$, acceleration and a satisfied total NEC coexist when
\begin{align}
-\rho_{\rm m}\leq p_{\rm ze}<-\frac{\rho_{\rm m}}{3}.
\label{eq:dust-acc-nec}
\end{align}
This follows from $\mathcal I_{\rm tot}=\rho_{\rm m}+p_{\rm ze}$ and $\mathcal M_{\rm tot}=\rho_{\rm m}+3p_{\rm ze}$, and illustrates the distinction between individual and total properties.

The conservation restriction is immediate. A separately conserved component with $\rho_{\rm ze}\equiv0$ obeys $3Hp_{\rm ze}=0$ by Eq.~\eqref{cont}. It therefore cannot maintain nonzero pressure on an interval with $H\neq0$, irrespective of the other components or spatial curvature. If it is the only source in spatially flat GR, Eq.~\eqref{fe:00} further forces $H\equiv0$, and Eq.~\eqref{fe:11} then requires $p_{\rm ze}=0$. Allowing spatial curvature does not alter this conclusion for a universe containing only this source. Conservation with nonzero pressure first requires $H=0$. The constraint $3(H^2+K/a^2)=0$ then gives $K=0$, and the spatial field equation consequently implies $p_{\rm ze}=0$. By contrast, an isolated zero-crossing of a separately conserved density satisfies
\begin{align}
p_*=-\frac{\dot\rho_*}{3H_*},\qquad H_*\neq0.
\label{eq:isolated-crossing}
\end{align}
The density can cross zero without energy exchange. A smooth, simple negative-to-positive crossing in an expanding background has $\dot\rho_*>0$ and hence $p_*=\mathcal I_*<0$. The restriction derived above concerns persistence at zero, not passage through it.

The nonzero pressure of a persistently zero-density source can instead be sustained by intercomponent energy--momentum exchange, of the kind studied extensively for interacting dark sectors~\cite{Bolotin:2013jpa,Wang:2016lxa,Wang:2024vmw},
\begin{align}\label{sum-emt-int}
\nabla_\mu T_{(i)}^{\mu\nu}&=\mathcal Q_{(i)}^\nu,\\
\label{sum-Q}
\sum_i \mathcal Q_{(i)}^\mu&=0.
\end{align}
Here Eq.~\eqref{sum-Q} arises from conservation of the total EMT. For a perfect fluid satisfying the pressure-only EMT
\begin{align}
T_{\rm ze}^{\mu\nu}=p_{\rm ze}h_{\rm ze}^{\mu\nu},\qquad
h_{\rm ze}^{\mu\nu}=g^{\mu\nu}+u_{\rm ze}^\mu u_{\rm ze}^\nu,
\label{eq:exact-pressure-only}
\end{align}
its divergence gives the covariant relation
\begin{align}
\mathcal Q_{\rm ze}^{\nu}
=\Theta_{\rm ze}p_{\rm ze}u_{\rm ze}^{\nu}
+{\rm D}_{\rm ze}^{\nu}p_{\rm ze}
+p_{\rm ze}A_{\rm ze}^{\nu}.
\label{eq:pressure-only-transfer}
\end{align}
The first term supplies the energy balance; the remaining terms give the pressure-gradient and acceleration contributions to momentum transfer. This relation follows from the assumed EMT without additional constitutive assumptions, but it is not itself a closure law for the independent matter variables. At the FLRW background level, it reduces to $\mathcal Q_{\rm ze}=3Hp_{\rm ze}$. For a general perfect fluid, isotropy therefore requires the background transfer to have the following form
\begin{align}
\mathcal Q_{(i)}^\mu=\mathcal Q_i(t)u^\mu,
\end{align}
implying that in the comoving frame, the spatial components of the interaction vector vanish, $\mathcal Q_{(i)}^j=0$. This is an energy transfer without momentum transfer at the exact homogeneous level; the latter may be present in perturbations. The component-wise continuity equation~\eqref{sum-emt-int} thus becomes
\begin{align} \label{cont:species}
\dot{\rho}_i+3H(\rho_i+p_i)=\mathcal Q_i,
\end{align}
and summing over all source components gives the total continuity equation~\eqref{cont-total}. 

We proceed by considering a two-component system consisting of only one standard perfect fluid source, labeled by ${\rm s}$, and an accompanying zero-density source with nonvanishing pressure. For this specific consideration, Eqs.~\eqref{fe:00} and \eqref{fe:11} transform into
\begin{align}
    \label{zero-fe00}
    3 H^2&= \kappa\,\rho_{\rm s},\\
    \label{zero-fe11}
-2 \dot{H}-3 H^2&=\kappa (p_{\rm s}+p_{\rm ze}),
\end{align}
while the continuity equation~\eqref{cont-total} becomes
\begin{equation} \label{zero-cont}
\dot{\rho}_{\rm s} + 3H(\rho_{\rm s} + p_{\rm s})  + 3H p_{\rm ze}= 0.
\end{equation}
Next, from Eq.~\eqref{sum-Q}, we write $\mathcal Q_{\rm s}=-\mathcal Q_{\rm ze}\equiv \mathcal Q$ and the corresponding equations~\eqref{cont:species} are
\begin{align}
    \label{eq:zero-int1}
    \dot{\rho}_{\rm s} + 3H(\rho_{\rm s} + p_{\rm s}) &= +\mathcal Q,\\
    \label{eq:zero-int2}
    3H\,p_{\rm ze} &= -\mathcal Q,
\end{align}
Under this convention, $\mathcal Q>0$ transfers energy into the standard component and $\mathcal Q<0$ transfers energy out of it. In an expanding background, negative $p_{\rm ze}$ requires $\mathcal Q>0$, and positive $p_{\rm ze}$ requires $\mathcal Q<0$. The transfer balances the pressure term while $\rho_{\rm ze}$ remains zero; it does not describe depletion of a stored comoving energy density for that source. Equation~\eqref{eq:zero-int2} is well defined also at an isolated $H=0$ instant, where finite pressure implies $\mathcal Q=0$. The divided expression $p_{\rm ze}=-\mathcal Q/(3H)$ should only be used for $H\neq0$.\footnote{As a simple static model, two zero-density perfect fluid contributions can have equal and opposite pressures, $p_{{\rm ze},1}=-p_{{\rm ze},2}$. Their total EMT vanishes, and the spatially flat background is static. This cancellation is an EMT-level construction, not a specification of microscopic constituents.}

For a separately conserved component, ${\rm d}\rho/{\rm d}a=-3\mathcal I/a$. In an expanding phase, positive $\mathcal I$ decreases the signed density and negative $\mathcal I$ increases it. With an interaction written as $\mathcal Q_{\rm s}=3Hf(\rho_i,p_i)$, the corresponding expression is given by
\begin{equation}
\frac{{\rm d}\rho_{\rm s}}{{\rm d}a}=-\frac{3}{a}\left[\mathcal I_{\rm s}-f(\rho_i,p_i)\right].
\end{equation}
The sign of $\mathcal I_{\rm s}$ alone then no longer determines dilution. These scale factor equations apply on monotonic background histories; cosmic time should be used through a bounce or turnaround.

The conservation argument also applies to an effective perfect fluid sector whenever it is separately conserved and used to realize zero density in a nonstatic FLRW description. Its applicability is determined by these assumptions, rather than by naming a Jordan or Einstein frame. A conformal transformation or a different split of the gravitational equations can change the definitions and exchange of the effective sectors, and those must be assessed in the chosen description.

To close the homogeneous two-component system, one can specify the standard source EoS $p_{\rm s}(\rho_{\rm s})$ and the pressure $p_{\rm ze}$, or equivalently the exchange rate compatible with Eq.~\eqref{eq:zero-int2}. Alternatively, one can specify the expansion history and reconstruct the source variables once the standard source EoS is fixed. This freedom also makes clear why the background split does not uniquely identify physical constituents. We next examine its connections to established creation and dissipative-pressure descriptions before constructing explicit backgrounds.

\subsection{Relation to Hoyle's Creation Field}
The steady-state universe, proposed by Bondi, Gold, and Hoyle in 1948, is based on the perfect cosmological principle, according to which the Universe is homogeneous not only in space but also in time. In an expanding universe, this requires the continuous creation of pressureless matter in order to compensate for the dilution caused by the expansion and thereby maintain a constant matter density~\cite{Bondi:1948qk,Hoyle:1948zz}. The resulting steady-state model is spatially flat and exponentially expanding, $a(t)\propto e^{Ht}$ with $H={\rm const.}$, and is therefore de Sitter-like at the background level. Hoyle implemented this idea by modifying the EFE through the introduction of a creation tensor $C_{\mu\nu}$~\cite{Hoyle:1948zz}, as follows
\begin{equation}\label{fe:hoyle}
G_{\mu\nu}-C_{\mu\nu}=\kappa T_{\mu\nu}.
\end{equation}
In this framework, the ordinary matter sector is not conserved independently owing to the continuous creation of matter. This is analogous to the component-wise energy exchange stated above. The creation tensor is defined as
\begin{equation} \label{c-field}
C_{\mu\nu}=\partial_\nu C_\mu-\Gamma^\alpha_{\nu\mu}C_\alpha,
\end{equation}
where $\Gamma^\alpha_{\nu\mu}$ denotes the Christoffel symbol associated with the spacetime metric $g_{\mu\nu}$. The creation vector is chosen as follows
\begin{equation}
C_\mu=\left(3/\mathcal{A},0,0,0\right),
\label{vector}
\end{equation}
where $\mathcal{A}$ is a constant. We use $\mathcal{A}$ in place of the symbol ``$a$'' adopted in Hoyle's original paper~\cite{Hoyle:1948zz} to avoid confusion with the scale factor $a$. With the choice given in Eq.~\eqref{vector}, Eq.~\eqref{c-field} becomes
\begin{equation}
C_{\mu\nu}=-\Gamma^0_{\nu\mu}C_0.
\end{equation}
For the RW metric in cosmic time, $g_{00}=-1$ and $g_{0i}=0$, so $\Gamma^0_{00}=0$ and $C_{00}=0$. The nonvanishing components are instead the spatial ones,
\begin{equation} \label{eq:Cij}
C_{ij}=-\frac{3}{2\mathcal{A}}\partial_0 g_{ij}.
\end{equation}
A key feature of the $C_{\mu\nu}$ tensor is that, within a spatially homogeneous RW spacetime, its temporal component vanishes, $C_{00}=0$, and therefore makes no effective energy density contribution to the Friedmann equation~\cite{Hoyle:1948zz}. This distinguishes the 1948 construction from the later Hoyle--Narlikar C-field theory, in which the creation field carries a negative energy density~\cite{Hoyle:1963}. Equation~\eqref{fe:hoyle} shows that moving $-C_{\mu\nu}$ from the left-hand side to the right-hand side is equivalent to introducing an effective material source with EMT $T_{\mu\nu}^{\rm(C)}=C_{\mu\nu}/\kappa$, characterized by vanishing energy density but nonvanishing pressure. However, the spatial components of $C_{\mu\nu}$, which determine the effective pressure of this source, are not arbitrary in Hoyle's model but are fixed by the spacetime geometry, see Eq.~\eqref{eq:Cij}. This is the sense in which Hoyle's creation field provides a specific realization of the zero-density source; the assumptions giving the exact background correspondence are presented in Sec.~\ref{sec:ssh}.

\subsection{Relation to Particle Creation Processes and Bulk Viscous Fluids}

In cosmological models with gravitationally induced particle creation, the macroscopic effects of particle production are represented by an effective creation pressure that supplements the equilibrium thermodynamic pressure of the cosmic fluid. On a spatially flat, homogeneous and isotropic background, the relevant Friedmann equations are $3 H^2= \kappa\,\rho_{\rm s}$ and $-2 \dot{H}-3 H^2=\kappa (p_{\rm s}+p_{\rm c})$, which have the same form as Eqs.~\eqref{zero-fe00} and~\eqref{zero-fe11}.

For adiabatic particle creation, characterized by constant entropy per particle, the creation pressure reads
\begin{align}
    \label{pres:creation}
    p_{\rm c}=-\frac{\Gamma}{3 H}(\rho_{\rm s}+p_{\rm s}),
\end{align}
as follows from the thermodynamics of open systems~\cite{Calvao:1991wg,Paliathanasis:2016dhu,Schiavone:2026agq}. Here, $\Gamma$ denotes the particle creation rate, with $\Gamma>0$ corresponding to particle production, while its functional form is specified phenomenologically.

Within GR, the macroscopic effect of particle creation is incorporated into the EMT of the particle-producing fluid through the additional creation pressure $p_{\rm c}$. Although particle creation is a nonequilibrium thermodynamic process, homogeneity and isotropy allow its contribution to the background dynamics to be represented by an additional isotropic pressure. The effective EMT therefore takes the perfect fluid form, with energy density $\rho_{\rm s}$ and effective pressure $p_{\rm eff}=p_{\rm s}+p_{\rm c}$. Its covariant conservation thus yields~\cite{Schiavone:2026agq}
\begin{align}
    \dot{\rho}_{\rm s}+3H(\rho_{\rm s}+p_{\rm s}+p_{\rm c})=0.
\end{align}
From the two-component perspective adopted here, the creation-pressure term may be represented as an algebraically separate contribution to the effective EMT, with $T_{\mu\nu}^{({\rm c})}=p_{\rm c}h_{\mu\nu}$. This contribution has vanishing comoving energy density, $\rho_{\rm c}=0$, and nonzero pressure $p_{\rm c}$. It can therefore be mapped onto the zero-density source considered in the present framework through the identification $p_{\rm ze}=p_{\rm c}$. Using Eq.~\eqref{eq:zero-int2}, the corresponding interaction term is then
\begin{align}
\mathcal Q=\Gamma(\rho_{\rm s}+p_{\rm s}).
\end{align}
The correspondence established here holds at the level of the homogeneous background dynamics and does not imply an identical physical interpretation. In particle-creation models, $p_{\rm c}$ is a nonequilibrium pressure contribution associated with the particle-producing fluid and is determined by Eq.~\eqref{pres:creation}. In the present framework, by contrast, $p_{\rm ze}$ is assigned to a zero-density interacting contribution and need not be associated with particle production. A specific microscopic interpretation cannot be inferred from the effective EMT alone and would require additional physical input.

An analogous effective pressure correspondence arises in bulk viscous cosmology. In the standard first-order description, bulk viscosity contributes a dissipative pressure $\Pi$ in addition to the equilibrium pressure, so that
\begin{align}
    p_{\rm eff}=p_{\rm s}+\Pi, \qquad \Pi=-3H\zeta,
\end{align}
where $\Pi$ is the bulk viscous pressure and $\zeta\geq0$ is the corresponding viscosity coefficient. For an FLRW background, the spacetime symmetries, namely spatial homogeneity and isotropy, exclude heat flux and anisotropic stresses, so the dissipative sector is fully characterized by the bulk viscous pressure~\cite{Zimdahl:1996fj}. The identification $p_{\rm ze}=\Pi$ therefore renders the homogeneous background dynamics of the bulk viscous description formally equivalent to those of the present zero-density source framework. Their physical interpretations, however, remain distinct: $\Pi$ is a nonequilibrium pressure associated with an imperfect fluid, whereas $p_{\rm ze}$ is assigned to a separate zero-density contribution in our decomposition, without specifying dissipative microphysics. Together with the relation $p_{\rm ze}=p_{\rm c}$ established above, this identification implies $p_{\rm c}=\Pi$, recovering the familiar background-level correspondence between particle creation and bulk viscosity. The two processes are nevertheless not thermodynamically equivalent~\cite{Lima:1992np}.

\subsection{Background correspondences and their domains}
\label{sec:background-map}

The pressure-only representation organizes several established background models, but the component densities and microscopic mechanisms need not agree. Table~\ref{tab:background-map} summarizes the identifications used here and in Sec.~\ref{sec:particular}. In particular, for dust, $p_{\rm ze}=\alpha\rho_{\rm s}$ corresponds to $\Gamma=-3\alpha H$, while constant $p_{\rm ze0}$ corresponds to $\Gamma=-\kappa p_{\rm ze0}/H$ where $H\neq0$. The latter gives the creation cold dark matter (CCDM) background of Ref.~\cite{Lima:2009ic}; broader creation histories and their early-time consistency have been studied in Ref.~\cite{Steigman:2008bc}. For $H>0$ and $\rho_{\rm s}+p_{\rm s}>0$, genuine particle production ($\Gamma>0$), or nonnegative first-order bulk viscosity, requires $p_{\rm ze}<0$. Positive-pressure solutions below remain valid interacting EMT backgrounds, but they do not describe positive-rate particle creation or positive-viscosity expansion under these identifications.

\begin{table*}[t]
\caption{Background correspondences and their domains in spatially flat GR, with $\rho_{\rm ze}=0$ and $\mathcal Q=-3Hp_{\rm ze}$. The density $\rho_{\rm s}$ is that of the standard component in the pressure-only representation. Conditions on creation and viscosity refer to $H>0$; the table does not identify the underlying microscopic theories or their perturbations. The running vacuum model uses a different dust density, made explicit in Eq.~\eqref{eq:running-vacuum-map}.}
\label{tab:background-map}
\small
\renewcommand{\arraystretch}{1.23}
\begin{tabular}{@{}p{0.300\textwidth}p{0.235\textwidth}p{0.415\textwidth}@{}}
\toprule
Description & Pressure-only identification & Domain and interpretation \\
\midrule
Hoyle's 1948 creation tensor~\cite{Hoyle:1948zz}
& $p_{\rm ze}=-3H/(\kappa\mathcal A)$
& Dust standard source; $\mathcal A>0$. Its steady-state solution has $H=1/\mathcal A$ and $p_{\rm ze}=-\rho_{\rm s}$. \\
Adiabatic particle creation~\cite{Calvao:1991wg,Steigman:2008bc}
& $p_{\rm ze}=-\Gamma(\rho_{\rm s}+p_{\rm s})/(3H)$
& $\mathcal Q=\Gamma(\rho_{\rm s}+p_{\rm s})$. For positive inertial density, $\Gamma>0$ gives negative pressure. \\
Constant-pressure CCDM~\cite{Lima:2009ic,Ramos:2014dba}
& $p_{\rm ze}=p_{\rm ze0}<0$, $p_{\rm s}=0$
& $\Gamma=-\kappa p_{\rm ze0}/H$. A conventional dust plus positive-$\Lambda$ history requires $\rho_{\rm s0}+p_{\rm ze0}>0$. \\
First-order bulk viscosity~\cite{Zimdahl:1996fj,Li:2009mf,Velten:2011bg}
& $p_{\rm ze}=-3H\zeta$
& $\zeta\geq0$ implies $p_{\rm ze}\leq0$. A dissipative pressure law is additional constitutive input. \\
Vacuum term $\Lambda=3\beta H^2$~\cite{Carvalho:1991ut,Basilakos:2009wi}
& $p_{\rm ze}=-\beta\rho_{\rm s}$, $p_{\rm s}=0$
& $\alpha=-\beta$ and $\rho_{\rm m}^{(\Lambda)}=(1-\beta)\rho_{\rm s}$. Positive dust and positive vacuum require $0<\beta<1$. \\
Simple-graduated dark energy~\cite{Acquaviva:2021jov}
& $p_{\rm s}=-\rho_{\rm s}$, $p_{\rm ze}=p_{\rm ze0}$
& The combined source has constant $\mathcal I_{\rm tot}=p_{\rm ze0}$; use the standard source with $\rho_{\rm s}=3H^2/\kappa\geq0$. \\
Scale-independent EMSG, dust 

background~\cite{Akarsu:2023nyl}
& $p_{\rm ze}=\alpha\rho_{\rm s}$, $p_{\rm s}=0$
& The dust background equations coincide; the added pressure is an effective modification rather than an independently chosen material constituent. \\
\bottomrule
\end{tabular}
\end{table*}

The role of the perturbative treatment is already apparent in these established models. For the CCDM construction, Ref.~\cite{Ramos:2014dba} finds a $\Lambda$CDM degeneracy when the effective sound speed is negligible and the clustering density contrast is defined appropriately. In the bulk viscous models of Refs.~\cite{Li:2009mf,Velten:2011bg}, perturbations instead impose restrictions not apparent from the expansion history alone. These contrasting examples motivate the unnormalized perturbative formulation of Sec.~\ref{sec:perturbation} and the action-level calculation of Appendix~\ref{app:growth}; they do not establish perturbative equivalence in general or rule out pressure-only representations.

\section{Particular Models With a Zero Energy Density Source}
\label{sec:particular}
Having established the interacting FLRW framework and its relation to existing cosmological constructions, we now turn to explicit background solutions for a two-component universe composed of a standard source and a zero energy density source satisfying $\rho_{\rm ze}=0$. We assume that the standard component is barotropic, $p_{\rm s}=w\rho_{\rm s}$ with constant EoS parameter $w$, and close the system by choosing a form for the pressure $p_{\rm ze}$ of the zero-density source. Through Eq.~\eqref{eq:zero-int2}, this choice also fixes the interaction kernel as $\mathcal Q=-3Hp_{\rm ze}$.
Before considering particular forms of $p_{\rm ze}$, it is useful to examine how this pressure modifies the evolution of the standard component. Under the above assumptions, the total continuity equation~\eqref{zero-cont} can be recast in the integral form to give
\begin{align}
\label{sol:rho_s}
\rho_{\rm s} \exp\left(
3\int\frac{{\rm d}a}{a}\,
\frac{p_{\rm ze}}{\rho_{\rm s}}
\right)
=
\mathcal{C}_0\,a^{-3(1+w)},
\end{align}
where $\mathcal{C}_0$ is an integration constant. The exponential factor quantifies the departure of $\rho_{\rm s}$ from its usual noninteracting dilution law. In particular, if the standard component is desired to retain its conventional scaling $\rho_{\rm s}\propto a^{-3(1+w)}$, this factor must remain constant, which can only be satisfied when $p_{\rm ze}=0$. 

Thus, the nonvanishing pressure of the zero-density source necessarily modifies the dilution of the standard source through the interaction between the two components. Because $\rho_{\rm s}$ itself enters the integrand, Eq.~\eqref{sol:rho_s} is an implicit representation of the continuity equation~\eqref{zero-cont}. An explicit evolution for $\rho_{\rm s}(a)$ is obtained only after specifying the ratio $p_{\rm ze}/\rho_{\rm s}$, or more nontrivially $p_{\rm ze}$ itself. A complementary approach is to characterize the model directly by specifying its expansion history. Using $p_{\rm s}=w\rho_{\rm s}$, Eqs.~\eqref{zero-fe00} and \eqref{zero-fe11} can then be combined to express the pressure of the zero-density source in terms of $H(a)$ as
\begin{align}
\kappa p_{\rm ze}
=
-2aH\frac{{\rm d}H}{{\rm d}a}-3(1+w)H^2,
\end{align}
 and
\cref{zero-fe00} further gives $\rho_{\rm s}=3H^2/\kappa$.

These relations show that a given expansion history $H(a)$ uniquely determines the corresponding $\rho_{\rm s}(a)$ and $p_{\rm ze}(a)$. We develop this inverse construction and its observational interpretation in Sec.~\ref{sec:inference}. In what follows, we instead consider several choices of $p_{\rm ze}$ and study the resulting interaction kernels and background evolutions, classifying the models according to the pressures of the zero-density source.

\subsection{Hoyle's steady-state universe}
\label{sec:ssh}

To make the connection with the discussion of Hoyle's creation tensor explicit, we first recover Hoyle's steady-state universe within the present framework. Choosing the standard component to be pressureless matter, $p_{\rm m}=0$, the background equations of Hoyle's model become
\begin{align}
3H^2 &= \kappa\rho_{\rm m},
\label{eq:hoyle-fe00}
\\
-2\dot H-3H^2+\frac{3}{\mathcal A}H &= 0.
\label{eq:hoyle-fe11}
\end{align}
Comparing Eq.~\eqref{eq:hoyle-fe11} with the second Friedmann equation of the two-component system, viz., 
$-2\dot H-3H^2=\kappa p_{\rm ze}$ obtained from Eq.~\eqref{zero-fe11},
identifies the pressure of the zero-density source as
\begin{align}
p_{\rm ze} = -\frac{3}{\kappa\mathcal A}H.
\label{eq:hoyle-pn}
\end{align}
The interaction kernel is consequently given by
\begin{align}
\mathcal Q=-3Hp_{\rm ze} = \frac{9}{\kappa\mathcal A}H^2.
\label{eq:hoyle-Q}
\end{align}
For an expanding universe with $\mathcal A>0$, one has $p_{\rm ze}<0$ and $\mathcal Q>0$, which corresponds, according to our sign convention, to an energy transfer from the zero-density source to the matter component.
The steady-state condition requires the Hubble parameter to remain constant. Setting $\dot H=0$ in Eq.~\eqref{eq:hoyle-fe11} gives the nontrivial expanding solution $H=1/{\mathcal A}.$
The scale factor therefore evolves exponentially as
$a(t)\propto e^{t/{\mathcal A}}$.
Equation~\eqref{eq:hoyle-fe00} gives the constant matter density, while Eq.~\eqref{eq:hoyle-pn} determines the corresponding pressure of the zero-density source:
\begin{align}
\quad p_{\rm ze}
=-\rho_{\rm m}= -\frac{3}{\kappa\mathcal A^2}.
\label{eq:hoyle-steady-pressure}
\end{align}
The same condition follows directly from the total continuity equation~\eqref{zero-cont}, which becomes
$\dot\rho_{\rm m} +
3H\left(\rho_{\rm m}+p_{\rm ze}\right) =0$ in the present case.
For a constant matter density, $\dot\rho_{\rm m}=0$, and an expanding background, $H>0$, this equation requires $p_{\rm ze}=-\rho_{\rm m}$. The interaction kernel then reduces to $\mathcal Q=3H\rho_{\rm m}$.
Thus, in Hoyle's steady-state universe, the continuous transfer of energy to the matter component exactly compensates for its dilution due to cosmic expansion. The Universe can therefore expand exponentially while maintaining a constant matter density.

\subsection{Proportional-Pressure Model}

\label{sec:prop-pres}

The formal solution~\eqref{sol:rho_s} does not determine a unique background evolution unless the ratio $p_{\rm ze}/\rho_{\rm s}$ is specified. Different choices for this ratio therefore define distinct cosmological scenarios and lead to different expansion histories. We investigate the simplest case, in which $p_{\rm ze}/\rho_{\rm s}$ is constant. In Hoyle's steady-state solution discussed above, the requirement of a constant pressureless matter density fixes this ratio to $p_{\rm ze}/\rho_{\rm m}=-1$. We generalize this relation by allowing an arbitrary constant ratio without restricting the standard component to pressureless matter. Accordingly, we consider a barotropic standard source with constant EoS parameter, and take the pressure of the zero-density source to be proportional to $\rho_{\rm s}$:
\begin{equation}
p_{\rm s}=w\,\rho_{\rm s}, \qquad  \qquad p_{\rm ze}=\alpha\,\rho_{\rm s},
\end{equation}
with $\alpha$ being a constant. Through Eq.~\eqref{eq:zero-int2}, this choice also fixes the interaction kernel to $\mathcal Q=-3H\alpha\rho_{\rm s}$. The continuity equation~\eqref{zero-cont} thus becomes
\begin{equation}\label{eq:model1-cont}
    \dot{\rho}_{\rm s}+3H\!\left(1+w+\alpha\right)\rho_{\rm s}=0.
\end{equation}
Normalizing the scale factor at the present time as $a(t_0)=a_0=1$, the general solution for this case reads
\begin{align}
\rho_{\rm s}= \rho_{\rm s0}\,a^{-3(1+w+\alpha)},
\label{generalRho1}
\end{align}
where a subscript $0$ denotes the present-day value. The Friedmann equations~\eqref{zero-fe00}--\eqref{zero-fe11} take the following forms:
\begin{align} \label{model1-e00}
    3 H^2&=\kappa\rho_{\rm s},\\
    \label{model1-e11}
    -2 \dot{H}-3 H^2&=\kappa (w+\alpha) \rho_{\rm s}.
\end{align}
Equation~\eqref{generalRho1} implies that $\rho_{\rm s}$ is always positive for $\rho_{\rm s0}>0$. The condition $w+\alpha=-1$ makes $\rho_{\rm s}$ constant and gives rise to an exponentially expanding de Sitter background. For pressureless matter, $w_{\rm m}=0$, this requires $\alpha=-1$ and reproduces Hoyle's steady-state solution. For radiation, $w_{\rm r}=1/3$, the corresponding value is $\alpha=-4/3$. The conventional vacuum energy case is recovered for $w_{\rm vac}=-1$ and $\alpha=0$, for which the zero-density source has vanishing pressure, so that its EMT vanishes. We also note that, for $w_{\rm m}=0$, the homogeneous FLRW background equations of the present model coincide with those obtained in scale-independent energy-momentum squared gravity (EMSG)~\cite{Akarsu:2023nyl}.

For $w_{\rm m}=0$, the same total density and pressure also arise in the flat time dependent vacuum model with $\Lambda=3\beta H^2$ and constant $G$~\cite{Carvalho:1991ut,Basilakos:2009wi}. The component map is
\begin{align}
\alpha=-\beta,\qquad
\rho_{\Lambda(H)}=\beta\rho_{\rm s},\qquad
\rho_{\rm m}^{(\Lambda)}=(1-\beta)\rho_{\rm s}.
\label{eq:running-vacuum-map}
\end{align}
Here $\rho_{\rm m}^{(\Lambda)}$, not $\rho_{\rm s}$, is the dust density coupled to the time varying vacuum. It obeys $\dot\rho_{\rm m}^{(\Lambda)}+3H\rho_{\rm m}^{(\Lambda)}=3\beta H\rho_{\rm m}^{(\Lambda)}$, whereas our standard component has $\mathcal Q=-3H\alpha\rho_{\rm s}$. Positive conventional dust requires $\alpha>-1$; positive vacuum density further requires $\alpha<0$. The $\alpha=-1$ limit is pure vacuum in that representation but a nonzero constant dust density in our interacting split. For $\alpha<-1$, the component map gives negative $\rho_{\rm m}^{(\Lambda)}$ and loses the ordinary dust interpretation. Thus, matching $H(a)$ does not identify the constituent densities or the transfer functions.

The background dynamics of the proportional-pressure model are governed entirely by the combination $w+\alpha$. Depending on whether this quantity is greater than, equal to, or less than minus unity, the solution space is partitioned into three distinct regions:
\begin{enumerate}[nosep,wide,label=(\roman*)]
   \item The case $w+\alpha>-1$: Choosing the origin of cosmic time at the big bang, such that $a(0)=0$, the scale factor and the energy density of the standard source evolve as
    \begin{equation}
        \begin{aligned}
    a(t)&=\left(\frac{t}{t_0}\right)^{\frac{2}{3(1+w+\alpha)}},\\
        \rho_{\rm s}(t)&=\rho_{\rm s0}\left(\frac{t}{t_0}\right)^{-2}.
        \end{aligned}
    \end{equation}
    Here, $t_0$ denotes the present age of the Universe and is related to the present-day Hubble parameter $H_0$ by
    \begin{equation}
        t_0=\frac{2}{3(1+w+\alpha)H_0}.
    \end{equation}

    From the second Friedmann equation~\eqref{model1-e11}, $\ddot a/a=-(\kappa/6)\rho_{\rm s}[1+3(w+\alpha)]$, we observe that the Universe undergoes decelerated expansion for $w+\alpha>-1/3$ and accelerated expansion for $-1<w+\alpha<-1/3$. At the boundary value $w+\alpha=-1/3$, the scale factor grows linearly with cosmic time, $a(t)=t/t_0$, leading to an expansion with no acceleration $\ddot a=0$. Also, the special case $p_{\rm s}=-p_{\rm ze}$, corresponding to $w+\alpha=0$, gives $a(t)=(t/t_0)^{2/3}$ and $\rho_{\rm s}=\rho_{\rm s0}a^{-3}$. Thus, at the background level, the combined system reproduces the expansion history of a spatially flat matter-dominated universe, even though the standard source need not itself be pressureless; its pressure is exactly cancelled by that of the zero-density source.

   \item The case $w+\alpha=-1$: In this case, the continuity equation~\eqref{eq:model1-cont} gives $\dot{\rho}_{\rm s}=0$, so the energy density of the standard source remains constant. Eq.~\eqref{model1-e00} then yields a constant Hubble parameter,
$H=H_0=\sqrt{\kappa\rho_{\rm s0}/3}$.
   The resulting de Sitter-like solution is represented by
   \begin{equation}
   a(t)=e^{H_0(t-t_0)},
   \qquad
   \rho_{\rm s}(t)=\rho_{\rm s0}=\mathrm{const.}
   \end{equation}

    \item The case $w+\alpha<-1$: In this regime, the effective EoS is phantom-like, i.e., $w_{\rm eff}=w+\alpha<-1$ since the energy density of the standard source~\eqref{generalRho1} is always positive. The expanding solution can be written as
    \begin{equation}
        \begin{aligned}
            a(t)&=\left(\frac{t_{\rm c}-t}{t_{\rm c}-t_0}\right)^{\frac{2}{3(1+w+\alpha)}},\\
            \rho_{\rm s}(t)&=\rho_{\rm s0}
            \left(\frac{t_{\rm c}-t}{t_{\rm c}-t_0}\right)^{-2},
        \end{aligned}
    \end{equation}
    where the time of finite future singularity is given by
    \begin{equation}
        t_{\rm c}=t_0-\frac{2}{3(1+w+\alpha)H_0}.
    \end{equation}
    Since $w+\alpha<-1$, one has $t_{\rm c}>t_0$. As $t\to t_{\rm c}^{-}$, the scale factor diverges, and the Universe reaches a big rip singularity at a finite future cosmic time~\cite{Caldwell:1999ew,Caldwell:2003vq}. The density $\rho_{\rm s}\propto a^{-3(1+w+\alpha)}$ grows during expansion in this regime because the exponent is positive. This phantom-like evolution does not require the standard source itself to have an intrinsically phantom EoS, $w<-1$. Instead, it arises from the pressure of the zero-density source and the associated interaction, which drive the effective EoS parameter below $-1$. For a pressureless standard source, $w=0$, the phantom regime requires $\alpha<-1$. We restrict our attention to the expanding universe, $H>0$, and do not consider the corresponding contracting solutions in this model.

\end{enumerate}

\subsection{Constant-Pressure Model}
\label{sec:lcdm}

As a second example, we consider a model comprising a barotropic standard source with a constant EoS parameter interacting with a zero-density source of constant nonzero pressure, $p_{\rm ze}=p_{\rm ze0}$:
\begin{align}  \label{const-pres}
  p_{\rm s} = w \rho_{\rm s},  \qquad p_{\rm ze} = p_{\rm ze0} = \text{const}.
\end{align}
The Friedmann equations~\eqref{zero-fe00} and~\eqref{zero-fe11} for these choices become
\begin{align}
    3H^2
    &=\kappa\rho_{\rm s},
    \label{fe1_consp}\\
    -2\dot H-3H^2
    &=\kappa\left(w\rho_{\rm s}+p_{\rm ze0}\right).
    \label{fe2_consp}
\end{align}
Accordingly, the continuity equation~\eqref{zero-cont} is given by
\begin{align}
\label{cont_consp}
    \dot{\rho}_{\rm s} + 3H\left[(1+w)\rho_{\rm s} + p_{\rm ze0}\right] &= 0.
\end{align}
For $w\neq-1$, integrating it, we find
\begin{align}
\label{rho_s_zero_lcdm}
\rho_{\rm s} =\left(\rho_{\rm s0}
+ \frac{p_{\rm ze0}}{1+w} \right) a^{-3(1+w)}
- \frac{p_{\rm ze0}}{1+w}.
\end{align}
Interestingly, the solution~\eqref{rho_s_zero_lcdm} retains the usual barotropic scaling,
$a^{-3(1+w)}$, but the constant pressure $p_{\rm ze0}$ modifies the
coefficient of the evolving term and introduces an additional constant
contribution to $\rho_{\rm s}$. Furthermore, at the background level, this constant term acts as an effective cosmological constant contribution. In particular, if one chooses the standard source to be the familiar pressureless matter, i.e., $w_{\rm m}=0$, the
resulting expansion history reproduces that of a spatially flat
$\Lambda$CDM model
\begin{align}
    \frac{H^2}{H^2_0}=
    (1+\mathcal{P}_{\rm ze0})a^{-3}- \mathcal{P}_{\rm ze0}
\end{align}
where $\mathcal P_{\rm ze0}\equiv p_{\rm ze0}/\rho_{\rm cr0}$ and $\rho_{\rm cr0}=3H_0^2/\kappa=\rho_{\rm s0}$ from Eq.~\eqref{fe1_consp}. The corresponding flat-$\Lambda$CDM parameters are
\begin{align}
\Omega_{{\rm m}0}^{\Lambda{\rm CDM}}&=1+\mathcal P_{\rm ze0},\\
\Omega_{\Lambda}^{\Lambda{\rm CDM}}&=-\mathcal P_{\rm ze0}.
\label{eq:lcdm-parameters}
\end{align}
The usual positive-dust, positive-vacuum case has $-1<\mathcal P_{\rm ze0}<0$. In the interacting representation, the entire background density is assigned to the standard component, so its fractional density is unity. This is a property of the source split and should not be interpreted as a prediction that the observationally inferred, separately conserved matter fraction is unity.

The possible degeneracy extends beyond the background for a particular realization. Suppose that $p_{\rm ze}=-\rho_\Lambda$ is constant throughout spacetime and that both contributions share one four-velocity. Then, the total EMT is
\begin{align}
T_{\rm tot}^{\mu\nu}
&=\rho_{\rm m}u^\mu u^\nu-\rho_\Lambda h^{\mu\nu}\nonumber\\
&=(\rho_{\rm m}-\rho_\Lambda)u^\mu u^\nu-\rho_\Lambda g^{\mu\nu}.
\label{eq:lcdm-exact-split}
\end{align}
Defining $\tilde{\rho}_{\rm d}=\rho_{\rm m}-\rho_\Lambda$, total conservation gives $\nabla_\mu(\tilde{\rho}_{\rm d}u^\mu u^\nu)=0$. Wherever $\tilde{\rho}_{\rm d}\geq0$, the total EMT is thus exactly that of conserved dust plus vacuum energy. This stronger equivalence relies on the assumptions of common four-velocity and constant-pressure along spacetime; a constant background pressure alone does not impose it on linear perturbations or on the scalar field reconstructions below. The same constant-pressure dust background was obtained in CCDM cosmology~\cite{Lima:2009ic}. Its perturbative degeneracy with $\Lambda$CDM for a negligible effective sound speed and a consistently defined clustering contrast was analyzed in Ref.~\cite{Ramos:2014dba}. Those results define the necessary constraints beyond background matching; a different pressure or momentum-transfer function need not share them. For the covariant dust--scalar completion of Appendix~\ref{app:action}, the same background is realized with a scalar-mediated dust force $G_{\rm eff}/G=1-6\Omega_\Lambda(a)$, which changes the sign of the dark-matter growth rate at low redshift; see Sec.~\ref{app:constant-pressure-growth}.

For completeness, we consider the exceptional case $w_{\rm vac}=-1$, which is not contained in the general solution derived below. The standard component
then has a vacuum-form EMT but need not be separately conserved. Solving the continuity equation \eqref{cont_consp} for this case yields
\begin{align}
\rho_{\rm vac}= \rho_{\rm vac0}-3p_{\rm ze0}\ln a.
\end{align}
We note that this logarithmic evolution of the standard source for $w=-1$ coincides
with that of simple graduated dark energy (simple-gDE)~\cite{Acquaviva:2021jov}, a special $\lambda=0$ case of
graduated dark energy (gDE)~\cite{Akarsu:2019hmw}. Here the equivalence concerns the total background density and pressure, $\rho_{\rm tot}=\rho_{\rm vac}$ and $p_{\rm tot}=-\rho_{\rm vac}+p_{\rm ze0}$, so $\mathcal I_{\rm tot}=p_{\rm ze0}$ is constant. In the spatially flat two-component model, only the portion with $\rho_{\rm vac}=3H^2/\kappa\geq0$ is physical. For this constant inertial density case, the evolution can also be written directly in terms of cosmic time, as 
\begin{align}
H(t)&=H_0-\frac{\kappa p_{\rm ze0}}2(t-t_0),\\
a(t)&=\exp\!\left[H_0(t-t_0)-\frac{\kappa p_{\rm ze0}}4(t-t_0)^2\right].
\label{eq:simple-gde-time}
\end{align}
For $p_{\rm ze0}<0$, $H$ and $a$ grow without bound only as $t\to\infty$, while $\dot H$ is a positive constant. This is the little sibling of the big rip behavior~\cite{Bouhmadi-Lopez:2014cca} of constant negative inertial density~\cite{Acquaviva:2021jov}; it is distinct from the finite-time big rip in the proportional-pressure case. For $p_{\rm ze0}>0$, an initially expanding branch reaches $H=0$ and turns around. The restriction $\rho_{\rm s}=3H^2/\kappa\geq0$ still applies to this pure two-component model.

After discussing these two special values of $w$, we proceed to derive the background evolution for a general constant $w$. Setting $\mathcal{P}_{\rm ze0}=-0.7$ yields the benchmark flat $\Lambda$CDM values $\Omega_{\rm m0}^{\Lambda\mathrm{CDM}}=0.3$ and $\Omega_{\Lambda0}^{\Lambda\mathrm{CDM}}=0.7$. We use this value only as an illustrative benchmark; the analytic solutions below retain arbitrary $\mathcal P_{\rm ze0}$ in their stated domains. The evolution equation for $H$ can be obtained by combining the first and second Friedmann equations \eqref{fe1_consp} and \eqref{fe2_consp}, as follows
\begin{align}
\label{Hdot}
\dot H = -\frac{3}{2}(1+w)H^2
-\frac{3}{2}H_0^2\mathcal{P}_{\rm ze0}.
\end{align}
\begin{figure*}[t]
\centering
\includegraphics[width=0.98\textwidth]{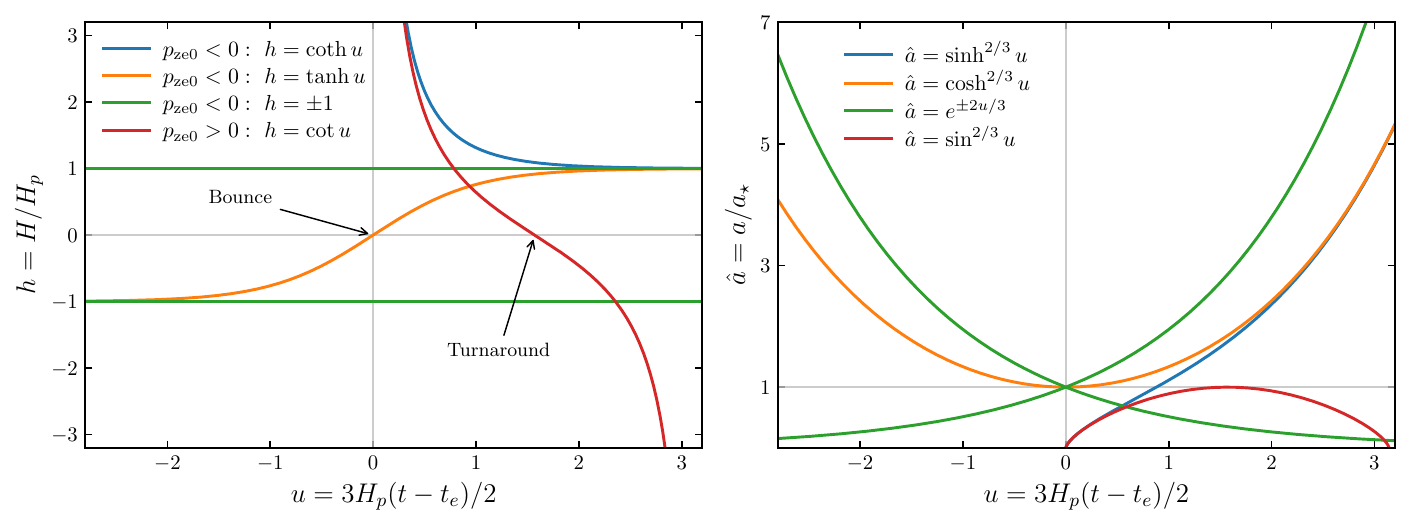}
\caption{Constant-pressure solutions with dust as the standard component ($w_{\rm m}=0$): $h=H/H_p$ (left) and $\hat a=a/a_\star$ (right), against $u=3H_p(t-t_e)/2$, where $H_p=\sqrt{\kappa|p_{\rm ze0}|/3}$. Matching colors correspond to the same solution branch. Negative pressure gives $(h,\hat a)=(\coth u,\sinh^{2/3}u)$ for the expanding big bang ($u>0$), $(\tanh u,\cosh^{2/3}u)$ for the bounce, and $(\pm1,e^{\pm2u/3})$ for the de Sitter limits. Positive pressure gives $(\cot u,\sin^{2/3}u)$ on $0<u<\pi$, with turnaround at $u=\pi/2$. The independent positive normalization $a_\star$ is the bounce minimum, the recollapse maximum, or the de Sitter value at $t_e$; for the big bang branch it is the integration amplitude. Each solution has its own origin $t_e$: the big bang time for the singular solutions, the bounce time, or an arbitrary reference time for de Sitter. The curves represent distinct cosmologies, not successive phases of one universe. The right panel approaches the singular $a\to0$ limits from within their physical domains; the corresponding divergences of $H$ lie outside the vertical range of the left panel.}
\label{fig:pressure-branches}
\end{figure*}

A negative pressure of the zero-density source
$\mathcal{P}_{\rm ze0}<0$ is of particular interest, since it corresponds to a repulsive contribution
to the background dynamics. The character of the solutions is then determined by the sign of $1+w$. We restrict our attention to the case $w>-1$, and define $\Delta\equiv \frac{3}{2}H_0\sqrt{-\mathcal{P}_{\rm ze0}(1+w)}$.
The general solution of Eq.~\eqref{Hdot} can then be written as
\begin{align}
H(t)=H_0\sqrt{-\frac{\mathcal{P}_{\rm ze0}}{1+w}}\,
\frac{\mathcal{C}_1e^{\Delta t}-\mathcal{C}_2e^{-\Delta t}}
{\mathcal{C}_1e^{\Delta t}+\mathcal{C}_2e^{-\Delta t}},
\end{align}
where $\mathcal{C}_1$ and $\mathcal{C}_2$ are integration constants. Further integration of $H=\dot a/a$ gives the corresponding scale factor,
\begin{align}
a(t)=\left[\mathcal{C}_1e^{\Delta t}+\mathcal{C}_2e^{-\Delta t}\right]^{\frac{2}{3(1+w)}}.
\end{align}
For this choice of parameter ranges, namely
$\mathcal{P}_{\rm ze0}<0$ and $w>-1$, the general solution admits three distinct cosmological branches. Depending on the integration constants, $\mathcal{C}_1\mathcal{C}_2<0$ yields a hyperbolic sine solution with a big bang/big crunch-type singularity, $\mathcal{C}_1\mathcal{C}_2>0$ yields a hyperbolic
cosine solution describing a nonsingular bounce, and $\mathcal{C}_1\mathcal{C}_2=0$ corresponds to the limiting de Sitter solution. The three branches in the $w>-1$ case can equivalently be distinguished by the sign of the
coefficient of the evolving term in
\cref{rho_s_zero_lcdm}. In particular,
$\mathcal{P}_{\rm ze0}>-(1+w)$,
$\mathcal{P}_{\rm ze0}=-(1+w)$, and
$\mathcal{P}_{\rm ze0}<-(1+w)$ correspond, respectively, to the big bang branch, the de Sitter solution, and the bouncing branch. For the benchmark value $\mathcal{P}_{\rm ze0}=-0.7$, the condition for the
$\sinh$ big bang solution reduces to $w>-0.3$.

For $p_{\rm ze0}<0$ and $w>-1$, the bouncing branch has a negative coefficient of $a^{-3(1+w)}$ in Eq.~\eqref{rho_s_zero_lcdm}. Its nonnegative-density domain is $a\geq a_{\rm b}$, where $a_{\rm b}$ is the zero of $\rho_{\rm s}(a)$; the solution does not extend to smaller scale factors with ordinary positive-density matter. For the bouncing branch, we define
\begin{align}
H_p = \sqrt{-\frac{\kappa p_{\rm ze0}}{3(1+w)}}>0,
\;\,
u = \frac{3}{2}(1+w)H_p(t-t_{\rm b}),
\end{align}
and then obtain
\begin{align}
H&=H_p\tanh u,\quad
a=a_{\rm b}\cosh^{2/[3(1+w)]}u.
\label{eq:explicit-bounce}
\end{align}
At $t=t_{\rm b}$, $H=\rho_{\rm s}=0$ while $\dot H=-\kappa p_{\rm ze0}/2>0$ and $\mathcal I_{\rm tot}=p_{\rm ze0}<0$. The homogeneous curvature remains finite, so the solution describes a regular bounce, which is nevertheless accompanied by total NEC violation. This is a background solution of the chosen interacting EMT system, not a proof that a microscopic matter realization is regular or stable at $\rho_{\rm s}=0$.

We now focus on the expanding hyperbolic sine solution, which is the one that most closely resembles the standard cosmological evolution. In line with this, we assume that the Universe originates from a big bang singularity at a finite past
cosmic time, with $a\to0$, and subsequently evolves toward an accelerated phase at late times. Choosing the origin of cosmic time at the big bang, $t=t_{\rm BB}$, and defining $\vartheta\equiv H_0(t-t_{\rm BB})$, the $\sinh$ solution can be written as follows
\begin{align}
H(\vartheta)=H_0\sqrt{-\frac{\mathcal{P}_{\rm ze0}}{1+w}}\,
\coth\!\left(\frac{\Delta}{H_0}\vartheta\right).
\end{align}
Hence, the evolution of the scale factor is modified by
$\mathcal{P}_{\rm ze0}$ as
\begin{align}
a(\vartheta)=\left[-1-\frac{1+w}
{\mathcal{P}_{\rm ze0}}\right]^{\frac{1}{3(1+w)}}
\sinh^{\frac{2}{3(1+w)}}\!\left(\frac{\Delta}{H_0}\vartheta\right).
\end{align}
In the early-time limit, $\vartheta\rightarrow0^+$, one has
$a(\vartheta)\rightarrow0$ alongside $H(\vartheta)\rightarrow\infty$,
demonstrating the big bang singularity. At late times,
$H(\vartheta)\rightarrow H_0\sqrt{-\mathcal{P}_{\rm ze0}/(1+w)}$ and $\rho_{\rm s}\rightarrow-p_{\rm ze0}/(1+w)$, so that the expansion asymptotically approaches a de Sitter state.

For $\mathcal{P}_{\rm ze0}>0$, the corresponding solutions can be obtained
directly from the above results by the replacement $ \Delta \rightarrow i\Delta_+$
where $\Delta_+
    \equiv
    \frac{3}{2}H_0
    \sqrt{\mathcal{P}_{\rm ze0}(1+w)},$
so that using $\sinh(ix)=i\sin x$ and $\coth(ix)=-i\cot x$, the hyperbolic
solutions are replaced by
\begin{align}
    H(\vartheta)
    =
    H_0\sqrt{\frac{\mathcal{P}_{\rm ze0}}{1+w}}\,
    \cot\left(\frac{\Delta_+}{H_0}\vartheta\right),
\end{align}
and
\begin{align}
    a(\vartheta)
    =
    \left(1+
        \frac{1+w}
        {\mathcal{P}_{\rm ze0}}
    \right)^{\frac{1}{3(1+w)}}
    \sin^{\frac{2}{3(1+w)}}\left(
        \frac{\Delta_+}{H_0}\vartheta
    \right).
\end{align}
For $\mathcal{P}_{\rm ze0}>0$ and $w>-1$, the coefficient of the evolving
term in \cref{rho_s_zero_lcdm} is always positive, while the constant
contribution to $H^2$ is negative. This implies that $H$ necessarily vanishes at the finite scale factor
\begin{align}
    a_{\rm max}
    =
    \left(1+
        \frac{1+w}
        {\mathcal{P}_{\rm ze0}}
    \right)^{\frac{1}{3(1+w)}}.
\end{align}
Thus, unlike the $\mathcal{P}_{\rm ze0}<0$ case, the only physical branch is a recollapsing solution, evolving
from a big bang at $\vartheta=0$ to a maximum scale factor at
$\Delta_+\vartheta/H_0=\pi/2$, followed by a big crunch at
$\Delta_+\vartheta/H_0=\pi$. The total lifetime of the Universe is therefore
\begin{align}
    t_{\rm BC}-t_{\rm BB}
    =
    \frac{\pi}{\Delta_+}
    =
    \frac{2\pi}
    {3H_0\sqrt{\mathcal{P}_{\rm ze0}(1+w)}}.
\end{align}

Figure~\ref{fig:pressure-branches} compares these exact histories for $w_{\rm m}=0$ without fixing a present-day parameter fit. The left panel identifies expansion and contraction through the sign of $H$, while the right panel displays the corresponding scale-factor evolution. In particular, $H=0$ marks a minimum of $a$ on the negative-pressure bouncing branch but a maximum on the positive-pressure recollapsing branch. At either regular extremum, Eqs.~\eqref{fe1_consp} and~\eqref{fe2_consp} give $\rho_{\rm m}=0$ and $\ddot a/a=\dot H=-\kappa p_{\rm ze0}/2$. The sign of the nonzero pressure therefore distinguishes a bounce from a turnaround even when the instantaneous total energy density vanishes. The constituent condition $\rho_{\rm ze}\equiv0$ holds throughout each plotted history; the additional vanishing of $\rho_{\rm m}$, and hence of the total density, occurs only at the bounce or turnaround.

Combining these histories with the field equations also distinguishes the repulsive character of the sector from acceleration of the total background. For the negative-pressure expanding $\sinh$ branch, the total inertial density is positive even when the expansion accelerates. In the dust case, $\mathcal I_{\rm tot}/|p_{\rm ze0}|=h^2-1$ and $\mathcal M_{\rm tot}/|p_{\rm ze0}|=h^2-3$, with $h=H/H_p$: $1<h<\sqrt3$ is therefore an accelerating, total-NEC-satisfying regime. Sector repulsion holds throughout the evolution, while total acceleration begins only after the matter term becomes sufficiently small.

The background solutions above use a perfect fluid EMT form without a thermodynamic constitutive model. A microscopic fluid interpretation would additionally have to specify its state variables, pressure, and energy exchange functions. We now examine scalar field descriptions as a different route to realizing the same backgrounds.

\section{Scalar Field Realizations of the Zero Energy Density Source}
\label{sec:scalar-realizations}
We now reconstruct scalar field counterparts of the proportional-pressure and constant-pressure models above. The canonical, phantom and power-law constructions impose zero density along a particular homogeneous trajectory. A separate question is whether the rest-frame density can vanish identically as a function of the scalar variables. Within $P(X,\phi)$ theories, this selects a square-root kinetic structure. We keep these two constructions distinct.

The scalar kinetic variable is $X\equiv-\tfrac12 g^{\mu\nu}\nabla_\mu\phi\nabla_\nu\phi$. Throughout this section, $X>0$ so that the field gradient defines a timelike congruence. We reconstruct the scalar Lagrangian and the homogeneous exchange needed to sustain each pressure law. This determines a consistent background description, but it does not yet specify an interaction action or the distribution of interaction stress between the components. Appendix~\ref{app:action} completes this step for two dust--scalar examples: an exponential mass and potential realize proportional pressure, whereas a different mass function and a constant potential realize constant pressure. Keeping the reconstruction and the specification of the interaction separate allows us to identify which conclusions follow from the background alone and which require the dynamics of a particular model.

For a scalar field described by the Lagrangian density $\mathcal{L}^{(\phi)}(X, \phi)$, the EMT associated with the scalar sector is given by
\begin{align}
    \label{eq:sf-emt}
    T_{\mu\nu}^{(\phi)}=\mathcal{L}^{(\phi)}_{,X}\nabla_\mu\phi\nabla_\nu\phi+\mathcal{L}^{(\phi)} g_{\mu\nu}.
\end{align}
Here we define the subscript $_{,X}=\partial /\partial X$. For a timelike gradient, $X>0$, Eq.~\eqref{eq:sf-emt} takes perfect-fluid form with four-velocity
\begin{align}
    \label{eq:sf-four-velocity}
    u_\mu=\pm\frac{\nabla_\mu\phi}{\sqrt{2X}}, \qquad u_\mu u^\mu=-1,
\end{align}
where the sign is chosen such that $u^\mu$ is future-directed~\cite{Diez-Tejedor:2013nwa}. For a discussion of the formal correspondence between scalar fields and effective perfect fluids, see also Refs.~\cite{Faraoni:2012hn,Semiz:2012zz}. Using Eq.~\eqref{eq:sf-four-velocity}, the scalar field EMT~\eqref{eq:sf-emt} takes the perfect fluid form in Eq.~\eqref{eq:emt-pf} with the energy density and pressure of the scalar field given by
\begin{align}
\label{eq:sf-canonical-density}
    \rho_\phi&=2X\mathcal{L}^{(\phi)}_{,X}-\mathcal{L}^{(\phi)}, \\
    \label{eq:sf-canonical-pressure}
    p_\phi&=\mathcal{L}^{(\phi)}.
\end{align}
The same scalar field EMT obeys the identity
\begin{align}
\nabla_\mu T_{(\phi)}^{\mu\nu}
=\left[\nabla_\mu(\mathcal L^{(\phi)}_{,X}\nabla^\mu\phi)
+\mathcal L^{(\phi)}_{,\phi}\right]\nabla^\nu\phi.
\label{eq:scalar-transfer-direction}
\end{align}
The energy-momentum transfer vector is therefore aligned with the scalar field gradient. A general covariant fluid interaction does not necessarily have this form. The reconstructions considered below satisfy this condition at the background level. Any perturbative completion retaining the scalar EMT~\eqref{eq:sf-emt} must satisfy the same identity; a phenomenological fluid interaction that violates it does not admit this scalar representation.

\subsection{Canonical/Phantom Scalar Field}

The Lagrangian density of the canonical/phantom scalar field is given by
\begin{equation}
\label{eq:sf-can-lagrangian}
\mathcal{L}^{(\phi)}=\varepsilon X-V(\phi),
\end{equation}
where $\varepsilon=\pm1$ controls the sign of the kinetic term, with $\varepsilon=+1$ describing a canonical scalar and $\varepsilon=-1$ a phantom scalar with a negative kinetic term. For the homogeneous field $\phi=\phi(t)$, the kinetic scalar simplifies to $X=\dot{\phi}^{2}/2$, and we therefore use $\dot{\phi}$ rather than $X$ in the remainder of this subsection. Doing so, the corresponding energy density~\eqref{eq:sf-canonical-density} and pressure~\eqref{eq:sf-canonical-pressure} associated with the Lagrangian density~\eqref{eq:sf-can-lagrangian} become
\begin{align}
\label{eq:sf-can-rho}
\rho_\phi&=\varepsilon\frac{\dot{\phi}^{2}}{2}+V(\phi),\\
\label{eq:sf-can-p}
p_\phi&=\varepsilon\frac{\dot{\phi}^{2}}{2}-V(\phi).
\end{align}
Additionally, for a \emph{dynamical} scalar field, viz., $\dot{\phi} \neq 0$, exchanging energy with the standard source at the rate $\mathcal Q$, substituting Eqs.~\eqref{eq:sf-can-rho} and \eqref{eq:sf-can-p} into the component-wise continuity equation \eqref{cont:species} gives the
modified Klein--Gordon equation
\begin{align}
    \label{klein_ingeneral}
    \varepsilon \left[\ddot{\phi}+3 H \dot{\phi}\right]+\frac{{\rm d}V}{\rm{d}\phi}=-\frac{\mathcal Q}{\dot\phi}.
\end{align}
Note that in the absence of energy exchange, $\mathcal Q=0$, Eq.~\eqref{klein_ingeneral} immediately reduces to the standard Klein--Gordon case.

Having established the interaction framework for the canonical/phantom scalar field, we now proceed with the realization of the zero-density source within this description. Imposing the condition for the scalar field energy density~\eqref{eq:sf-can-rho} to vanish, i.e., $\rho_\phi=0$, along the cosmological
background yields
\begin{align}
    \label{zero-pot}
    V(\phi)=-\varepsilon\,\frac{\dot{\phi}^2}{2},
\end{align}
and accordingly, the scalar field pressure becomes
\begin{align}
    \label{zero-phi-pres}
    p_\phi=\varepsilon\,\dot{\phi}^{2}=-2 V(\phi).
\end{align}
The background Friedmann equations and the corresponding total continuity equation are still given by Eqs.~\eqref{zero-fe00}--\eqref{zero-cont}, respectively, but the pressure of the zero-density source $p_{\rm ze}$ is replaced by the scalar field pressure given by Eq.~\eqref{zero-phi-pres}.

Also, combining Eq.~\eqref{zero-phi-pres} with Eq.~\eqref{eq:zero-int2} gives the required interaction kernel
\begin{align} \label{phi-kernel}
\mathcal Q=-3\varepsilon H\dot{\phi}^{2}.
\end{align}
With this homogeneous energy exchange function, the condition $\rho_\phi=0$, equivalent to Eq.~\eqref{zero-pot}, can be maintained along the reconstructed cosmological trajectory with nonzero pressure, provided $\dot\phi\neq0$. We emphasize that Eq.~\eqref{zero-pot} holds only along the reconstructed background trajectory and should therefore be interpreted as an on-shell condition, rather than as an identity of the scalar field Lagrangian. An identity-level condition will be considered separately within the $k$-essence class.

The need for energy exchange to sustain the zero-density condition in a nonstatic universe ($H\neq0$) can also be seen directly from the Klein--Gordon equation. Consider first the noninteracting limit, $\mathcal Q=0$, of Eq.~\eqref{klein_ingeneral}. Taking the time derivative of Eq.~\eqref{zero-pot} gives
\begin{align}
\label{time-deriv-pot}
\frac{{\rm d}V}{{\rm d}\phi}=-\varepsilon\ddot{\phi},
\end{align}
for $\dot{\phi}\neq0$, as required for $p_\phi\neq0$. Substituting this relation into the $\mathcal Q=0$ limit of Eq.~\eqref{klein_ingeneral} yields
$3H\varepsilon\dot{\phi}=0$. Since $\dot{\phi}\neq0$, it follows that $H=0$. This reproduces, within the scalar field description, the
result established in Sec.~\ref{sec:flrw}: a component with
zero energy density and nonvanishing pressure cannot be
separately conserved in a nonstatic FLRW background.

With the exchange function~\eqref{phi-kernel}, the Klein--Gordon equation~\eqref{klein_ingeneral} reduces to
\begin{align}
    \label{Klein_g_zero}
    \varepsilon \ddot{\phi}+\frac{{\rm d}V}{{\rm d}\phi}=0,
\end{align}
which is equivalent to Eq.~\eqref{time-deriv-pot}. Conversely, differentiating Eq.~\eqref{zero-pot} and inserting the resulting relation into Eq.~\eqref{klein_ingeneral} directly recovers the interaction kernel~\eqref{phi-kernel}. Thus, the vanishing energy density condition~\eqref{zero-pot}, the interaction kernel~\eqref{phi-kernel}, and the reduced Klein--Gordon equation~\eqref{Klein_g_zero} form a mutually consistent description of the zero-density scalar field. Preserving the zero-density trajectory does not, by itself, make it a dynamically selected solution. If Eq.~\eqref{phi-kernel} is assumed to hold away from the zero-density trajectory as well, the continuity equation gives $\dot\rho_\phi=0$ for neighboring solutions. An initially vanishing density remains zero, but is not dynamically preferred over other constant values. If the exchange is specified only on the reconstructed trajectory, its off-trajectory behavior remains undetermined.

The sign of the pressure also has consequences for the scalar kinetic term. Equation~\eqref{zero-phi-pres} gives positive pressure with a canonical field and negative potential, whereas negative pressure requires the phantom kinetic sign. In the latter case, a luminal intrinsic sound speed does not remove the ghost associated with the negative time-kinetic coefficient; Sec.~\ref{sec:perturbation} derives this statement and its generalization to $P(X,\phi)$ theories. We now reconstruct the two pressure laws separately. The resulting field trajectories and potentials provide the starting point for the covariant interactions and dynamical tests in Appendix~\ref{app:action}. In particular, the homogeneous saddle found there for the proportional-pressure model distinguishes the existence of the reconstructed solution from its selection by the evolution of nearby initial conditions.

\subsubsection{Realization of the Proportional-Pressure Model} \label{sf:prop-pres}

We first consider the proportional-pressure model, introduced in Sec.~\ref{sec:prop-pres}, for which the scalar field pressure is identified as     $p_\phi=\alpha\rho_{\rm s}$. Using this relation and Eq.~\eqref{zero-pot}, obtained by
imposing $\rho_\phi=0$ along the homogeneous background solution, we arrive at
\begin{align}
    \label{eq:sf-prop-phi-dot}
    \dot{\phi}^{2}&=\frac{\alpha}{\varepsilon}\rho_{\rm s}, \\
    V(\phi)&=-\frac{\alpha}{2}\rho_{\rm s}. \label{eq:sf-prop-v}
\end{align}
Since $\rho_{\rm s}$ given by Eq.~\eqref{generalRho1} is always positive for $\rho_{\rm s0}>0$, the existence of a real-valued
scalar field solution requires $\alpha/\varepsilon>0$, implying ${\rm sgn}(\alpha)={\rm sgn}(\varepsilon)$. Therefore,
$\alpha>0$ requires $\varepsilon=+1$, corresponding to a canonical kinetic term with a negative potential, whereas $\alpha<0$ requires $\varepsilon=-1$,
corresponding to a phantom kinetic term with a positive potential.

Since the scale factor evolutions $a(t)$ have already been obtained in Sec.~\ref{sec:prop-pres}, it is convenient to use $a$ as the time variable instead of cosmic time $t$ in the following analysis. The evolution of the scalar field as a function of the scale factor can then be determined from the background dynamics. Inserting the first Friedmann equation~\eqref{model1-e00} into Eq.~\eqref{eq:sf-prop-phi-dot}, and using the relation $\dot\phi=aH\,{\rm d}\phi/{\rm d}a$ eliminates both $\rho_{\rm s}$ and $H$, yielding

\begin{equation}
    \label{eq:sf-prop-phi}
    \phi(a)
    =\phi_0\pm
    \sqrt{\frac{3\alpha}{\varepsilon\kappa}}\ln a,
\end{equation}
where $\phi_0\equiv\phi(a=1)$ is the present-day value of the scalar field. To express the potential explicitly as a function of $\phi$, we use the background evolution of the standard source given in Eq.~\eqref{generalRho1}. Eliminating $a$ in favor of $\phi$ through
Eq.~\eqref{eq:sf-prop-phi}, Eq.~\eqref{eq:sf-prop-v} gives
\begin{equation}
\label{eq:sf-prop-potential}
V(\phi)=-\frac{\alpha}{2}\rho_{\rm s0}
        \exp\left[\mp\eta(\phi-\phi_0)\right],
\end{equation}
where we define the parameter
\begin{equation}
\eta \equiv 3(1+w+\alpha)
        \sqrt{\frac{\varepsilon\kappa}{3\alpha}}.
\end{equation}

Eqs.~\eqref{eq:sf-prop-phi} and
\eqref{eq:sf-prop-potential} reproduce the background dynamics of the proportional-pressure model within the FLRW framework. In particular,
the proportional relation $p_\phi=\alpha\rho_{\rm s}$ leads to a logarithmic evolution of the scalar field with the scale factor and an exponential potential $V(\phi)$. Moreover, one obtains $\phi(t)$ by substituting $a(t)$ of the relevant background dynamics into Eq.~\eqref{eq:sf-prop-phi}. Thus, the zero-density source in this model admits a canonical/phantom scalar field realization with its kinetic character determined by the sign of $\alpha$. In the canonical case, $V<0$ and, for nonzero $\eta$, $V_{,\phi\phi}=\eta^2V<0$. A nonconstant negative exponential potential, if extended over the full field range, is unbounded from below in one direction. This should be distinguished from the wrong-sign kinetic term: the properties of the potential alone do not determine the stability of the interacting system. Negative-potential cosmologies are discussed in Ref.~\cite{Felder:2002jk}, while a specific coupling is examined explicitly in Appendix~\ref{app:action}. The $\eta=0$ constant-potential limit, for which $V_{,\phi\phi}=0$, is excluded from this result.

\subsubsection{Realization of the Constant-Pressure Model} \label{sec:const-pres}

We next turn to the constant pressure model, described by $p_\phi=p_{\rm ze0}={\rm const.}$, of Sec.~\ref{sec:lcdm}. Combining this relation with Eq.~\eqref{zero-pot} gives
\begin{align}
    \label{eq:sf-const-phi-dot}
    \dot{\phi}^{2}&=\frac{p_{\rm ze0}}{\varepsilon},  \\
    V(\phi)&=-\frac{p_{\rm ze0}}{2} \label{eq:sf-const-v}.
\end{align}
As in the proportional-pressure case, the requirement of a real-valued scalar field fixes the relative signs of the pressure and kinetic term. Equation~\eqref{eq:sf-const-phi-dot} requires
$p_{\rm ze0}/\varepsilon>0$, and hence ${\rm sgn}(p_{\rm ze0})={\rm sgn}(\varepsilon)$. Thus, $p_{\rm ze0}>0$ implies a canonical kinetic term, whereas $p_{\rm ze0}<0$ implies a phantom kinetic term, with the potential having the opposite sign. In particular, the $w_{\rm m}=0$ case reproducing the standard $\Lambda$CDM background has $p_{\rm ze0}<0$ and is therefore described by a phantom scalar field.

The scalar field evolution follows directly by integrating
Eq.~\eqref{eq:sf-const-phi-dot}. Using the sign relation
above, the integration yields
\begin{align}
    \phi(t)
    =
    \phi_0
    \pm
    \sqrt{\left|p_{\rm ze0}\right|}\,(t-t_0),
\label{eq:phi_const_t}
\end{align}
where $\phi_0=\phi(t_0)$. The field has a constant speed in field space for either sign of $p_{\rm ze0}$. The two signs of $\dot\phi$ give the same homogeneous density and pressure. Its linear dependence on cosmic time is independent of $w$ and of the background dynamics; the dependence on the cosmological solution enters only when one expresses $t$ in terms of $a$. This inverse is branch-dependent near a bounce or turnaround, whereas Eq.~\eqref{eq:phi_const_t} remains regular there. In particular, for $w>-1$, the negative-pressure solution can continue toward an asymptotic de Sitter state, while the positive-pressure solution evolves over the finite lifetime of the recollapsing background. For the negative-pressure dust branch, Appendix~\ref{app:constant-pressure-growth} constructs a mass coupling compatible with this constant potential and field velocity. That example allows the $\Lambda$CDM background correspondence to be tested at the level of dust perturbations, using the interaction fixed by the action rather than a separate phenomenological prescription.

\subsection{\texorpdfstring{$k$}{k}-essence}
\label{sec:sf-general}

We next consider a scalar field Lagrangian density with a general kinetic dependence~\cite{Armendariz-Picon:2000ulo,Garriga:1999vw},
\begin{align}
\mathcal L^{(\phi)}=P(X,\phi),
\label{eq:sf-action-general}
\end{align}
whose rest-frame density and pressure are
\begin{align}
\rho_\phi&=2XP_{,X}-P,
\label{eq:sf-general-density}\\
p_\phi&=P.
\label{eq:sf-general-pressure}
\end{align}
 Requiring zero density along one trajectory constrains these functions only on that trajectory. Requiring it as an identity on a timelike domain is stronger: at each fixed $\phi$, one must solve
\begin{align}
2XP_{,X}-P=0
\label{eq:sf-zero-condition}
\end{align}
throughout an open range of $X>0$. The nontrivial solution is
\begin{align}
P(X,\phi)=A(\phi)\sqrt X,
\label{eq:sf-square-root}
\end{align}
where $A$ is a function of the field. This gives
\begin{align}
\rho_\phi\equiv0,\qquad p_\phi=A(\phi)\sqrt X.
\label{eq:sf-square-root-rho-p}
\end{align}
Nonzero pressure requires $A\neq0$ on the field interval under consideration. This is an identity for the scalar rest-frame density on the timelike domain, rather than a cancellation between the two terms in Eq.~\eqref{eq:sf-general-density} imposed only on a homogeneous solution. It does not extend the perfect fluid construction through $X=0$, where the field congruence becomes undefined.

The relation to the cuscuton can be made explicit. In our sign convention, its Lagrangian density is~\cite{Afshordi:2006ad,Afshordi:2007yx}
\begin{align}
P_{\rm cus}(X,\phi)=\mu^2\sqrt X-V(\phi),
\end{align}
where a conventional factor of $\sqrt2$ is absorbed into $\mu^2$. Its rest-frame density is $V$, so the potential-free case has zero density. Moreover, on a patch where $A$ is nonzero and has fixed sign, we define a new scalar
\begin{align}
\chi(\phi)=\int^\phi\frac{|A(\tilde{\phi})|}{\mu^2}\,{\rm d}\tilde{\phi},
\qquad \mu^2>0.
\label{eq:cuscuton-redefinition}
\end{align}
Since $X^{(\chi)}=A^2X^{(\phi)}/\mu^4$, Eq.~\eqref{eq:sf-square-root} becomes
\begin{align}
A(\phi)\sqrt{X^{(\phi)}}
=\operatorname{sgn}(A)\,\mu^2\sqrt{X^{(\chi)}},
\label{eq:cuscuton-equivalence}
\end{align}
where $X^{(\phi)}$ and $X^{(\chi)}$ are the kinetic terms associated with $\phi$ and $\chi$, respectively. A field-dependent, nonzero $A$ therefore belongs locally to the same kinetic class as a signed potential-free cuscuton. The equivalence extends to the coupled action when every interaction involving $\phi$ is rewritten in terms of $\chi$. Comparing only the kinetic terms while specifying different interactions would not establish equivalence of the two theories. The redefinition is limited to intervals on which $A$ is nonzero and has fixed sign; it breaks down at a zero or sign change of $A$. Square-root scalar structures also arise in the holographic pressure-only description of Ref.~\cite{Compere:2011dx}.

The identity $\rho_\phi\equiv0$ does not remove the field equation. Direct variation of the uncoupled square-root action gives
\begin{align}
\nabla_\mu\!\left(\frac{A}{2\sqrt X}\nabla^\mu\phi\right)
+A_{,\phi}\sqrt X=0.
\label{eq:cuscuton-zero-eom}
\end{align}
On a time-oriented patch with $u^\mu=-\nabla^\mu\phi/\sqrt{2X}$, one has $u^\mu\nabla_\mu\phi=\sqrt{2X}$. The two terms containing $A_{,\phi}$ cancel, leaving $A\nabla_\mu u^\mu=0$. Choosing the opposite gradient orientation gives the same constraint. Thus, nonzero $A$ forces $\Theta=3H=0$ in the homogeneous, uncoupled case, reproducing the conservation restriction even for a field-dependent coefficient. A nonstatic FLRW realization with nonzero pressure therefore still needs the energy exchange stated in Eq.~\eqref{eq:zero-int2}. The vanishing principal time-kinetic coefficient of the square-root model and its perturbative implications are discussed in Sec.~\ref{sec:perturbation}.

As a distinct generalization of the trajectory-level construction, consider the power-law family~\cite{Serish:2025ian}
\begin{align}
P(X,\phi)&=f(\phi)\left[X^\nu-V(\phi)\right],
\label{sf:power-law}
\end{align}
leading to the scalar field energy density
\begin{align}
\rho_\phi&=f(\phi)\left[(2\nu-1)X^\nu+V(\phi)\right].
\label{rho-power-law}
\end{align}
The case $\nu=1$, $f=1$ is canonical, and $\nu=1/2$, $V=0$ gives the square-root model with $f=A$. For the following reconstructions we set $f=1$ and take $\nu>1/2$, for which the intrinsic kinetic coefficients and squared sound speed are positive on $X>0$. An overall positive coefficient of $X^\nu$ is absorbed into the field normalization. With $\hbar=c=1$, this convention assigns mass dimension $2/\nu-1$ to the power-law field, so that $[X^\nu]=[V]=4$. 
Alternatively, one may assign mass dimension unity to the field and write the kinetic term as $M^{4(1-\nu)}X^\nu$, with $M$ a fixed positive mass scale. These are equivalent normalizations of the same fixed-$\nu$ theory; they do not correspond to comparing fields with different mass dimensions at the same numerical value of $\phi$.
 Imposing zero-density condition along the background gives
\begin{align}
V(\phi)&=-(2\nu-1)X^\nu,
\label{eq:powerlaw-Vzero}\\
p_\phi&=2\nu X^\nu=-\frac{2\nu}{2\nu-1}V(\phi)>0.
\label{eq:powerlaw-pzero}
\end{align}
Consequently, with a positive normalization, this family realizes only positive pressure: $\alpha>0$ in the proportional-pressure model and $p_{\rm ze0}>0$ in the constant-pressure model. It does not supply the negative-pressure backgrounds of the phantom reconstruction.

For the proportional-pressure model $p_\phi=\alpha\rho_{\rm s}$, we define the following parameters
\begin{align}
b=3(&1+w+\alpha),\qquad r=\frac{b(\nu-1)}{2\nu},
\label{eq:powerlaw-exponents}\\
&\tilde{C}_\nu=\pm\frac{\sqrt2}{H_0}
\left(\frac{\alpha\rho_{\rm s0}}{2\nu}\right)^{1/(2\nu)}.
\label{eq:powerlaw-C}
\end{align}
In the expanding case, $H=H_0a^{-b/2}$ and $\dot\phi=H_0\tilde{C}_\nu a^{-b/(2\nu)}$. Integration therefore yields
\begin{align}
\phi(a)-\phi_0=\tilde{C}_\nu
\begin{cases}
(a^r-1)/r,&r\neq0,\\
\ln a,&r=0.
\end{cases}
\label{eq:powerlaw-phi}
\end{align}
Writing $V_0=-(2\nu-1)\alpha\rho_{\rm s0}/(2\nu)$, the generic solution with $r\neq0$ has
\begin{align}
V(\phi)=V_0
\left[1+\frac{r}{\tilde{C}_\nu}(\phi-\phi_0)\right]^{-2\nu/(\nu-1)},
\label{eq:powerlaw-potential}
\end{align}
on the positive-bracket domain corresponding to $a>0$. The logarithmic limit has two possibilities. For $\nu=1$, the potential is $V_0\exp[-b(\phi-\phi_0)/\tilde{C}_1]$, reproducing the canonical result given in Eqs.~\eqref{eq:sf-prop-phi} and~\eqref{eq:sf-prop-potential}. For $b=0$, the background is de Sitter and $X$ and $V$ are constant; $\phi-\phi_0=\tilde{C}_\nu\ln a$ for every allowed $\nu$. Thus logarithmic field evolution is not confined to the linear kinetic case. For the constant-pressure model with $p_{\rm ze0}>0$, the counterparts of Eqs.~\eqref{eq:phi_const_t} and~\eqref{eq:sf-const-v} are
\begin{align}
\phi(t)&=\phi_0\pm
\left(\frac{2^{\nu-1}p_{\rm ze0}}{\nu}\right)^{1/(2\nu)}(t-t_0),
\label{eq:powerlaw-constphi}\\
V(\phi)&=-\frac{2\nu-1}{2\nu}p_{\rm ze0}.
\label{eq:powerlaw-constV}
\end{align}
The field remains linear in cosmic time and the potential is constant, while their coefficients depend on $\nu$.

The canonical, phantom and power-law constructions realize zero density on particular homogeneous trajectories. The square-root construction instead enforces it as an identity in the timelike scalar rest frame. Both should be distinguished from stealth configurations, in which a nontrivial field has an entirely vanishing EMT~\cite{Ayon-Beato:2004nzi,Ayon-Beato:2013bsa,Blanco:2025kjb,Aguilar-Perez:2026jus}. Here the pressure contributes to gravitational evolution even though the sector supplies no background density in the Hamiltonian constraint. Its effect on the expansion also changes the evolution of the interacting standard component. These background correspondences motivate, but do not uniquely determine, a perturbative treatment.

\section{Linear Perturbations: Regular Variables and Physical Completion}
\label{sec:perturbation}

A vanishing background density does not prevent a source from contributing to linear perturbations. The relevant quantities are perturbations of its EMT, which can be used directly without division by $\bar\rho_{\rm ze}$. We retain the unnormalized variables of Ref.~\cite{Akarsu:2026pia} and extend their conservation equations to interacting sources. This formulation separates three questions: whether the variables remain regular at zero density, which constitutive relations or field equations close their evolution, and whether the resulting model is dynamically stable. We first establish the general equations and then examine the intrinsic kinetic properties of the scalar realizations. The explicit interactions in Appendix~\ref{app:action} supply applications in which the exchange and dust-growth equations are fixed by the same action.

A related issue with formulating regular perturbations arises when crossing the phantom divide. The parametrized post-Friedmann~(PPF) treatment of Ref.~\cite{Fang:2008sn} provides an approach for evolving dark energy perturbations across $w=-1$, where a conventional component rest frame can become ill-defined as $\rho+p$ vanishes. Here $\bar\rho_{\rm ze}=0$ but $\overline{\mathcal I}_{\rm ze}=\bar p_{\rm ze}\neq0$: the zero-density condition does not coincide with the NEC boundary. Regular variables are useful in both settings, but the PPF framework does not by itself determine the pressure and energy-momentum transfer perturbations of the interacting system considered here.

\subsection{Unnormalized variables and a regular energy-momentum transfer frame}

To retain the general-gauge formulation of the interacting fluid equations~\cite{Kodama:1984ziu,Malik:2002jb,Malik:2004tf,Valiviita:2008iv}, we use the standard scalar-perturbed FLRW metric
\begin{equation}
\begin{aligned}
{\rm d}s^2=a^2(\tau)\big\{&-(1+2\Psi){\rm d}\tau^2+2\partial_j B\,{\rm d}\tau\,{\rm d}x^j\\
&+[(1-2\Phi)\delta_{j\ell}+2\partial_j\partial_\ell E]
{\rm d}x^j{\rm d}x^\ell\big\}.
\end{aligned}
\label{eq:pert-metric}
\end{equation}
Here ${\rm d}\tau={\rm d}t/a$, $\mathcal H=a^{-1}\partial_\tau a$, and $\partial_\tau$ denotes differentiation with respect to conformal time. In Newtonian gauge, $B=E=0$, this is exactly the potential convention of Ref.~\cite{Akarsu:2026pia}. We use $i$ for component labels and $j,\ell$ for spatial indices.

For a nonzero Fourier mode with $\nabla^2\to-k^2$, the definitions are given by
\begin{align}
\delta T^0{}_{j(i)}&=\partial_j\mathcal J_i,\qquad Q_i=-k^2\mathcal J_i,
\label{eq:momentum-potential}\\
\pi^j{}_{\ell(i)}&=-\frac32\left(\hat k^j\hat k_\ell-\frac13\delta^j{}_\ell\right)\Pi_i.
\label{eq:anisotropic-potential}
\end{align}
These definitions use the EMT directly, without division by $\bar\rho_i$ or $\overline{\mathcal I}_i=\bar\rho_i+\bar p_i$. When a regular component velocity exists, one has $\mathcal J_i=\overline{\mathcal I}_i(v_i+B)$ and $Q_i=\overline{\mathcal I}_i\theta_i$, with $\theta_i=-k^2(v_i+B)$. Similarly, $\Pi_i=\overline{\mathcal I}_i\sigma_i$ for a component with nonzero inertial mass density, where $\sigma_i$ is normalized as in Ref.~\cite{Ma:1995ey}. The homogeneous scalar mode belongs to the background variation and is treated separately, rather than by dividing by $k^2$.

Choosing a unit timelike reference congruence $u_{\rm r}^\mu$, the energy-momentum transfer vector can be decomposed  as
\begin{align}
\mathcal Q_{(i)}^\mu&=\mathcal Q_i u_{\rm r}^\mu+F_{(i)}^\mu,
\qquad u_{{\rm r}\mu}F_{(i)}^\mu=0,
\label{full-kernel}\\
\mathcal Q_i&=\bar{\mathcal Q}_i+\delta\mathcal Q_i,\qquad
F_{(i)}^\mu=a^{-1}(0,\partial^jf_i).
\label{eq:transfer-perturbations}
\end{align}
At first order, $u_{\rm r}^\mu=a^{-1}(1-\Psi,\partial^jv_{\rm r})$. Total conservation implies $\sum_i\bar{\mathcal Q}_i=\sum_i\delta\mathcal Q_i=\sum_i f_i=0$, up to a consistently fixed spatially homogeneous ambiguity in $f_i$. For the perturbation equations below, we choose the reference normal to the constant-$\tau$ slices, for which $v_{\rm r}+B=0$. Next, define the momentum-transfer divergence
\begin{align}
\mathcal S_i=-k^2[\bar{\mathcal Q}_i(v_{\rm r}+B)+f_i],\qquad
\sum_i\mathcal S_i=0.
\label{eq:transfer-force}
\end{align}
Changing the reference congruence changes the scalar decomposition, not the vector $\mathcal Q_{(i)}^\mu$ or the physical exchange. This choice is convenient since it does not require an energy-frame velocity. Where a total energy-frame velocity exists, it obeys
\begin{align}
\overline{\mathcal I}_{\rm tot}(v_{\rm tot}+B)=\sum_i\mathcal J_i.
\label{eq:total-frame-degeneracy}
\end{align}
At $\overline{\mathcal I}_{\rm tot}=0$, as in the exact de Sitter backgrounds above, this relation no longer defines a regular $v_{\rm tot}$. Even if the right-hand side vanishes, the vacuum-form background EMT does not select a unique energy-frame velocity. Neither a total nor a component energy-frame division is needed in the formulation below.

Expanding $\nabla_\mu T_{(i)}^{\mu\nu}=\mathcal Q_{(i)}^\nu$ at first order gives
\begin{equation}
\begin{aligned}
\partial_\tau\delta\rho_i={}&-3\mathcal H(\delta\rho_i+\delta p_i)
+3\overline{\mathcal I}_i\partial_\tau\Phi-Q_i\\
&+k^2\overline{\mathcal I}_i(\partial_\tau E-B)
+a(\delta\mathcal Q_i+\bar{\mathcal Q}_i\Psi),
\end{aligned}
\label{eq:pert-energy-A}
\end{equation}
\begin{align}
\partial_\tau Q_i=-4\mathcal H Q_i
+k^2(\delta p_i+\overline{\mathcal I}_i\Psi-\Pi_i)+a\mathcal S_i.
\label{eq:pert-momentum-A}
\end{align}
In the source-free Newtonian-gauge limit, these equations reduce to the conservation equations of Ref.~\cite{Ma:1995ey} once the normalizations by $\bar\rho_i$ and $\overline{\mathcal I}_i$ are undone, in the potential convention adopted in Ref.~\cite{Akarsu:2026pia}. The general-gauge form retains the shear combination $\partial_\tau E-B$, and summing over components cancels the transfer terms and recovers total conservation.

To obtain a predictive system, the pressure response and the energy and momentum transfer perturbations must be specified consistently. In an action-based realization, the same covariant interaction determines the background exchange and its perturbations. This consistency requirement is distinct from stability: a consistently closed interacting model can still develop large-scale instabilities that are not apparent from its background evolution~\cite{Valiviita:2008iv,He:2008si,Gavela:2010tm,LopezHonorez:2010esq,Clemson:2011an}. The appropriate test is therefore to derive the perturbative equations for the chosen interaction and examine their solutions, rather than infer stability from a regular background.

\subsection{Background zero density and the first-order condition}

For the zero-density component, $\bar\rho_{\rm ze}\equiv0$ and $3\mathcal H\bar p_{\rm ze}=a\bar{\mathcal Q}_{\rm ze}$. Under an infinitesimal temporal gauge shift $T$,
\begin{align}
\widetilde{\delta\rho}_i=\delta\rho_i-(\partial_\tau\bar\rho_i)T.
\label{eq:density-gauge-shift}
\end{align}
Thus $\delta\rho_{\rm ze}$ is gauge invariant because $\partial_\tau\bar\rho_{\rm ze}=0$~\cite{Malik:2004tf}. In contrast, an isolated density zero-crossing, with a generally nonzero density derivative, does not imply the same result.

Imposing zero energy density only on the background permits $\delta\rho_{\rm ze}\neq0$: the component can contribute to the perturbed density even though its homogeneous density vanishes. Linearization must be assessed using the perturbation hierarchy of the underlying fields, EMTs, and metric, not $\delta\rho_{\rm ze}/\bar\rho_{\rm ze}$. Normalizing by the nonzero $\bar p_{\rm ze}=\overline{\mathcal I}_{\rm ze}$ is possible on the stated domain, but is not needed for Eqs.~\eqref{eq:pert-energy-A}--\eqref{eq:pert-momentum-A}.\footnote{Inertial mass density normalization has also been used for smooth sign-changing dark energy perturbations~\cite{Bouhmadi-Lopez:2026vyc}. Such a normalization is regular only where its denominator is nonzero; the unnormalized EMT variables used here do not require this division.}

Alternatively, imposing $\delta\rho_{\rm ze}=0$ as a first-order condition turns the energy equation into
\begin{equation}
\begin{aligned}
3\mathcal H\delta p_{\rm ze}-3\bar p_{\rm ze}\partial_\tau\Phi
+Q_{\rm ze}&-k^2\bar p_{\rm ze}(\partial_\tau E-B)\\
&=a(\delta\mathcal Q_{\rm ze}+\bar{\mathcal Q}_{\rm ze}\Psi),
\end{aligned}
\label{eq:zero-pert-constraint}
\end{equation}
while the momentum equation remains
\begin{align}
\partial_\tau Q_{\rm ze}=-4\mathcal H Q_{\rm ze}
+k^2(\delta p_{\rm ze}+\bar p_{\rm ze}\Psi-\Pi_{\rm ze})
+a\mathcal S_{\rm ze}.
\label{eq:zero-euler}
\end{align}
The constraint~\eqref{eq:zero-pert-constraint} must be preserved by the pressure and energy-momentum transfer perturbations. Vanishing density through first order does not establish the full nonlinear identity~\eqref{eq:exact-pressure-only}. If that stronger perfect fluid pressure-only identity is separately imposed, its intrinsic anisotropic stress vanishes and its divergence is fixed by Eq.~\eqref{eq:pressure-only-transfer}. The corresponding transfer perturbations must then be calculated from this divergence. The more general first-order system may retain $\Pi_{\rm ze}\neq0$ when an imperfect perturbative sector is specified.

For the scalar field EMT of Sec.~\ref{sec:scalar-realizations}, the transfer vector is parallel to the field gradient by Eq.~\eqref{eq:scalar-transfer-direction}. Substituting this into Eq.~\eqref{full-kernel} gives
\begin{align}
f_\phi=\bar{\mathcal Q}_\phi(v_\phi-v_{\rm r})
\label{eq:scalar-momentum-transfer}
\end{align}
at first order. The perturbation of the scalar field energy-transfer rate is likewise determined by the underlying coupling. Appendix~\ref{app:action} provides an example in which the coupling is specified covariantly. For the square-root scalar, $\rho_\phi\equiv0$ on its timelike domain is an identity for the scalar field rest-frame density. Consequently, $\delta\rho_\phi=0$ at first order about a homogeneous solution. This should not be confused with an exact coordinate condition $T^0{}_0=0$ in an arbitrary frame: a relative tilt generates a nonzero density at second order in the velocity, as follows already from Eq.~\eqref{eq:rho-gen}.

\subsection{Rest-frame pressure response}

Table~\ref{tab:scalar-realizations} summarizes how the implementation of zero density differs from the kinetic character of the scalar field realization. The background constraint, the intrinsic characteristic speed, and the stability of the coupled theory are separate issues. The coefficients entering this comparison follow directly from the scalar field action, as we will show.

\begin{table*}[t]
\caption{Scalar field realizations on a timelike domain $X>0$. $K^{(\phi)}=P_{,X}+2XP_{,XX}$ is the intrinsic time-kinetic coefficient. The first three rows impose $\rho_\phi=0$ along a background trajectory, not as an identity of the action. The signs and speeds refer to the considered scalar principal structure; none provides a full coupled-theory stability criterion. Nonzero pressure in nonstatic FLRW requires energy exchange in every row.}
\label{tab:scalar-realizations}
\centering\small
\renewcommand{\arraystretch}{1.35}
\begin{tabular}{@{}p{0.16\textwidth}p{0.20\textwidth}p{0.16\textwidth}p{0.16\textwidth}p{0.22\textwidth}@{}}
\toprule
Scalar Lagrangian & Zero-density condition & Pressure & $K^{(\phi)}$; $c_{s,\phi}^2$ & Physical qualification \\
\midrule
$X-V$ & $V=-X$ & $2X>0$ & $1$; $1$ & Positive intrinsic kinetic and gradient terms; no attractor implied. \\
$-X-V$ & $V=X$ & $-2X<0$ & $-1$; $1$ & Ghost despite a luminal characteristic speed. \\
$X^\nu-V$, $\nu>1/2$ & $V=-(2\nu-1)X^\nu$ & $2\nu X^\nu>0$ & $\nu(2\nu-1)X^{\nu-1}$;\newline $(2\nu-1)^{-1}$ & Positive pressure only; field normalization as in Sec.~\ref{sec:sf-general}. \\
$A(\phi)\sqrt X$ & $\rho_\phi\equiv0$ (identity)\newline for $X>0$ & $A(\phi)\sqrt X$ & $0$; no propagating\newline scalar speed & Constrained scalar; locally a signed potential-free cuscuton for $A\neq0$. \\
\bottomrule
\end{tabular}
\end{table*}

When a regular component rest frame exists, the density and pressure transformations are~\cite{Malik:2004tf,Valiviita:2008iv}
\begin{align}
\delta\rho_i&=\delta\rho_i|_{\rm rf}-(\partial_\tau\bar\rho_i)(v_i+B),
\label{gauge-rho}\\
\delta p_i&=\delta p_i|_{\rm rf}-(\partial_\tau\bar p_i)(v_i+B).
\label{gauge-pres}
\end{align}
For a phenomenological zero-density sector, one may specify a finite response coefficient and an additional rest-frame pressure fluctuation through
\begin{align}
\delta p_{\rm ze}|_{\rm rf}
=c_{{\rm eff},{\rm ze}}^2\delta\rho_{\rm ze}|_{\rm rf}
+\delta p_{{\rm ex},{\rm ze}}.
\label{eq:general-rest-closure}
\end{align}
In an arbitrary gauge, this becomes
\begin{align}
\delta p_{\rm ze}=c_{{\rm eff},{\rm ze}}^2\delta\rho_{\rm ze}
-(\partial_\tau\bar p_{\rm ze})(v_{\rm ze}+B)
+\delta p_{{\rm ex},{\rm ze}}.
\label{eq:zero-pressure-general}
\end{align}
These response terms must be specified by the underlying model; the introduction of a coefficient $c_{{\rm eff},{\rm ze}}^2$ alone does not determine a propagating sound speed. For $\delta\rho_{\rm ze}=0$, the equations reduce to
\begin{align}
\delta p_{\rm ze}|_{\rm rf}&=\delta p_{{\rm ex},{\rm ze}},\nonumber\\
\delta p_{\rm ze}&=-(\partial_\tau\bar p_{\rm ze})(v_{\rm ze}+B)
+\delta p_{{\rm ex},{\rm ze}}.
\label{eq:zero-pressure-constrained}
\end{align}
Thus, setting the rest-frame pressure fluctuation to zero is an additional assumption. The usual background adiabatic sound speed $(\partial_\tau\bar p_{\rm ze})/(\partial_\tau\bar\rho_{\rm ze})$ is undefined, including the $0/0$ case for the constant-pressure model. The conventional decomposition in terms of 
 $c_s^2-c_a^2$ therefore does not provide a basis for imposing this condition.

\subsection{Intrinsic scalar propagation and the negative-pressure obstruction}
\label{sec:kinetic-obstruction}

The action of a scalar field determines its intrinsic response directly. For timelike $P(X,\phi)$, the rest-frame slicing has $\delta\phi|_{\rm rf}=0$, giving
\begin{align}
\delta\rho_\phi|_{\rm rf}&=K^{(\phi)}\delta X|_{\rm rf},\qquad
\delta p_\phi|_{\rm rf}=P_{,X}\delta X|_{\rm rf},
\label{eq:kessence-rest-variations}\\
K^{(\phi)}&\equiv P_{,X}+2XP_{,XX}.
\label{eq:kessence-K}
\end{align}
For $K^{(\phi)}\neq0$, the intrinsic rest-frame sound speed is~\cite{Garriga:1999vw}
\begin{align}
c_{s,\phi}^2=\frac{P_{,X}}{K^{(\phi)}}.
\label{eq:kessence-sound}
\end{align}
Neither the intrinsic response nor the characteristic speed involves normalization by $\rho_\phi$. The scalar-only principal part of the quadratic action in cosmic time reads
\begin{equation}
S^{(2)}_{\phi,\rm prin}=\frac12\int {\rm d}t\,{\rm d}^3x\,a^3
\left[K^{(\phi)}(\delta\dot\phi)^2
-\frac{P_{,X}}{a^2}(\boldsymbol\nabla\delta\phi)^2\right].
\label{eq:scalar-principal-action}
\end{equation}
Equivalently, the field equation's principal tensor is $P_{,X}g^{\mu\nu}-P_{,XX}\nabla^\mu\phi\nabla^\nu\phi$. In the local rest frame its temporal and spatial coefficients are $-K^{(\phi)}$ and $P_{,X}$; the high-frequency dispersion relation is $\omega^2=c_{s,\phi}^2 k_{\rm phys}^2$. These describe the intrinsic scalar sector presented here; lower-derivative forces, metric constraints, and kinetic mixing with other fields must be included to assess the complete coupled theory.

Provided that the interaction does not change this principal scalar structure, the nondegenerate no-ghost and gradient conditions are $K^{(\phi)}>0$ and $P_{,X}>0$. For zero density, one has
\begin{align}
\rho_\phi=0\quad\Longrightarrow\quad p_\phi=2XP_{,X}.
\label{eq:zero-pressure-kinetic-sign}
\end{align}
On $X>0$, Eq.~\eqref{eq:zero-pressure-kinetic-sign} gives $p_\phi<0$ if and only if $P_{,X}<0$. If $K^{(\phi)}>0$, the gradient term is unstable; if $K^{(\phi)}<0$, the mode is a ghost even when the sound-speed quotient is positive. Thus, \emph{a nondegenerate single $P(X,\phi)$ scalar with a timelike field gradient cannot simultaneously have zero density, negative pressure, and positive kinetic and gradient coefficients when its principal structure is unchanged by the interaction}. The restriction follows from the same NEC/stability relation that appears in the scalar phantom-crossing literature~\cite{Vikman:2004dc,Hu:2004kh,Kunz:2006wc}. Here the zero-density condition makes the pressure sign alone determine the sign of $P_{,X}$. The positive-pressure canonical realization is not excluded by this argument. A degenerate constrained scalar requires a different analysis, because the absence of an ordinary scalar wave mode is not itself a ghost or a gradient instability.

For $P=\varepsilon X-V$, one has $P_{,X}=K^{(\phi)}=\varepsilon$ and $c_{s,\phi}^2=1$. The negative-pressure realization therefore illustrates why a positive sound-speed quotient is not sufficient for stability: both principal coefficients have the phantom sign. For the power-law model with a positive kinetic term, one has
\begin{align}
P_{,X}=\nu X^{\nu-1},\qquad K^{(\phi)}=\nu(2\nu-1)X^{\nu-1},
\end{align}
so that $c_{s,\phi}^2=1/(2\nu-1)>0$ for $\nu>1/2$. Requiring this intrinsic characteristic cone not to exceed the metric light cone would further restrict $\nu\geq1$; that requirement is distinct from positivity of the kinetic coefficients. A scalar field cone wider than the metric light cone does not imply a causal inconsistency; the causal structure and admissible initial-value problem should be assessed independently~\cite{Babichev:2007dw}. For $P=A(\phi)\sqrt X$, however, $K^{(\phi)}\equiv0$. The usual sound-speed quotient~\eqref{eq:kessence-sound} is not the speed of an ordinary propagating mode. The timelike field equation is constrained, so the interacting system must be treated as a degenerate theory~\cite{Afshordi:2006ad,Afshordi:2007yx}. Its rest-frame fluctuations are
\begin{align}
\delta\rho_\phi|_{\rm rf}=0,\qquad
\delta p_\phi|_{\rm rf}=\frac{A(\phi)}{2\sqrt X}\delta X|_{\rm rf}.
\label{eq:square-root-pressure-pert}
\end{align}
The pressure perturbation is determined by the field equation along with the energy-momentum exchange equation; it is not an independently specified entropy perturbation. In particular, the singular sound-speed quotient should not be interpreted as an independently propagating infinite-speed signal.

Table~\ref{tab:scalar-realizations} summarizes what the zero-density condition fixes, and what remains to be specified, for each scalar class. A Boltzmann implementation requires $\delta p_i$, $\Pi_i$, $\delta\mathcal Q_i$, and $\mathcal S_i$ from consistent constitutive relations or field equations, together with initial conditions that preserve any imposed density constraint. The regularity of the variables makes this calculation possible; it does not predetermine its outcome.

The dust--scalar models of Appendix~\ref{app:action} provide explicit examples of this distinction. Once $m(\phi)$ and $V(\phi)$ are specified, the interaction determines both the force on the dust and the change in its velocity damping. For the proportional-pressure model, these effects give $G_{\rm eff}/G=1+6\alpha$ and exclude a growing power-law dust mode on the accelerating zero-density trajectory in the quasistatic regime. The homogeneous saddle established in Sec.~\ref{app:homogeneous} concerns a different perturbation sector. The constant-pressure model in Sec.~\ref{app:constant-pressure-growth} instead has time-dependent growth coefficients: its perturbation mode that initially grows during matter domination reaches a maximum at low redshift and then decays. These calculations connect the general formulation to particular interactions, while a full stability analysis of either coupled theory also requires its quadratic action and constraints.

\section{Zero Energy Density Source in Generic Spacetimes}

\label{sec:anisotropic}

The symmetries of RW spacetime imply vanishing shear and exclude momentum density and anisotropic stresses at the background level. Relaxing isotropy allows these quantities to be nonzero, and the $1{+}3$ covariant conservation equations then show how energy flux and anisotropic stress can contribute to the evolution of a zero-density source without requiring energy-momentum transfer with another component. We illustrate this explicitly with exact Bianchi Type~I solutions. In this setting, the pressure contribution need not be balanced by an intercomponent exchange, since the imperfect stresses provide additional terms in the conservation equations. Bianchi~I backgrounds and anisotropic-source dynamics have been studied extensively~\cite{Jacobs:1968,Koivisto:2008ig}; our focus here is the corresponding zero-density sector.

To make this generalization explicit, we decompose the EMT relative to a timelike congruence $u^\mu$ as

\begin{align}
T_{\mu\nu}=\rho u_{\mu} u_{\nu} +q_{\mu} u_{\nu} +u_{\mu} q_{\nu}+p h_{\mu\nu}  +\pi_{\mu\nu},
\end{align}
where $q^\mu u_\mu=0$ and $\pi_{\mu\nu}u^\nu=0$, with $\pi^\mu{}_{\mu}=0$. The quantities are defined relative to the specified unit timelike congruence $u^\mu$. If $q^\mu\neq0$, this is not an energy eigenframe, so ``zero density'' here means the zero projection $T_{\mu\nu}u^\mu u^\nu$ in that congruence. The perfect fluid case has $q^\mu=\pi_{\mu\nu}=0$. For the total source, with any cosmological constant included in its EMT, the projected Gauss constraint is given by~\cite{Ellis:1998ct,Tsagas:2007yx,Ellis2012}
\begin{align} \label{hamiltonian}
3H^2 - \sigma^2 + \omega^2 +\frac{1}{2}\,{}^{(3)}\!R= \kappa\rho,
\end{align}
where $H=\Theta/3$, $\sigma^2=\sigma_{\mu\nu}\sigma^{\mu\nu}/2$, and $\omega^2=\omega_{\mu\nu}\omega^{\mu\nu}/2$. Here $\sigma_{\mu\nu}={\rm D}_{(\mu}u_{\nu)}-\Theta h_{\mu\nu}/3$ and $\omega_{\mu\nu}={\rm D}_{[\mu}u_{\nu]}$ are the shear and vorticity tensors. The scalar $^{(3)}R$ is the curvature scalar of the projected $1{+}3$ spatial connection; it is the intrinsic curvature of orthogonal spatial hypersurfaces when $\omega_{\mu\nu}=0$. Flux and anisotropic stress do not enter this constraint explicitly. For a separately conserved source, the continuity equation is~\cite{Ellis:1998ct}
\begin{align} \label{cont-generic}
\dot{\rho}+3H (\rho+p)+D_{\mu} q^{\mu}  +2 A_{\mu} q^{\mu} +\sigma_{\mu\nu}\pi^{\mu\nu} =0.
\end{align}

For a source satisfying $\rho\equiv0$ relative to $u^\mu$, the EMT becomes
\begin{align}
T_{\mu\nu}^{(\rm ze)}=q_{\mu} u_{\nu} +u_{\mu} q_{\nu} +p h_{\mu\nu} +\pi_{\mu\nu},
\end{align}
If this is the only source, Eq.~\eqref{hamiltonian} and its conservation equation reduce to
\begin{align}\label{GFREqn-1}
    3H^2 &=  \sigma^2 - \omega^2 -\frac{1}{2}\,{}^{3}\!R, \\
    3H p&=-D_{\mu} q^{\mu}  -2 A_{\mu} q^{\mu} -\sigma_{\mu\nu}\pi^{\mu\nu}. \label{GFREqn-2}
\end{align}
Equation~\eqref{GFREqn-2} shows that a nonzero isotropic pressure is compatible with $\rho=0$ when the pressure term is balanced by the energy-flux divergence or the shear--stress contribution. This conservation relation also applies to a separately conserved zero-density component in a multicomponent spacetime, but the pure-source constraint~\eqref{GFREqn-1} must then be replaced by the constraint with the total density.

For the spatially homogeneous, orthogonal models considered below, we use the geodesic normal congruence in synchronous cosmic time, so $\omega_{\mu\nu}=A_\mu=0$. The equations then read
\begin{align}
    3H^2&=\sigma^2-\frac{1}{2}\,{}^{3}\!R, \\
    3H p&=-D_{\mu} q^{\mu}-\sigma_{\mu\nu}\pi^{\mu\nu}.
\end{align}

In what follows, we investigate the simplest anisotropic generalization of spatially flat RW spacetime metric, namely the Bianchi~I spacetime, in the presence of the zero-density source only. In this particular example, we have
\begin{align} \label{bianchiI}
    ^{3}R=0 \quad\text{and}\quad q^{\mu}=0.
\end{align}
In this case, the isotropic pressure can remain nonzero,
\begin{align}
  p_{\rm ze}=-\frac{1}{3H}\sigma_{\mu\nu}\pi^{\mu\nu},
\end{align}
for $H\neq0$. Thus no intercomponent exchange is needed in this anisotropic example. For a diagonal stress and shear, a nonzero trace-free anisotropic stress has at least two nonzero directional eigenvalues. Its contraction with the shear must be nonzero to balance a nonzero $3Hp$. For $H>0$, negative isotropic pressure requires $\sigma_{\mu\nu}\pi^{\mu\nu}>0$; for a contracting congruence the sign relation reverses.

Outside FLRW, the sign of $\rho+3p$ still fixes the sign of the corresponding Ricci contribution, but does not alone determine the mean acceleration. For the geodesic, irrotational Bianchi~I congruence,
\begin{align}
\frac{\ddot s}{s}=-\frac{2}{3}\sigma^2-\frac{\kappa}{6}(\rho+3p),
\label{eq:bianchi-raychaudhuri}
\end{align}
where $s^3$ is the volume scale factor. Shear therefore opposes mean acceleration even when the pressure contribution is repulsive.

\subsection{Bianchi Type~I Framework}

We now make the above discussion explicit by specializing to the Bianchi Type~I geometry. Accordingly, we replace the isotropic RW metric with the Bianchi Type~I metric~\cite{Collins:1973lda,Ellis:1998ct}
\begin{align} \label{bianchi}
{\rm d}s^2=-{\rm d}t^2+A^2(t)\,{\rm d}x^2+B^2(t)\,{\rm d}y^2+C^2(t)\,{\rm d}z^2,
\end{align}
and the perfect fluid EMT with its anisotropic generalization
\begin{align}
 T^\mu{}_{\nu}={\rm diag}(-\rho,p_x,p_y,p_z),
\end{align}
where $p_x$, $p_y$, and $p_z$ are the pressures along the $x$-, $y$-, and $z$-axes, respectively, as measured by observers comoving with the fluid. The average scale factor is defined as $s=(ABC)^{1/3}$. The average Hubble parameter is then given by $H=\dot{s}/s=\frac{1}{3}(H_x+H_y+H_z)$, where $H_x=\dot{A}/A$, $H_y=\dot{B}/B$, and $H_z=\dot{C}/C$ are the directional Hubble parameters.

The isotropic pressure in the $1{+}3$ decomposition is $p=(p_x+p_y+p_z)/3$. For this diagonal EMT, the NEC requires $\rho+p_j\geq0$ in each principal direction $j=x,y,z$. At $\rho=0$ it becomes $p_j\geq0$ for all three pressures. The sign of the mean combination $\rho+p$ alone is therefore not a complete NEC classifier once anisotropic stress is present.

For the Bianchi Type~I geometry, the Hamiltonian constraint~\eqref{hamiltonian} and the continuity equation~\eqref{cont-generic} reduce to
\begin{align}
    H_xH_y+H_yH_z+H_zH_x&=\kappa \rho, \label{BI00} \\
    \dot{\rho}+3H\rho+H_xp_x+H_yp_y+H_zp_z&=0. \label{BIcont}
\end{align}
Furthermore, the anisotropy of the expansion can be characterized by the shear scalar
\begin{equation}
    \label{shear_bianchi}
    \sigma^2=\frac{1}{6}\left[(H_x-H_y)^2+(H_y-H_z)^2+(H_z-H_x)^2\right].
\end{equation}
Using this definition, Eq.~\eqref{BI00} can be written as
\begin{equation}
    3H^2=\kappa\rho+\sigma^2.
\end{equation}

We now restrict the analysis to the case $\rho=0$, so that this component alone drives the Bianchi Type~I dynamics. Under this assumption, Eq.~\eqref{BI00} and the spatial components of the EFE take the following form
\begin{align}
    H_xH_y+H_yH_z+H_zH_x&=0, \label{BI00-zero}\\
    -\dot{H}_y-\dot{H}_z-H^2_y-H^2_z-H_yH_z&=\kappa p_x, \label{BI11-zero}\\
    -\dot{H}_z-\dot{H}_x-H^2_z-H^2_x-H_zH_x&=\kappa p_y, \label{BI22-zero}\\
    -\dot{H}_x-\dot{H}_y-H^2_x-H^2_y-H_xH_y&=\kappa p_z. \label{BI33-zero}
\end{align}
The constraint fixes the relative anisotropy,
\begin{align}
3H^2=\sigma^2,\qquad
\frac{\sigma^2}{3H^2}=1\quad(H\neq0).
\label{eq:normalized-shear}
\end{align}
The shear may decrease in absolute magnitude while remaining as large relative to the mean expansion as required by Eq.~\eqref{eq:normalized-shear}. Such a solution does not approach an isotropic FLRW state in the normalized-shear sense. In the standard diagonal, Hubble-normalized shear variables~\cite{Wainwright:1997}, the constraint is $\Sigma_+^2+\Sigma_-^2=\Sigma^2\equiv\sigma^2/(3H^2)=1$ wherever $H\neq0$. The shear coordinates therefore lie on the same unit circle as those of the vacuum Kasner family. Nonzero directional stresses nevertheless distinguish the sourced dynamics: the shared circle in the shear projection does not make every solution a vacuum Kasner geometry or a vacuum equilibrium point. This gives a direct connection to the isotropization question studied by Collins and Hawking~\cite{Collins:1972tf}, with the obstruction here following from the zero-density Hamiltonian constraint itself. Under the same zero-density condition, Eq.~\eqref{BIcont} reduces to
\begin{align}
H_xp_x+H_yp_y+H_zp_z=0. \label{BIcont-zero}
\end{align}
The system contains six unknown functions, $A$, $B$, $C$, $p_x$, $p_y$, and $p_z$. Of Eqs.~\eqref{BI00-zero}--\eqref{BIcont-zero}, only four are independent, since the continuity equation follows from the EFE as a consequence of the twice-contracted Bianchi identity. Thus, two additional relations are required to close the system.

For isotropic nonzero pressure, $p_x=p_y=p_z=p\neq0$, Eq.~\eqref{BIcont-zero} requires $H_x+H_y+H_z=0$. Together with Eq.~\eqref{BI00-zero}, this gives $H_x^2+H_y^2+H_z^2=0$, so all directional Hubble parameters vanish. The spatial equations then give $p=0$, a contradiction. Anisotropic stress is therefore required for a nontrivial pure zero-density Bianchi~I solution. The same constraint rules out a nontrivial constant-volume solution maintained by compensating expansion and contraction. A single nonzero directional pressure is also insufficient. If $p_y=p_z=0$ and $p_x\neq0$ on an interval, Eq.~\eqref{BIcont-zero} gives $H_x=0$, and the constraint gives $H_yH_z=0$. On a subinterval with $H_z=0$, Eq.~\eqref{BI33-zero} then implies $\dot H_y+H_y^2=0$, which makes $p_x=0$ by Eq.~\eqref{BI11-zero}; the other case is symmetric. Thus setting two directional pressures to zero forces the third to vanish for a regular solution. The remaining vacuum class includes the Kasner geometries and the static limit. Nonvacuum zero-density solutions require at least two nonzero directional pressures.

\subsubsection{Generalized Kasner}

To generalize the Kasner solution on $t>0$, we take two of the directional Hubble parameters as $H_x=n_1/t$ and $H_y=n_2/t$. For $n_1+n_2\neq0$, Eq.~\eqref{BI00-zero} then implies $H_z\propto t^{-1}$, leading to the constraint
\begin{align}
  n_1 n_2+n_2 n_3+ n_3 n_1=0,
\end{align}
where we have used $H_z=\frac{n_3}{t}$.\footnote{For the special case $n_1+n_2=0$, Eq.~\eqref{BI00-zero} implies $n_1=n_2=0$, while $H_z$ remains arbitrary. We do not consider this more general case for $H_z$, which is static in the $x$- and $y$-directions with arbitrary $H_z(t)$; instead, for the purpose of generalizing the Kasner solution, we restrict attention to the power-law case $H_z\propto t^{-1}$.}
Defining $\sum_j n_j=N$, the generalized Kasner exponents satisfy
\begin{align}
n_1+n_2+n_3 &= N,  \label{constraint}\\
n_1^2+n_2^2+n_3^2 &= N^2.\label{constraint-sqr}
\end{align}
where $N$ is an arbitrary constant. For $N=1$, Eqs.~\eqref{constraint} and \eqref{constraint-sqr} recover the standard Kasner conditions corresponding to the vacuum solution. Since $s\propto t^{N/3}$ and hence $ABC\propto t^N$, the parameter $N$ determines the power-law behavior of both the average scale factor and the volume scale factor. The corresponding average Hubble parameter is $H=N/(3t)$, and the relation $3H^2=\sigma^2$ gives the shear scalar $\sigma^2=N^2/(3t^2)$. Thus, for $N\neq0$, the generalized Kasner solution has nonvanishing shear, which decreases as $t^{-2}$. For $N>0$, the average scale factor increases with $t$, corresponding to an overall expansion of the spatial volume, whereas for $N<0$ it decreases, corresponding to an overall contraction.

For $N=0$, Eq.~\eqref{constraint-sqr} requires $n_1=n_2=n_3=0$, yielding the trivial static universe, consistent with the discussion above. Using Eqs.~\eqref{constraint} and \eqref{constraint-sqr}, the directional pressures obtained from Eqs.~\eqref{BI11-zero}--\eqref{BI33-zero} are

\begin{align}
p_x=\frac{(1-N)(N-n_1)}{\kappa t^2}, \\
p_y=\frac{(1-N)(N-n_2)}{\kappa t^2},   \\
p_z=\frac{(1-N)(N-n_3)}{\kappa t^2}.
\end{align}
The pressure signs provide a simple classification of the expanding family. For $N>0$, Eq.~\eqref{constraint-sqr} implies $n_i\leq N$. Hence $0<N<1$ gives nonnegative directional pressures and satisfies the NEC and weak energy condition, while $N>1$ gives nonpositive pressures with nonzero negative entries and violates the NEC. Since $s\propto t^{N/3}$, mean accelerated expansion requires $N>3$ and lies in the latter branch. This stronger bound reflects the shear term in Eq.~\eqref{eq:bianchi-raychaudhuri}.

Note that for $N=1$, one has $p_x=p_y=p_z=0$, as in the standard Kasner solution, while the departure of the directional pressures from the vacuum Kasner case is controlled by the factor $1-N$. Taking the trace of the EFE, $R=-\kappa T$, where $R$ is the Ricci scalar and $T=T^{\mu}{}_{\mu}$ is the trace of the EMT, gives
\begin{align}
R=\frac{2N (N-1)}{t^2}.
\end{align}
This highlights an important distinction between the generalized solution and the standard Kasner solution. Since the latter is a vacuum solution, its Ricci scalar vanishes, $R=0$. In the generalized case, however, the Ricci scalar is nonzero for $N\neq\{0,1\}$, showing that the zero-density source is not equivalent to vacuum despite its vanishing energy density.
Nevertheless, since the spatial hypersurfaces of the Bianchi Type~I spacetime are flat, the three-dimensional Ricci scalar remains zero, ${}^{3}R=0$; see Eq.~\eqref{bianchiI}.

For generalized solutions with $N\neq0,1$, the Ricci scalar diverges as $t\rightarrow0$, indicating a curvature singularity. For $N=1$, although $R=0$, the full spacetime curvature is generally nonzero and the Kasner singularity at $t=0$ persists, except at the flat Kasner points. Thus, the vanishing of the Ricci scalar alone does not imply the absence of a curvature singularity.

In the orthonormal frame, the independent curvature components are $R_{\hat0\hat i\hat0\hat i}=-(\dot H_i+H_i^2)$ and $R_{\hat i\hat j\hat i\hat j}=H_iH_j$ for $i\neq j$, up to the common Riemann-sign convention. Squaring and summing gives $4\sum_i(\dot H_i+H_i^2)^2+4\sum_{i<j}H_i^2H_j^2$. With $H_i=n_i/t$, the full curvature invariant is
\begin{align}
R_{\mu\nu\alpha\beta}R^{\mu\nu\alpha\beta}
=\frac4{t^4}\left[\sum_i n_i^2(n_i-1)^2+\sum_{i<j}n_i^2n_j^2\right].
\label{eq:kasner-kretschmann}
\end{align}
It vanishes only for the static exponents or a permutation of $(1,0,0)$, and makes the exceptional flat Kasner points explicit.

A convenient parametrization of Eqs.~\eqref{constraint} and \eqref{constraint-sqr} is obtained by introducing an angular parameter $\varphi$ in addition to $N$~\cite{Ellis:1998ct,Ellis2012,Groen:2007zz}:
\begin{align}
n_1&=\frac{N}{3}\left(1+2\cos\varphi\right), \label{par1}\\
n_2&=\frac{N}{3}\left[1+2\cos\left(\varphi-\frac{2\pi}{3}\right)\right], \label{par2}\\
n_3&=\frac{N}{3}\left[1+2\cos\left(\varphi+\frac{2\pi}{3}\right)\right]. \label{par3}
\end{align}
The parameter $\varphi$ determines how the directional expansion and contraction are distributed among the three spatial directions. Together with $N$, it provides a two-parameter parametrization of the generalized Kasner family.

A simple subclass of particular interest is the generalized locally rotationally symmetric (LRS) Kasner case, for which $n_2=n_3=n$, where $n$ is a constant. For $n\neq0$, substituting this relation into Eqs.~\eqref{constraint}--\eqref{constraint-sqr} gives\footnote{For the special case $n=0$, the constraint is satisfied independently of $n_1$, giving $n_2=n_3=0$ with arbitrary $n_1$.}
\begin{align}
 n_1=-\frac{n}{2}\;,\; N= \frac{3n}{2}.
\end{align}
This case corresponds to $\varphi=\pi$ in the parametrization above. For this subclass, Eq.~\eqref{shear_bianchi} gives the shear scalar $\sigma^2=3n^2/(4t^2)$. The corresponding directional pressures are as follows
\begin{align}
p_x&=\frac{n(2-3n)}{\kappa t^2},\\
p_y&=p_z=\frac{n(2-3n)}{4\kappa t^2},
\end{align}
so that $p_x=4p_y=4p_z$. For $n=2/3$, the pressures vanish and the exponents reduce to $n_1=-1/3$ and $n_2=n_3=2/3$, recovering the standard LRS Kasner vacuum solution. For $n\neq \{0,2/3\}$, this family departs from the vacuum LRS Kasner case and is supported by anisotropic directional pressures.

\subsubsection{Exponential Expansion}

For the zero energy density case, we now consider a second class of solutions by taking two of the directional Hubble parameters to be constant, $H_x=\beta_1$ and $H_y=\beta_2$. For $\beta_1+\beta_2\neq0$, Eq.~\eqref{BI00-zero} implies that $H_z=\beta_3$ is also constant, so that the directional scale factors evolve exponentially. Equations~\eqref{BI00-zero}--\eqref{BI33-zero} then reduce to

\begin{align}
    \label{sys_eq1}
    \beta_1\beta_2+\beta_2\beta_3+\beta_1\beta_3&=0,\\
    \label{sys_eq2}
    -\beta_2^2-\beta_3^2-\beta_2\beta_3&=\kappa p_x,\\
    \label{sys_eq3}
    -\beta_3^2-\beta_1^2-\beta_1\beta_3&=\kappa p_y,\\
    \label{sys_eq4}
    -\beta_1^2-\beta_2^2-\beta_1\beta_2&=\kappa p_z.
\end{align}
Combining Eqs.~\eqref{sys_eq1}--\eqref{sys_eq4} gives
\begin{equation}
    \beta_1^2+\beta_2^2+\beta_3^2 =-\dfrac{\kappa}{2}(p_x+p_y+p_z),
\end{equation}
which implies $p_x+p_y+p_z\leq0$.

To obtain a simple anisotropic solution, we impose partial isotropy, $\beta_2=\beta_3$. Equation~\eqref{sys_eq1} then gives $\beta_2(\beta_2+2\beta_1)=0$. For $\beta_2\neq0$, this yields $\beta_1=-\beta_2/2$. Substituting this relation into Eqs.~\eqref{sys_eq2}--\eqref{sys_eq4}, we obtain the directional pressures satisfying

\begin{equation}
    p_x=4p_y=4p_z=-\dfrac{3\beta_2^2}{\kappa}.
\end{equation}
For $H_y=H_z=\beta_2>0$, the $y$- and $z$-directions exhibit de Sitter-like exponential expansion, whereas $H_x=-\beta_2/2<0$ describes contraction along the $x$-direction. Nevertheless, the average Hubble parameter is $H=\beta_2/2>0$, so that the spatial volume expands despite the contraction along one direction. For this solution, Eq.~\eqref{shear_bianchi} gives $\sigma^2=3\beta_2^2/4$, so the shear remains nonzero and constant. Hence, the exponential expansion does not approach isotropy. At the same time, the relation $3H^2=\sigma^2$ shows that, for $\rho=0$, the nonvanishing average expansion is necessarily accompanied by nonzero shear. Thus, this solution provides an explicit example of an expanding Bianchi Type~I universe with zero energy density and persistent anisotropy.

\subsubsection{LRS Bianchi Type~I Subclass}

We now consider the LRS Bianchi Type~I subclass of the metric~\eqref{bianchi} by setting $B=C$, so that $H_y=H_z$ and $p_y=p_z$. Equations~\eqref{BI00-zero}--\eqref{BI33-zero} then reduce to
\begin{align}
    (2H_x+H_y)H_y&=0,  \label{lrs00}\\
    -2\dot{H}_y-3H^2_y&=\kappa p_x,\label{lrs11}\\
    -\dot{H}_x-\dot{H}_y-H^2_x-H^2_y-H_xH_y&=\kappa p_y. \label{lrs22}
\end{align}
The continuity equation~\eqref{BIcont-zero} becomes
\begin{align}   \label{lrscont}
    H_xp_x+2H_yp_y=0.
\end{align}
Only three of Eqs.~\eqref{lrs00}--\eqref{lrscont} are independent, so one additional relation is required to close the system. Equation~\eqref{lrs00} admits two solutions: (i) $H_y=0$ and (ii) $H_y=-2H_x$. We consider them separately.

\begin{enumerate}[nosep,wide,label=(\roman*)]
    \item For $H_y=0$, Eq.~\eqref{lrs11} gives $p_x=0$, while Eq.~\eqref{lrs22} reduces to
    \begin{align}
        -\dot{H}_x-H_x^2=\kappa p_y.
    \end{align}
    In this case, the spacetime is static in the $y$- and $z$-directions, while the $x$-direction may evolve.  Equation~\eqref{shear_bianchi} gives $\sigma^2=H_x^2/3$.  The zero-density source has a nonvanishing pressure in general except for the two cases $H_x=0$ or $H_x=1/(t+\mathcal{C}_4)$, where $\mathcal{C}_4$ is an arbitrary integration constant.
    \item For $H_y=-2H_x\neq0$, the continuity equation~\eqref{lrscont} gives $H_x(p_x-4p_y)=0$. Since $H_x\neq0$, it follows that $p_x=4p_y$. Equations~\eqref{lrs11} and \eqref{lrs22} then reduce to
    \begin{align}
        \dot{H}_x-3 H_x^2=\kappa p_y.
    \end{align}
    For this case, Eq.~\eqref{shear_bianchi} gives $\sigma^2=3H_x^2$. In this solution, expansion in the $x$-direction accompanies contraction in the $y$- and $z$-directions, and conversely. The zero-density source has nonvanishing pressure in general except for the case $H_x=\frac{1}{\mathcal{C}_5-3 t}$ with $\mathcal{C}_5$ being an arbitrary constant.
\end{enumerate}

The LRS condition leaves one function undetermined. A directional-pressure choice, for example for $p_y$, or a specified directional expansion history $H_x$, is therefore needed to obtain a particular solution with nonvanishing pressure. These solutions are exact geometric realizations of conserved anisotropic EMTs with zero density. They do not by themselves specify microscopic matter variables, their constitutive equations, or the stability of an anisotropic medium. In particular, a single homogeneous timelike $P(X,\phi)$ scalar has isotropic spatial stress and cannot supply these directional pressures. An anisotropic field configuration or a different material sector is required; the Bianchi examples are not continuations of the homogeneous scalar reconstructions of Sec.~\ref{sec:scalar-realizations}.

\section{Cosmological Interpretation and Connections to Dark-Energy Inference}
\label{sec:inference}

\begin{figure}[!htbp]
\centering
\includegraphics[width=\columnwidth]{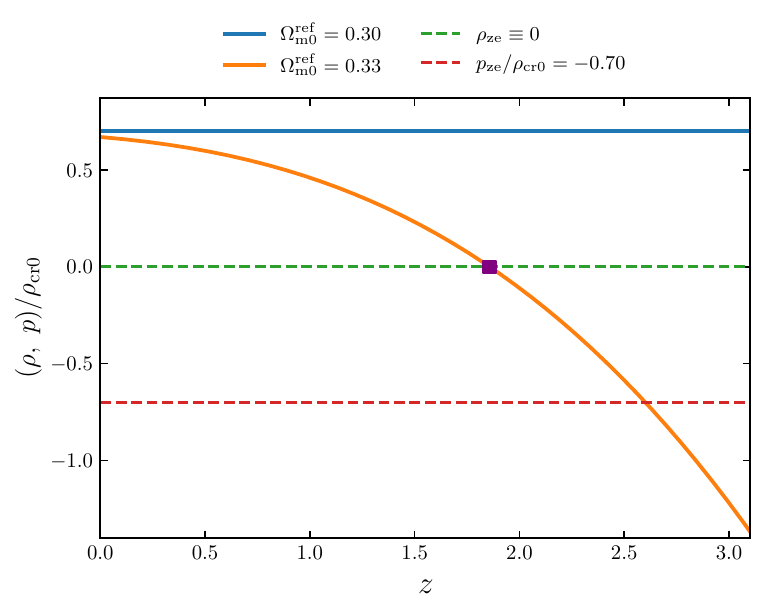}
\caption{A persistently zero source density and a zero-crossing inferred dark energy density, both rescaled by $\rho_{\rm cr0}$, for the fixed expansion history $H^2/H_0^2=0.3(1+z)^3+0.7$. In the chosen interacting two-source system, the pressure-only source has $\rho_{\rm ze}\equiv0$ and $p_{\rm ze}/\rho_{\rm cr0}=-0.7$. The two inferred dark energies are obtained by subtracting conserved reference matter sectors with $\Omega_{\rm m0}^{\rm ref}=0.30$ (solid blue curve) and $\Omega_{\rm m0}^{\rm ref}=0.33$ (solid orange curve) from the full background. The former corresponds to the cosmological constant with $\Omega_{\Lambda0}^{\rm ref}=0.7$ whereas the density of the latter crosses zero at $z_\dagger\simeq1.86$ marked by the purple square. The only difference between the inferred dark energy densities (solid curves) is the subtracted reference matter density, which scales as $\Omega_{\rm m0}^{\rm ref} (1+z)^3$. Neither the zero-density source nor the actual matter interacting with it undergo a density sign transition. The values used here are illustrative and do not represent competing fits or imply that matter density constraints can be ignored.}
\label{fig:inferred-crossing}
\end{figure}

Returning to spatially flat FLRW cosmology, we ask what an expansion history determines about the sources that produce it. The background correspondences of Sec.~\ref{sec:particular} show why this question requires an explicit choice of constituent densities and their conservation laws. We first relate the interacting matter plus pressure-only system to a reference description with conserved matter and an inferred dark-energy remainder. Figure~\ref{fig:inferred-crossing} illustrates how the remainder can cross zero even though the density of the pressure-only constituent remains identically zero. We then examine what additional information is supplied by perturbations, using the completed scalar models as examples. These background relations and dynamical tests are not observational fits; they clarify how the interacting construction relates to the separately conserved, density-sign-aware diagnostics of Ref.~\cite{Akarsu:2026pia}.

\subsection{Reconstructed pressure and standard expansion histories}

We consider pressureless matter as the standard source interacting with a zero-density source. As a reference, we write the late-time expansion history of a spatially flat universe as follows
\begin{align}
\frac{3H^2}{\kappa}
=\rho_{\rm m0}^{\rm ref}a^{-3}+\rho_{{\rm de}0}^{\rm ref} f(a),
\label{eq:inverse-f-history}
\end{align}
where the superscript ``ref'' labels the reference GR-like split, in which the matter obeys its usual background evolution, and the remainder is identified as a separately conserved, inferred (effective) dark energy sector. In the pressure-only source representation, the right-hand side of Eq.~\eqref{eq:inverse-f-history} equals the interacting matter density $\rho_{\rm m}\equiv\rho_{\rm s}$, see Eq.~\eqref{zero-fe00}. On an expanding interval where $\rho_{\rm m}>0$, the required pressure and interaction kernel are given by
\begin{align}
p_{\rm ze}&=-\rho_{{\rm de}0}^{\rm ref}
\left[f+\frac a3\frac{{\rm d}f}{{\rm d}a}\right],
\label{pres-de}\\
\mathcal Q&=-3Hp_{\rm ze}
=H\rho_{{\rm de}0}^{\rm ref}\left[3f+a\frac{{\rm d}f}{{\rm d}a}\right].
\label{eq:inverse-f-transfer}
\end{align}
Eq.~\eqref{eq:inverse-f-transfer} can be equivalently expressed as $\mathcal Q=H[a\,{\rm d}\rho_{\rm m}/{\rm d}a+3\rho_{\rm m}]$. The sign follows from the convention that $\mathcal Q>0$ transfers energy to the standard component. A prescribed differentiable signed $f(a)$ is admissible on intervals where the reconstructed total density is nonnegative. An isolated zero of $f$ does not make $p_{\rm ze}$ singular when its derivative remains finite. An abrupt behavior in the inferred density profile $f$ can be implemented through a specific regularization or distributional matching, but it is outside the smooth behavior class considered here.

The familiar cosmic expansion histories provide simple examples of this reconstruction. For $f=1$, one has $p_{\rm ze}=-\rho_{{\rm de}0}^{\rm ref}$, recovering the constant-pressure $\Lambda$CDM background in Sec.~\ref{sec:lcdm}. Defining $w_{\rm de}\equiv p_{\rm de}/\rho_{\rm de}$, an inferred dark energy with a constant EoS parameter $w_{\rm de}=w_{\star}={\rm const}.$ leads to
\begin{align}
f(a)=a^{-3(1+w_{\star})},\qquad p_{\rm ze}=w_{\star}\rho_{{\rm de}0}^{\rm ref}f.
\label{eq:inverse-wcdm}
\end{align}
For the Chevallier--Polarski--Linder (CPL) parametrization $w_{\rm de}(a)=w_0+w_a(1-a)$~\cite{Chevallier:2000qy,Linder:2002et}, Eqs.~\eqref{eq:inverse-f-history} and~\eqref{pres-de} yield
\begin{equation}
\begin{aligned}
f(a)&=a^{-3(1+w_0+w_a)}e^{-3w_a(1-a)},\\
p_{\rm ze}(a)&=[w_0+w_a(1-a)]\rho_{{\rm de}0}^{\rm ref}f(a).
\label{eq:inverse-cpl}
\end{aligned}
\end{equation}
These reconstructions are phenomenological equivalents of the considered dark energy models producing the same background evolution, but this does not imply action-level equivalence. Here $w_{\rm de}$ characterizes the inferred dark-energy sector wherever its density is nonzero. It is not an EoS parameter of the interacting zero-density constituent: since $\rho_{\rm ze}\equiv0$, the ratio $p_{\rm ze}/\rho_{\rm ze}$ is undefined throughout the evolution, and we do not assign $w_{\rm ze}$ to that source. The standard CPL density with positive present-day value, viz., $\rho_{{\rm de}0}^{\rm ref}>0$, remains positive for every finite $a>0$ as seen in Eq.~\eqref{eq:inverse-cpl}. Conversely, a directly specified $f(a)$ can attain negative values or even cross zero. At an isolated zero of $p_{\rm ze}$, the strict non-zero pressure condition of the zero-density source no longer holds. Extending a particular realization through the point (0,0) in the $(\rho_i,p_i)$ plane therefore requires a separate analysis of the background dynamics and its regularity.
Separately conserved constituents such as baryons and radiation, as well as other independently constrained sectors can be incorporated into the two-component interacting system described by Eqs.~\eqref{zero-fe00} and~\eqref{zero-fe11}. Denoting their total density and pressure by $\rho_{\rm ext}$ and $p_{\rm ext}$, respectively, the general reconstruction relations become
\begin{align}
\rho_{\rm s}&=\frac{3H^2}{\kappa}-\rho_{\rm ext},\label{eq:inverse-ext-rho}\\
p_{\rm ze}&=-\frac{2\dot H+3H^2}{\kappa}-p_{\rm ext}-p_{\rm s}.
\label{eq:inverse-ext-pres}
\end{align}
The pressure of each standard component must be included consistently as in Eq.~\eqref{eq:inverse-ext-pres}, and the reconstructed density must remain within the physical domain of that component. The microscopic coupling can then be restricted to an unknown dark component rather than applied to all pressureless matter. This allows us to assume no such coupling to observed baryons.

\subsection{Zero-crossing of the inferred density versus zero-density source}

To make the comparison in Fig.~\ref{fig:inferred-crossing} explicit, we proceed with the reference late-time expansion history given in Eq.~\eqref{eq:inverse-f-history}, which can be recast to define the inferred dark energy density as 
\begin{align}
\rho_{\rm de}^{\rm inf}(z)=\frac{3H^2(z)}{\kappa}
-\rho_{\rm m0}^{\rm ref}(1+z)^3,
\label{eq:inferred-density}
\end{align}
where radiation is neglected for this illustration. Its corresponding pressure is $p_{\rm de}^{\rm inf}=-(2\dot H+3H^2)/\kappa$ since the inferred dark energy, by construction, satisfies local energy-momentum conservation. When the standard source is pressureless matter, we have $p_{\rm de}^{\rm inf}=p_{\rm ze}$ but $\rho_{\rm de}^{\rm inf}\neq\rho_{\rm ze}$ in general. Under this assumption, the continuity equation of the inferred dark energy and the corresponding pressure of the zero-density source read 
\begin{align}
\rho_{\rm de}^{\rm inf'}-\frac{3}{1+z}(\rho_{\rm de}^{\rm inf}+p_{\rm ze})=0,
\label{eq:inferred-conservation}\\
p_{\rm ze}=\frac{2(1+z)HH'-3H^2}{\kappa}.
\label{eq:inverse-redshift}
\end{align}
where a prime denotes differentiation with respect to redshift, $'={\rm d}/{\rm d}z$. Note that in contrast to the effective dark energy, the constituents of the interacting system---actual pressureless matter and the zero-density source in this case---exchange energy. This explicitly demonstrates the background dark degeneracy~\cite{Kunz:2007rk}.

We fix the background expansion history as $H^2/H_0^2=0.3(1+z)^3+0.7$. Then, the zero-density source that can realize this history has $\rho_{\rm ze}\equiv0$ and constant pressure $p_{\rm ze}=-0.7\rho_{\rm cr0}$. Interpreting the same expansion with $\Omega_{\rm m0}^{\rm ref}=0.33$ instead yields
\begin{equation}
\begin{aligned}
\frac{\rho_{\rm de}^{\rm inf}}{\rho_{\rm cr0}}=0.7-0.03(1+z)^3,\\
z_\dagger=\left(\frac{0.7}{0.03}\right)^{1/3}-1\simeq1.86.
\label{eq:inferred-crossing-example}
\end{aligned}
\end{equation}
The inferred sector crosses zero while the pressure-only constituent in the chosen interacting system remains identically at zero. At the crossing,
\begin{align}
\mathcal I_{\rm de}^{\rm inf}(z_\dagger)&=p_{\rm ze},\qquad
\mathcal M_{\rm de}^{\rm inf}(z_\dagger)=3p_{\rm ze},
\end{align}
but away from it these inferred diagnostics differ from $\mathcal I_{\rm ze}$ and $\mathcal M_{\rm ze}$. The ratio $p_{\rm ze}/\rho_{\rm de}^{\rm inf}$ has a pole at the crossing without a singularity in $H$ or the EMT. Figure~\ref{fig:inferred-crossing} holds the expansion history fixed and changes only the conserved reference-matter subtraction. The inferred dark energy density can therefore cross zero while the pressure-only constituent remains identically zero; the crossing shown in Fig.~\ref{fig:inferred-crossing} is not a sign transition of that constituent.

The three figures describe successive aspects of the construction. Figure~\ref{fig:stress-plane} distinguishes the cosmological-constant EMT from a pressure-only EMT, even when their pressures agree. The negative-pressure expanding dust branch in Fig.~\ref{fig:pressure-branches} shows how a different assignment and evolution of the constituent densities can nevertheless reproduce a $\Lambda$CDM expansion history. Figure~\ref{fig:inferred-crossing} then demonstrates that a dark-energy density inferred by subtracting conserved reference matter need not coincide with the density of either interacting constituent. The distinction concerns what is being inferred, rather than the regularity of the measured expansion history.

This construction does not remove the observational constraints on matter. The adopted physical matter densities, their evolution, and the early-time calibration must be tested jointly. It shows instead why a zero-crossing of the density inferred after a particular subtraction is not a direct measurement of the energy density of an independently identified field. Conversely, the zero-density constituent condition is not a claim that the inferred dark energy density must vanish at all times, nor does this example establish that every physically sign-changing dark energy model is a subtraction artifact. Keeping the split explicit reconciles the present interacting construction with the separately conserved effective-sector diagnostics of Ref.~\cite{Akarsu:2026pia}.

\subsection{What distances and perturbations can test}

In a flat geometry and units $c=1$, the background distance measures involve $D_H=H^{-1}$ and $D_M(z)=\int_0^z {\rm d}\tilde z/H(\tilde z)$; BAO measure these relative to a sound-horizon calibration. A shared $H(z)$ gives shared late-time distances, but not automatically the same sound horizon, CMB spectra, or growth. The DR2 BAO analysis~\cite{DESI:2025DR2} and the later Lyman-$\alpha$ AP measurement at $z_{\rm eff}=2.33$~\cite{DESI:2026Lya} are therefore useful motivations for testing the intermediate-redshift behavior of a completed model. They are not evidence for the particular pressure-only decomposition presented here.

Unanchored supernova distances additionally constrain the shape of the late-time expansion history, so changing a BAO calibration alone need not relax the joint geometric restrictions~\cite{Pedrotti:2025ccw}. The distinction becomes sharper once the interacting density no longer obeys the usual dust dilution. Keeping a CMB-inferred matter density fixed while changing its subsequent exchange cannot be implemented simply by reusing $\rho_{\rm m}\propto a^{-3}$ at all epochs. The transfer history, any conserved baryons and radiation, and the calibrated sound horizon must be evolved consistently. Growth, gravitational lensing, and the CMB then probe the total density, momentum, and anisotropic-stress perturbations. For example, the linear GR constraints in Newtonian gauge are~\cite{Ma:1995ey}
\begin{equation}
\begin{aligned}
k^2\Phi+3\mathcal H(\partial_\tau\Phi+\mathcal H\Psi)
&=-4\pi Ga^2\delta\rho_{\rm tot},\\
k^2(\partial_\tau\Phi+\mathcal H\Psi)&=4\pi Ga^2\sum_i Q_i,\\
k^2(\Phi-\Psi)&=12\pi Ga^2\sum_i\Pi_i.
\label{eq:einstein-constraints}
\end{aligned}
\end{equation}
A source with $\bar\rho_{\rm ze}=0$ can thus affect metric perturbations through the allowed unnormalized variables and through its exchange with matter. Distinct closures can produce different growth and lensing at the same background. Where the total EMT is exactly related by the dust--vacuum rewriting of Sec.~\ref{sec:lcdm}, gravitational observables cannot distinguish the mere relabeling; physical discrimination requires a completion that differs in total stress or independently observable constituent physics.

The covariant realizations in Appendix~\ref{app:action} turn this distinction into a calculable example. The exponential proportional-pressure model already shows that a specified interaction can admit an accelerating background without a growing quasistatic power-law dust mode; the exponents are given in Eq.~\eqref{eq:growth-exponents}. The constant-pressure model of Sec.~\ref{app:constant-pressure-growth} addresses the more direct comparison with $\Lambda$CDM. It has exactly the same expansion history, but its scalar-dependent particle mass changes both the dust force and the velocity damping. In this realization, the scalar pressure is constant along the homogeneous solution, not throughout the perturbed spacetime, and the scalar and dust velocities need not coincide at first order. The assumptions required for the exact dust--vacuum rewriting are therefore not imposed. The resulting growth equation, Eq.~\eqref{eq:growth-lcdm-trajectory}, differs from that of conserved dust in the reference cosmology.

For the illustrative reference parameters $\Omega_{{\rm m}0}^{\Lambda{\rm CDM}}=0.3$ and $\Omega_\Lambda^{\Lambda{\rm CDM}}=0.7$, the mode chosen to grow during matter domination reaches a maximum near $z\simeq0.55$ and subsequently decays, while the reference $\Lambda$CDM dust mode continues to grow. The transition is not simultaneous with the onset of accelerated expansion: the selected mode is still growing when acceleration begins. This example distinguishes a true difference in constituent dynamics from the exact dust--vacuum relabeling discussed above. The shared background distances do not remove that difference, because the two models no longer have the same perturbed total EMT. The comparison applies to the specified trajectories and matter-sourced quasistatic regime; it is neither a likelihood constraint nor a statement about all solutions of either action.

The connection to the $H_0$, $S_8$, and growth-index tensions is therefore a possible direction for future model building, rather than a result of this paper. A complete likelihood analysis requires the covariant energy-momentum exchange, perturbative initial conditions, pressure response, and validity regime. The formulation above identifies regular variables and consistency conditions needed for that analysis, while avoiding an EoS parametrization that would discard the source at the outset.

\subsection{Vacuum offsets and the limitation of zero-density identities}

The square-root identity is also informative about the cosmological constant problem. Adding an allowed constant term gives
\begin{align}
P=A(\phi)\sqrt X-\Lambda_{\rm off},\qquad
\rho_\phi=\Lambda_{\rm off}.
\label{eq:vacuum-offset}
\end{align}
The inertial mass density remains $A(\phi)\sqrt X$ while the vacuum offset enters the Hamiltonian constraint. Therefore the algebraic cancellation of kinetic contributions does not eliminate an independent vacuum term. Whether a symmetry or a coupled constrained theory can preserve the zero-density structure against corrections remains an open question that we do not address here. This separates the present classification from mechanisms proposed to address vacuum energy gravitation~\cite{Weinberg:1988cp,Afshordi:2008xu,Aslanbeigi:2011si}, while identifying a concrete robustness test for future realizations.

\section{Conclusion}
\label{sec:Conclusion}

We have examined cosmological sources with vanishing comoving energy density and nonzero pressure. Such a source has a nonvanishing EMT with a zero timelike eigenvalue; it is not vacuum in the sense of $T_{\mu\nu}=0$. For a perfect fluid source, $\mathcal I_{\rm ze}=p_{\rm ze}$ and $\mathcal M_{\rm ze}=3p_{\rm ze}$ remain well defined, whereas the conventional EoS parameter is undefined because $\rho_{\rm ze}\equiv0$. Its pressure determines both its sector-level NEC character and its Raychaudhuri contribution. Neither statement, by itself, fixes the corresponding properties of a multicomponent universe.

The principal FLRW restriction follows from conservation. An isolated density zero is compatible with separate conservation, whereas maintaining $\rho_{\rm ze}\equiv0$ with $p_{\rm ze}\neq0$ throughout a nonstatic interval requires energy exchange. In the two-component convention used here, $\mathcal Q=-3Hp_{\rm ze}$, so pressure and background transfer cannot be chosen independently. For an exact perfect pressure-only EMT, its covariant divergence also fixes the relation between pressure gradients, acceleration, and momentum transfer. These identities supply consistency conditions, rather than an independently specified microscopic interaction.

The resulting background equations encompass several familiar cosmological constructions. Hoyle's creation tensor, particle creation pressure, and bulk viscous pressure can be represented as pressure-only contributions, while their physical mechanisms remain distinct. A pressure proportional to the standard component density gives power law, de Sitter, and big rip histories, with Hoyle's steady-state universe as a special case. Constant pressure yields a constant contribution in addition to the evolving term in the standard component density. For a pressureless standard component, the negative-pressure branch reproduces the flat-$\Lambda$CDM expansion history. With additional assumptions of a common four-velocity and a pressure constant throughout spacetime, the total EMT can in fact be rewritten exactly as conserved dust plus vacuum energy. Other perturbative completions need not share this equivalence. The constant-pressure family also includes bouncing solutions and, for positive pressure with $w>-1$, an expansion from a big bang to a maximum scale factor followed by recollapse.

The scalar realizations distinguish two different implementations of zero density. Canonical, phantom and power-law kinetic models enforce it along reconstructed homogeneous trajectories, rather than as a Lagrangian identity. The linear kinetic proportional-pressure model has a logarithmic field trajectory in the scale factor and an exponential potential, including its constant-potential de Sitter limit. Constant pressure gives a constant potential and a field linear in cosmic time. The positive-normalization power-law models considered here realize positive pressure and include a logarithmic de Sitter special case for any allowed kinetic exponent. Preserving a reconstructed zero-density trajectory does not, by itself, establish that the trajectory is an attractor.

Requiring $2XP_{,X}-P\equiv0$ on a timelike $P(X,\phi)$ domain instead selects $P=A(\phi)\sqrt X$. On a field interval where $A$ is nonzero and has fixed sign, a field redefinition reduces this kinetic sector to a signed potential-free cuscuton. Its degeneracy is distinct from an ordinary propagating scalar with a large sound speed. For the nondegenerate scalar class, the zero-density condition gives $p_\phi=2XP_{,X}$. Negative pressure is therefore incompatible with simultaneously positive intrinsic kinetic and gradient coefficients when the interaction does not change the principal kinetic structure. In particular, the phantom scalar reconstruction has a wrong-sign kinetic term despite its luminal intrinsic sound speed. These restrictions apply to the specified scalar class, not to every effective pressure-only sector or constrained completion.

At linear order, unnormalized density and momentum perturbations avoid division by a vanishing background density. A regular reference congruence also avoids defining the interaction through a total energy frame when $\mathcal I_{\rm tot}=0$. If zero density is imposed only on the background, $\delta\rho_{\rm ze}$ is an allowed, first-order gauge-invariant perturbation. Imposing $\delta\rho_{\rm ze}=0$ is an additional condition that turns the continuity equation into a constraint and restricts the admissible pressure and transfer perturbations. It does not imply an exact nonlinear pressure-only EMT or, without further closure, a vanishing rest-frame pressure perturbation. A specified perturbative model must supply these relations and satisfy the resulting constraints.

Relaxing isotropy changes the conservation restriction. Shear, energy flux, and anisotropic stress can support a separately conserved source with zero density relative to a specified congruence. The pure zero-density Bianchi Type~I solutions provide explicit examples: generalized Kasner, exponential, and LRS geometries are supported by nonzero directional pressures, while the constraint $3H^2=\sigma^2$ prevents an approach to isotropy in the normalized-shear sense wherever $H\neq0$. The generalized Kasner family also exhibits nonzero Ricci curvature away from its vacuum and static limits. These exact geometries demonstrate admissible stress--energy structures; identifying a stable microscopic anisotropic medium is a separate task.

The inverse-pressure construction connects these source models to cosmological inference. An expansion history can be represented using an interacting pressure-only constituent or using conserved reference matter and an inferred dark-energy remainder. A zero-crossing of that remainder is not necessarily a zero-crossing of the energy density of an independently identified field. The constituent densities and exchange laws must therefore be kept explicit when relating a background reconstruction to a physical model.

The mass-only coupling of a scalar with a linear kinetic term to number-conserving dust provides two examples in which the background and perturbations follow from specified actions. In the exponential proportional-pressure model, the reconstructed zero-density trajectory is a homogeneous saddle. Its leading quasistatic dust force satisfies $G_{\rm eff}/G=1+6\alpha$, and a growing power-law mode exists only for $\alpha>-1/6$. The accelerating trajectory, which requires $\alpha<-1/3$, has no such growing mode. The force strength is also constrained more generally within this action class: on a spatially flat two-component zero-density background, the same gradient-dominated reduction gives $G_{\rm eff}/G=1+6p_\phi/\rho_{\rm s}$. Thus, the background balance needed to maintain zero density constrains the dust force, not only the energy transfer.

A different choice of mass function and constant potential realizes the constant-pressure history. It reproduces the $\Lambda$CDM expansion while giving $G_{\rm eff}/G=1-6\Omega_\Lambda(a)$. For reference matter and vacuum fractions $0.3$ and $0.7$ today, the initially growing dust mode continues to grow briefly after the background starts accelerating, reaches a maximum near $z\simeq0.55$, and then decays. Force reversal, the onset of acceleration, and the end of growth are separate events. This provides a concrete example in which matching the background expansion does not match the clustering. The homogeneous saddle was established for the exponential model, not this constant-pressure realization; the growth calculations likewise concern selected trajectories and approximations, rather than all histories of the two actions. Both negative-pressure scalar realizations retain the intrinsic phantom kinetic problem.

A natural next step is to investigate matter-coupled realizations of the potential-free square-root scalar of Sec.~\ref{sec:sf-general}, for which $\rho_\phi\equiv0$ is an identity on the timelike domain. Its degeneracy places it outside the nondegenerate kinetic obstruction of Sec.~\ref{sec:kinetic-obstruction}, although its uncoupled field equation~\eqref{eq:cuscuton-zero-eom} enforces $\Theta=0$. A nonstatic cosmological realization would therefore require an interaction that supplies the exchange in Eq.~\eqref{eq:zero-int2} while preserving the constraint structure. Determining the resulting matter force and the stability of the coupled system would establish whether this particular scalar construction can support accelerated expansion without the problems found in the propagating examples.

The results therefore distinguish the existence of a zero-density source, its realization by fields and interactions, and its suitability as a cosmological model. The conservation analysis and exact isotropic and anisotropic solutions establish nontrivial stress structures for which the conventional EoS ratio $w=p/\rho$ is undefined. The scalar kinetic test and the explicit coupled models identify additional dynamical requirements, without excluding positive-pressure sources, constrained scalar sectors, or more general interactions. A physical realization must specify how its nonzero pressure is sustained, whether through intercomponent exchange or imperfect stresses, and test the resulting dynamics. The absence of comoving energy density does not settle those questions; the remaining stress--energy and its dynamics do.

\begin{acknowledgments}
\"{O}.A. acknowledges support from the
Turkish Academy of Sciences through the Outstanding Young Scientist
Award programme (T\"{U}BA-GEB\.{I}P). B.\"{O}. acknowledges support from the Istanbul Technical University Research Fund under Grant No. TDK-2025-47762. N.M.U. is supported by the
Scientific and Technological Research Council of T{\"u}rkiye
(T\"{U}B\.{I}TAK) through the 2218 National Postdoctoral Research
Fellowship Programme, Project No.~124C450. This article is based upon
work from COST Action CA21136, ``Addressing observational tensions in
cosmology with systematics and fundamental physics'' (CosmoVerse),
supported by COST (European Cooperation in Science and Technology).
\end{acknowledgments}

\appendix
\section{Covariant Dust--Scalar Realizations: Homogeneous Stability and Subhorizon Growth}
\label{app:action}

The reconstructions of Sec.~\ref{sec:scalar-realizations} determine scalar trajectories and potentials together with the homogeneous exchange needed to sustain zero density. We now complete two of those constructions by specifying covariant interactions. In both cases, the scalar controls the mass of number-conserving dust particles, so the same action fixes the background exchange, the scalar force, and the dust response to perturbations. We first choose exponential mass and potential functions to realize proportional pressure, examine the stability of that trajectory against homogeneous perturbations, and solve its leading subhorizon dust-growth equation. We then choose a different mass function and a constant potential to realize the constant-pressure $\Lambda$CDM background and determine how its clustering differs from that of conserved dust. The common action class facilitates the comparison; the two choices of functions define different models.

\subsection{Variable-mass dust and the proportional-pressure realization}

We take the classical action
\begin{equation}
\begin{aligned}
S={}&\int {\rm d}^4x\sqrt{-g}\left[\frac{R}{2\kappa}
+\varepsilon X-V(\phi)\right]\\
&-\sum_A\int m(\phi(x_A))\,{\rm d}s_A,
\label{eq:action-dust-scalar}
\end{aligned}
\end{equation}
where the sum is over particle worldlines, $\varepsilon=\pm1$, and $m(\phi)>0$ on the field interval under consideration. In the single-stream continuum limit, the particle sector is described by pressureless matter with conserved particle number, $\nabla_\mu(nu_{\rm s}^\mu)=0$, and density $\rho_{\rm s}=m(\phi)n$. Scalar-dependent particle masses are a standard way to implement a coupled dark sector~\cite{Amendola:1999er}; each choice of mass and potential functions below defines a particular realization.

Varying the scalar and particle sectors gives
\begin{equation}
\begin{aligned}
\varepsilon\Box\phi-V_{,\phi}&=n m_{,\phi},\\
\mathcal Q_{(\phi)}^\nu&=n m_{,\phi}\nabla^\nu\phi,
\qquad \mathcal Q_{(\rm s)}^\nu=-\mathcal Q_{(\phi)}^\nu.
\label{eq:action-transfer}
\end{aligned}
\end{equation}
Consequently, for a homogeneous field and common background velocity,
\begin{align}
\mathcal Q=\frac{{\rm d}\ln m}{{\rm d}\phi}\rho_{\rm s}\dot\phi,
\qquad n=n_0a^{-3},
\label{eq:action-background-Q}\\
\varepsilon(\ddot\phi+3H\dot\phi)+V_{,\phi}
=-\frac{{\rm d}\ln m}{{\rm d}\phi}\rho_{\rm s}.
\label{eq:action-KG}
\end{align}
These equations fix the energy exchange away from the reconstructed trajectory as well. The dust acceleration follows from the same vector,
\begin{align}
A_{\rm s}^\mu=-h_{\rm s}^{\mu\nu}\nabla_\nu\ln m.
\label{eq:dust-fifth-force}
\end{align}
It vanishes on the homogeneous background but need not vanish for perturbations. Thus, the component four-velocities are not required to coincide at first order.

For the proportional-pressure realization, take $p_\phi=\alpha\rho_{\rm s}$ on the target background. Let $\alpha\neq0$, $\alpha/\varepsilon>0$, and choose a signed nonzero constant
\begin{align}
c_\phi=\pm\sqrt{\frac{3\alpha}{\varepsilon\kappa}}.
\label{eq:action-c}
\end{align}
The model is specified by
\begin{align}
m(\phi)&=m_0\exp\!\left[-\frac{3\alpha}{c_\phi}(\phi-\phi_0)\right],\label{eq:action-mass}\\
V(\phi)&=-\frac{\alpha\rho_{\rm s0}}2
\exp\!\left[-\frac{3(1+\alpha)}{c_\phi}(\phi-\phi_0)\right],
\label{eq:action-reconstruction}
\end{align}
with $m_0n_0=\rho_{\rm s0}>0$. Direct substitution verifies the solution
\begin{equation}
\begin{aligned}
\phi-\phi_0&=c_\phi\ln a,\qquad
\rho_{\rm s}=\rho_{\rm s0}a^{-3(1+\alpha)},\\
3H^2&=\kappa\rho_{\rm s},\qquad
\rho_\phi=0,\qquad p_\phi=\alpha\rho_{\rm s}.
\label{eq:action-solution}
\end{aligned}
\end{equation}
Indeed $m=m_0a^{-3\alpha}$, $\mathcal Q=-3H\alpha\rho_{\rm s}$, and $V=-\varepsilon\dot\phi^2/2$. The Klein--Gordon equation is also satisfied, implying that the reconstructed action gives a consistent solution rather than merely specifying the background transfer rate. It applies on expanding intervals with positive $\rho_{\rm s}$. For $\alpha>0$ it uses the canonical scalar with negative potential; $\alpha<0$ uses the phantom scalar and inherits its kinetic problem. Baryons and radiation have not been included in this exact solution; the particles can be interpreted as a dark component for purposes of a further cosmological extension.

\subsection{Homogeneous stability of the proportional-pressure trajectory}
\label{app:homogeneous}

A reconstructed trajectory need not attract nearby solutions. We test this directly for the exponential model by allowing the homogeneous scalar density to depart from zero, while retaining the same $m(\phi)$ and $V(\phi)$. Introduce
\begin{align}
x=\frac{\sqrt\kappa\dot\phi}{\sqrt6 H},\qquad
y=\frac{\kappa V}{3H^2},\qquad
\Omega_{\rm s}=1-\varepsilon x^2-y,
\label{eq:action-phase-vars}
\end{align}
where $y$ is signed. This follows the autonomous-system approach for exponential potentials~\cite{Copeland:1997et,Amendola:1999er,Copeland:2006wr}, while retaining the sign of $V$. We define $\mathcal N=\ln a$ along with
\begin{align}
\lambda_{\rm d}=\frac{3(1+\alpha)}{c_\phi\sqrt\kappa},\qquad
\beta_{\rm d}=-\frac{3\alpha}{c_\phi\sqrt\kappa}.
\label{eq:action-slopes}
\end{align}
The homogeneous equations reduce to
\begin{align}
\frac{{\rm d}x}{{\rm d}\mathcal N}
&=-3x+\frac{\sqrt{3/2}}{\varepsilon}(\lambda_{\rm d} y-\beta_{\rm d}\Omega_{\rm s})
+\frac32x(1+\varepsilon x^2-y),\\
\frac{{\rm d}y}{{\rm d}\mathcal N}
&=y[-\sqrt6\lambda_{\rm d} x+3(1+\varepsilon x^2-y)].
\label{eq:action-autonomous}
\end{align}
The zero-density solution is a fixed point with
\begin{align}
x_* =\frac{\sqrt\kappa c_\phi}{\sqrt6},\qquad
y_*=-\frac\alpha2,\qquad \Omega_{{\rm s}*}=1,
\label{eq:action-fixed-point}
\end{align}
where $\varepsilon x_*^2=\alpha/2$. Linearizing in $(x,y)$ gives a Jacobian with
\begin{align}
\operatorname{tr}J=\frac32(\alpha-1),\qquad
\det J=-\frac92,
\label{eq:action-jacobian}
\end{align}
and its eigenvalues are
\begin{align}
\mu_\pm=\frac34\left[\alpha-1\pm\sqrt{(\alpha-1)^2+8}\right].
\label{eq:action-eigenvalues}
\end{align}
Their product is negative for every allowed $\alpha$, including the $\alpha=-1$ de Sitter case. The trajectory is therefore a saddle in the expanding homogeneous phase space, not an attractor. There is a growing homogeneous direction even in the canonical positive-pressure case.

The saddle demonstrates that this reconstructed background is an exact solution but is not selected as an attractor of the expanding homogeneous dynamics. The conclusion follows from the model defined by Eqs.~\eqref{eq:action-mass} and~\eqref{eq:action-reconstruction}, rather than from the conservation condition $\mathcal Q=-3Hp_\phi$ alone. An interaction beyond the mass-only form could agree on the target trajectory yet differ away from it and change its transverse stability. The growing homogeneous direction is also distinct from the intrinsic phantom ghost on the negative-pressure branch: the former describes evolution near a background solution, whereas the latter concerns the sign of a kinetic term. The full coupled quadratic action and constraints are needed to characterize all perturbative degrees of freedom, forces, and stability properties.

\subsection{Subhorizon dust growth in the proportional-pressure model}
\label{app:growth}

We next examine dust perturbations with nonzero comoving wave number $k$, rather than the homogeneous variations considered above. We use Newtonian gauge, $B=E=0$, with the metric convention of Sec.~\ref{sec:perturbation} and cosmic time. The scalar and single-stream dust have no linear anisotropic stress, so $\Phi=\Psi$. On scales well inside the Hubble radius, we seek the slowly varying scalar response sourced by the dust perturbations. This is the matter-sourced quasistatic part of the solution; independently excited scalar waves are not included. The calculation specializes the scalar-mediated force and growth analysis of Ref.~\cite{Amendola:2004Repulsive} to the reconstructed zero-density background~\eqref{eq:action-solution}.

For the exponential mass function, we define the constant coupling
\begin{align}
b\equiv\frac{{\rm d}\ln m}{{\rm d}\phi}=-\frac{3\alpha}{c_\phi},\qquad
\frac{b^2}{\kappa}=3\varepsilon\alpha,\qquad
b\dot\phi=-3\alpha H.
\label{eq:growth-coupling}
\end{align}
Because the particle mass fluctuates with the scalar field, the number-density and energy-density contrasts differ:
\begin{align}
\delta_n\equiv\frac{\delta n}{\bar n},\qquad
\delta_{\rm s}\equiv\frac{\delta\rho_{\rm s}}{\bar\rho_{\rm s}}
=\delta_n+b\,\delta\phi.
\label{eq:growth-contrasts}
\end{align}
Particle-number conservation provides a continuity equation directly for $\delta_n$. The scalar perturbation still contributes to $\delta_{\rm s}$ through the mass fluctuation, and its own density perturbation is allowed to be nonzero. We therefore do not impose $\delta\rho_\phi=0$: in this action, zero scalar density is a property of the background trajectory rather than an identity of the field theory.

To determine where spatial gradients dominate the scalar response, we evaluate the potential and mass-coupling derivatives on the trajectory:
\begin{equation}
\begin{aligned}
V_{,\phi\phi}=-&\frac92\varepsilon(1+\alpha)^2H^2,
\qquad \bar n m_{,\phi\phi}=9\varepsilon\alpha H^2,\\
m_{\rm eff,n}^2&\equiv\frac{V_{,\phi\phi}+\bar n m_{,\phi\phi}}{\varepsilon}
=-\frac92(1+\alpha^2)H^2.
\label{eq:growth-mass}
\end{aligned}
\end{equation}
Here $m_{\rm eff,n}^2$ is the coefficient in the scalar perturbation equation when $\delta n$ is retained as a source. It is not a mass eigenvalue of the complete coupled system. The gradient-dominated regime requires $k/a$ to exceed $|m_{\rm eff,n}|$, the background evolution rates, and the rates of the modes under consideration. At fixed finite $\alpha$, the hierarchy $k/(aH)\gg1+|\alpha|$ is sufficient for the solutions calculated below. Within that hierarchy, the scalar response follows the matter perturbation quasistatically, while the latter continues to evolve on the background timescale.

Neglecting the subleading time-derivative and metric-source terms in the perturbed scalar equation gives
\begin{equation}
\begin{aligned}
-\left[\varepsilon\frac{k^2}{a^2}+V_{,\phi\phi}
+\bar n m_{,\phi\phi}\right]\delta\phi
&\simeq b\bar\rho_{\rm s}\delta_n,\\
-\varepsilon\frac{k^2}{a^2}\delta\phi
&\simeq b\bar\rho_{\rm s}\delta_n
\simeq b\,\delta\rho_{\rm s}.
\label{eq:growth-scalar-qs}
\end{aligned}
\end{equation}
The second line uses gradient dominance. In this expansion $b\delta\phi$ is suppressed relative to $\delta_n$, and $\delta\rho_\phi$ is suppressed relative to $\delta\rho_{\rm s}$, by powers of $aH/k$ with coefficients depending on $\alpha$. Thus, the leading Poisson equation and mass perturbation read
\begin{align}
\frac{k^2}{a^2}\Psi&\simeq-\frac\kappa2\delta\rho_{\rm s},
\qquad
\delta\ln m=b\delta\phi\simeq\frac{2b^2}{\varepsilon\kappa}\Psi
=6\alpha\Psi.
\label{eq:growth-poisson}
\end{align}
Consequently, the coupled dust responds to
\begin{align}
\Psi_{\rm eff}&\equiv\Psi+\delta\ln m\simeq(1+6\alpha)\Psi,
\qquad \frac{G_{\rm eff}}{G}=1+6\alpha.
\label{eq:growth-Geff}
\end{align}
This is the effective force coefficient between coupled dust perturbations, not a change of the gravitational constant in the EFE. The scalar force is attractive for $\varepsilon=+1$ and repulsive for $\varepsilon=-1$, as in Ref.~\cite{Amendola:2004Repulsive}; the zero-density trajectory fixes its relative strength to $6\alpha$. Uncoupled matter and photons are not subject to this additional force.

For the physical peculiar velocity $\boldsymbol v$ and comoving spatial gradient $\boldsymbol\nabla$, Eq.~\eqref{eq:dust-fifth-force} gives
\begin{align}
&\dot{\boldsymbol v}+(H+b\dot\phi)\boldsymbol v
=-\frac1a\boldsymbol\nabla\Psi_{\rm eff},\\
&\dot\delta_n=-\frac1a\boldsymbol\nabla\!\cdot\boldsymbol v+3\dot\Phi
\simeq-\frac1a\boldsymbol\nabla\!\cdot\boldsymbol v.
\label{eq:growth-fluid}
\end{align}
The mass variation contributes to the velocity damping as well as to the force. Combining these equations with $\kappa\bar\rho_{\rm s}=3H^2$ yields
\begin{align}
\ddot\delta_n+(2-3\alpha)H\dot\delta_n
-\frac32(1+6\alpha)H^2\delta_n\simeq0.
\label{eq:growth-time}
\end{align}
Since $\dot H/H^2=-3(1+\alpha)/2$, Eq.~\eqref{eq:growth-time} in terms of $\mathcal N=\ln a$ has constant coefficients:
\begin{align}
\frac{{\rm d}^2\delta_n}{{\rm d}\mathcal N^2}
+\frac{1-9\alpha}{2}\frac{{\rm d}\delta_n}{{\rm d}\mathcal N}
-\frac32(1+6\alpha)\delta_n\simeq0.
\label{eq:growth-N}
\end{align}
At this order its independent modes, away from repeated roots, are $\delta_n\propto a^{s_\pm}$, where
\begin{align}
s_\pm=\frac{9\alpha-1\pm\sqrt{81\alpha^2+126\alpha+25}}4.
\label{eq:growth-exponents}
\end{align}
The $\alpha\to0$ limit recovers $s_\pm=(1,-3/2)$, although $\alpha=0$ itself is outside the nonzero-pressure reconstruction. The sum is $(9\alpha-1)/2$ and the product is $-3(1+6\alpha)/2$. A growing leading-order mode exists precisely for $\alpha>-1/6$; at $\alpha=-1/6$ the modes are constant and decaying. For $\alpha<-1/6$ the real roots are negative, or the complex pair has negative real part. The roots are complex for
\begin{align}
\frac{-7-2\sqrt6}{9}<\alpha<\frac{-7+2\sqrt6}{9};
\label{eq:growth-complex-range}
\end{align}
at the endpoints the second solution is $a^s\ln a$. For instance, $\alpha=0.1$ gives $G_{\rm eff}/G=1.6$ and $s_+\simeq1.524$, whereas $\alpha=-0.7$ gives $G_{\rm eff}/G=-3.2$ and $s_\pm\simeq-1.825\pm1.212i$.

The proportional-pressure background accelerates for $\alpha<-1/3$, entirely within the range in which no leading quasistatic power-law mode grows. An expanding universe that remains on this exact trajectory therefore cannot also contain a conventional era of growing dust perturbations through these modes. This conclusion concerns nonzero wave numbers in the regime just specified. It is separate from the homogeneous saddle of Sec.~\ref{app:homogeneous} and from the intrinsic phantom ghost, which test different aspects of the same realization.

The force calculation also explains which part of this restriction extends beyond the exponential model. For any spatially flat two-component solution of the same mass-only action class with $\rho_\phi\equiv0$, $\rho_{\rm s}>0$, and $p_\phi\neq0$, the background equations require $b\rho_{\rm s}\dot\phi=-3Hp_\phi$, $p_\phi=\varepsilon\dot\phi^2$, and $3H^2=\kappa\rho_{\rm s}$. Together they fix $b^2/(\varepsilon\kappa)=3p_\phi/\rho_{\rm s}$. Wherever the same gradient-dominated scalar response applies, the net force on the coupled dust consequently has $G_{\rm eff}/G=1+6p_\phi/\rho_{\rm s}$. Accelerated expansion requires $p_\phi/\rho_{\rm s}<-1/3$, which gives $G_{\rm eff}/G<-1$. Changing only the mass function therefore cannot independently reverse the force sign while preserving these assumptions.

The force relation is more general than the constant-coefficient growth solutions just obtained. For the proportional-pressure trajectory, number conservation and $\phi-\phi_0=c_\phi\ln a$ additionally fix $m(\phi)$ on the traversed interval through $m\propto a^{-3\alpha}$. When the pressure-to-density ratio varies, the growth coefficients become time dependent. Its perturbations must then be evolved on that history, rather than assigned the exponents of Eq.~\eqref{eq:growth-exponents}. The constant-pressure example below makes this distinction explicit. Different kinetic structures, more general interactions, or additional sources require their own force and stability analysis.

\subsection{Constant-pressure realization and its dust growth}
\label{app:constant-pressure-growth}

The constant-pressure background provides a particularly direct test of whether matching an expansion history also matches structure growth. Section~\ref{sec:lcdm} gives a branch with the flat-$\Lambda$CDM expansion, and Sec.~\ref{sec:const-pres} reconstructs a scalar with constant potential and constant field velocity for that branch. We now specify the compatible dust mass function. This is a second model within the action class~\eqref{eq:action-dust-scalar}, not another trajectory of the exponential model~\eqref{eq:action-mass}--\eqref{eq:action-reconstruction}. Define $\rho_\Lambda\equiv-p_{\rm ze0}>0$ and choose
\begin{equation}
\begin{aligned}
\varepsilon&=-1,\\
V(\phi)&=\frac{\rho_\Lambda}{2},\\
m(\phi)&=m_\star\cosh^2\!\left[\frac{\sqrt{3\kappa}}2(\phi-\phi_{\rm BB})\right],
\label{eq:constant-pressure-action}
\end{aligned}
\end{equation}
where $m_\star>0$. The constant $\rho_\Lambda$ sets the magnitude of the negative scalar pressure and equals the vacuum density in the reference $\Lambda$CDM description; it is not an additional constituent of the interacting model. Choose the positive-velocity solution of Eq.~\eqref{eq:phi_const_t}, $\phi-\phi_{\rm BB}=\sqrt{\rho_\Lambda}(t-t_{\rm BB})$, and write $u=\sqrt{3\kappa}(\phi-\phi_{\rm BB})/2$. The background solution then has $H=\sqrt{\kappa\rho_\Lambda/3}\,\coth u$ and $\rho_{\rm s}=\rho_\Lambda\coth^2u$, together with $\rho_\phi=0$ and $p_\phi=-\rho_\Lambda$.

The dust normalization is fixed by $m_\star n_0=\rho_{\rm s0}-\rho_\Lambda>0$ when $a(t_0)=1$. It gives $a^3=(\rho_{\rm s0}/\rho_\Lambda-1)\sinh^2u$, so that $\rho_{\rm s}=m(\phi)n_0a^{-3}$. Substitution also satisfies the Klein--Gordon equation and the dust energy balance. The negative-velocity solution is obtained by reflecting the field about $\phi_{\rm BB}$. These relations specify the scalar and dust evolution, not merely the pressure needed in the Friedmann equations.

The reconstruction can also be read directly from particle-number conservation. Since $n=n_0a^{-3}$, the mass along the chosen history must satisfy $m\propto\rho_{\rm s}a^3$, or $m(a)=m_0[(1+\mathcal P_{\rm ze0})-\mathcal P_{\rm ze0}a^3]$, where $m_0=m(\phi(t_0))$ and $-1<\mathcal P_{\rm ze0}<0$. Expressing $a$ in terms of the monotonic scalar solution yields the mass function in Eq.~\eqref{eq:constant-pressure-action}, together with the constant potential already found in Eq.~\eqref{eq:sf-const-v}. The ratio of scalar pressure to dust density is
\begin{align}
\frac{p_\phi}{\rho_{\rm s}}=-\Omega_\Lambda(a),\qquad
\Omega_\Lambda(a)\equiv\frac{\Omega_\Lambda^{\Lambda{\rm CDM}}}{\Omega_{{\rm m}0}^{\Lambda{\rm CDM}}a^{-3}+\Omega_\Lambda^{\Lambda{\rm CDM}}},
\label{eq:lcdm-trajectory-ratio}
\end{align}
where $\Omega_\Lambda(a)$ is the vacuum fraction of the reference history defined in Eq.~\eqref{eq:lcdm-parameters}. In the interacting model it is simply a convenient way to express the pressure-to-density ratio, not a new constituent fraction. The coupling is now field dependent, $b(\phi)={\rm d}\ln m/{\rm d}\phi=\sqrt{3\kappa}\tanh u$. On the solution it gives $b\dot\phi=3H\Omega_\Lambda(a)$. The general force relation derived above consequently becomes $G_{\rm eff}/G=1-6\Omega_\Lambda(a)$, provided that the gradient-dominated quasistatic conditions hold.

Those conditions can be checked using the functions defining this second model. With $V_{,\phi\phi}=0$ and the derivatives of the mass function in Eq.~\eqref{eq:constant-pressure-action}, the scalar mass coefficient is
\begin{align}
 m_{\rm eff,n}^2
 &=\frac{V_{,\phi\phi}+\bar n m_{,\phi\phi}}{\varepsilon}
 =-\frac92[1+\Omega_\Lambda(a)]H^2.
 \label{eq:constant-pressure-mass}
\end{align}
This differs from Eq.~\eqref{eq:growth-mass}, because changing the pressure history also changed the potential and mass function. Since $0<\Omega_\Lambda(a)<1$, the scale $\sqrt{|m_{\rm eff,n}^{2}|}$ remains of order $H$. We again retain the slowly varying scalar response sourced by matter, on scales for which $k/a$ dominates $|m_{\rm eff,n}|$ and the background and perturbation evolution rates. In that regime, the number-conservation and variable-mass Euler equations have the same form as before, but their background coefficients vary with $\Omega_\Lambda(a)$. Using $\dot H/H^2=-3[1-\Omega_\Lambda(a)]/2$ gives
\begin{align}
\frac{{\rm d}^2\delta_n}{{\rm d}\mathcal N^2}
+\frac{1+9\Omega_\Lambda(a)}{2}\frac{{\rm d}\delta_n}{{\rm d}\mathcal N}
-\frac32\left[1-6\Omega_\Lambda(a)\right]\delta_n\simeq0.
\label{eq:growth-lcdm-trajectory}
\end{align}
Unlike the proportional-pressure equation, Eq.~\eqref{eq:growth-lcdm-trajectory} has time-dependent coefficients. Its solutions are therefore not the power laws in Eq.~\eqref{eq:growth-exponents}. To determine the evolution of a mode that grows during matter domination, we must specify its initial conditions and follow it through the later change in the dust force and damping.

For a representative comparison, take $\Omega_{{\rm m}0}^{\Lambda{\rm CDM}}=0.3$ and $\Omega_\Lambda^{\Lambda{\rm CDM}}=0.7$. We work within the radiation-free two-component model and select its early matter-dominated growing solution by setting $\delta_n(a_{\rm i})=({\rm d}\delta_n/{\rm d}\mathcal N)_{a_{\rm i}}=a_{\rm i}$ at $a_{\rm i}=10^{-4}$. Repeating the integration from $a_{\rm i}=10^{-3}$ leaves the results unchanged at the precision quoted. The early initial time selects a limiting solution of this model; it does not incorporate the radiation era of the observed universe.

The growth rate $f\equiv {\rm d}\ln\delta_n/{\rm d}\mathcal N$ is approximately $0.80$, $0.43$, and $-0.96$ at $z=2$, $1$, and $0$, respectively. For conserved dust in the reference $\Lambda$CDM model, the corresponding rates are $0.96$, $0.87$, and $0.51$. Over the interval from $z=2$ to $z=0$, the growth-factor ratio is $\delta_n(0)/\delta_n(2)\simeq1.12$ in the coupled realization, compared with $2.37$ in the reference cosmology. The coupled model therefore first grows and then declines, rather than simply exhibiting a uniformly smaller positive growth rate. These results refer to the leading quasistatic number-density contrast; Eq.~\eqref{eq:growth-contrasts}, with the field-dependent $b(\phi)$, gives its relation to the energy-density contrast.

The sequence of events explains the turnover. The net dust force becomes repulsive at $\Omega_\Lambda(a)=1/6$, corresponding to $z\simeq1.27$. The background begins to accelerate later, at $\Omega_\Lambda(a)=1/3$ or $z\simeq0.67$. At that time the chosen perturbation still has a positive growth rate, $f\simeq0.14$. It reaches its maximum only at $z\simeq0.55$, after which $f<0$. The force enters a second-order evolution equation, so reversing its sign need not immediately reverse the first derivative of the density contrast. Instead, the growth rate evolves in response to the changing force and damping. Thus, the end of growth need not coincide with either force reversal or the onset of acceleration. The constant-pressure realization supplies a specific example of growth turnover and subsequent decay at an exactly $\Lambda$CDM background.

The two calculations test different realizations of the same zero-density stress structure. The proportional-pressure result concerns the power-law modes on the exponential model's accelerating trajectory. The constant-pressure result follows a mode selected during matter domination through a time-dependent background. The latter model retains the phantom kinetic sign, but its homogeneous and complete coupled stability have not been determined by the exponential-model saddle calculation. Matching the background pressure law, specifying an action, and testing the resulting dynamics remain separate steps.

The quasistatic calculation characterizes the leading scalar response sourced by dust, rather than the full spectrum of the coupled theory. Within the stated scale hierarchy, this response can be calculated for either sign of the scalar kinetic term. For $\varepsilon=-1$, however, the underlying action still contains a scalar degree of freedom with negative kinetic energy. Omitting independently excited scalar waves from the quasistatic particular solution does not remove that degree of freedom or establish stability. The growth results therefore identify an additional dynamical property of these realizations, distinct from their intrinsic ghost problem.

Both growth calculations describe the matter-sourced quasistatic regime. They omit independent scalar waves, horizon-scale evolution, nonlinear structures, additional baryons and radiation, and histories departing from the reconstructed trajectories. In particular, a constant $\alpha$ in the exponential model does not require every solution of that action to remain on its zero-density background. During accelerated expansion, a fixed comoving mode can leave the quasistatic regime, beyond which neither the power-law solutions nor the numerical integration used here applies. These restrictions define the scope of the comparison, not an observational exclusion of all zero-density sources. A different cosmological realization must be assessed using its own background, interaction, and perturbations.

\bibliographystyle{apsrev4-2_modified}
\bibliography{references}

\begin{thebibliography}{109}%
\makeatletter
\providecommand \@ifxundefined [1]{%
 \@ifx{#1\undefined}
}%
\providecommand \@ifnum [1]{%
 \ifnum #1\expandafter \@firstoftwo
 \else \expandafter \@secondoftwo
 \fi
}%
\providecommand \@ifx [1]{%
 \ifx #1\expandafter \@firstoftwo
 \else \expandafter \@secondoftwo
 \fi
}%
\providecommand \natexlab [1]{#1}%
\providecommand \enquote  [1]{``#1''}%
\providecommand \bibnamefont  [1]{#1}%
\providecommand \bibfnamefont [1]{#1}%
\providecommand \citenamefont [1]{#1}%
\providecommand \href@noop [0]{\@secondoftwo}%
\providecommand \href [0]{\begingroup \@sanitize@url \@href}%
\providecommand \@href[1]{\@@startlink{#1}\@@href}%
\providecommand \@@href[1]{\endgroup#1\@@endlink}%
\providecommand \@sanitize@url [0]{\catcode `\\12\catcode `\$12\catcode `\&12\catcode `\#12\catcode `\^12\catcode `\_12\catcode `\%12\relax}%
\providecommand \@@startlink[1]{}%
\providecommand \@@endlink[0]{}%
\providecommand \url  [0]{\begingroup\@sanitize@url \@url }%
\providecommand \@url [1]{\endgroup\@href {#1}{\urlprefix }}%
\providecommand \urlprefix  [0]{URL }%
\providecommand \Eprint [0]{\href }%
\providecommand \doibase [0]{https://doi.org/}%
\providecommand \selectlanguage [0]{\@gobble}%
\providecommand \bibinfo  [0]{\@secondoftwo}%
\providecommand \bibfield  [0]{\@secondoftwo}%
\providecommand \translation [1]{[#1]}%
\providecommand \BibitemOpen [0]{}%
\providecommand \bibitemStop [0]{}%
\providecommand \bibitemNoStop [0]{.\EOS\space}%
\providecommand \EOS [0]{\spacefactor3000\relax}%
\providecommand \BibitemShut  [1]{\csname bibitem#1\endcsname}%
\let\auto@bib@innerbib\@empty
%</preamble>
\bibitem [{\citenamefont {Akarsu}\ \emph {et~al.}(2026{\natexlab{a}})\citenamefont {Akarsu}, \citenamefont {De~Felice},\ and\ \citenamefont {Uzun}}]{Akarsu:2026pia}%
  \BibitemOpen
  \bibfield  {author} {\bibinfo {author} {\bibfnamefont {{\"O}.}~\bibnamefont {Akarsu}}, \bibinfo {author} {\bibfnamefont {A.}~\bibnamefont {De~Felice}},\ and\ \bibinfo {author} {\bibfnamefont {N.~M.}\ \bibnamefont {Uzun}},\ }\href@noop {} {\bibinfo {title} {{Defocusing dark energy: Raychaudhuri diagnostics beyond $w<-1/3$ and the phantom divide}}} (\bibinfo {year} {2026}{\natexlab{a}}),\ \Eprint {https://arxiv.org/abs/2607.18008} {2607.18008} \BibitemShut {NoStop}%
\bibitem [{\citenamefont {Hawking}\ and\ \citenamefont {Ellis}(2023)}]{Hawking:1973uf}%
  \BibitemOpen
  \bibfield  {author} {\bibinfo {author} {\bibfnamefont {S.~W.}\ \bibnamefont {Hawking}}\ and\ \bibinfo {author} {\bibfnamefont {G.~F.~R.}\ \bibnamefont {Ellis}},\ }\href {https://doi.org/10.1017/9781009253161} {\emph {\bibinfo {title} {{The Large Scale Structure of Space-Time}}}},\ Cambridge Monographs on Mathematical Physics\ (\bibinfo  {publisher} {Cambridge University Press},\ \bibinfo {year} {2023})\ \bibinfo {note} {50th anniversary edition; first published in 1973}\BibitemShut {NoStop}%
\bibitem [{\citenamefont {Akarsu}\ \emph {et~al.}(2026{\natexlab{b}})\citenamefont {Akarsu}, \citenamefont {Caruana}, \citenamefont {Dialektopoulos}, \citenamefont {Escamilla}, \citenamefont {Kahya},\ and\ \citenamefont {Levi~Said}}]{Akarsu:2026anp}%
  \BibitemOpen
  \bibfield  {author} {\bibinfo {author} {\bibfnamefont {{\"O}.}~\bibnamefont {Akarsu}}, \bibinfo {author} {\bibfnamefont {M.}~\bibnamefont {Caruana}}, \bibinfo {author} {\bibfnamefont {K.~F.}\ \bibnamefont {Dialektopoulos}}, \bibinfo {author} {\bibfnamefont {L.~A.}\ \bibnamefont {Escamilla}}, \bibinfo {author} {\bibfnamefont {E.~O.}\ \bibnamefont {Kahya}},\ and\ \bibinfo {author} {\bibfnamefont {J.}~\bibnamefont {Levi~Said}},\ }\href@noop {} {\bibinfo {title} {{Hints of sign-changing scalar field energy density and a transient acceleration phase at $z\sim 2$ from model-agnostic reconstructions}}} (\bibinfo {year} {2026}{\natexlab{b}}),\ \Eprint {https://arxiv.org/abs/2602.08928} {2602.08928} \BibitemShut {NoStop}%
\bibitem [{\citenamefont {G{\"o}k{\c{c}}en}\ \emph {et~al.}(2026)\citenamefont {G{\"o}k{\c{c}}en}, \citenamefont {Akarsu},\ and\ \citenamefont {Di~Valentino}}]{Gokcen:2026pkq}%
  \BibitemOpen
  \bibfield  {author} {\bibinfo {author} {\bibfnamefont {M.}~\bibnamefont {G{\"o}k{\c{c}}en}}, \bibinfo {author} {\bibfnamefont {{\"O}.}~\bibnamefont {Akarsu}},\ and\ \bibinfo {author} {\bibfnamefont {E.}~\bibnamefont {Di~Valentino}},\ }\bibfield  {title} {\bibinfo {title} {{Revisiting CPL with sign-switching density: To cross or not to cross the NECB}},\ }\href {https://doi.org/10.1016/j.dark.2026.102273} {\bibfield  {journal} {\bibinfo  {journal} {Phys. Dark Univ.}\ }\textbf {\bibinfo {volume} {52}},\ \bibinfo {pages} {102273} (\bibinfo {year} {2026})},\ \Eprint {https://arxiv.org/abs/2602.21169} {2602.21169} \BibitemShut {NoStop}%
\bibitem [{\citenamefont {Sahni}\ and\ \citenamefont {Shtanov}(2003)}]{Sahni:2002dx}%
  \BibitemOpen
  \bibfield  {author} {\bibinfo {author} {\bibfnamefont {V.}~\bibnamefont {Sahni}}\ and\ \bibinfo {author} {\bibfnamefont {Y.}~\bibnamefont {Shtanov}},\ }\bibfield  {title} {\bibinfo {title} {{Brane world models of dark energy}},\ }\href {https://doi.org/10.1088/1475-7516/2003/11/014} {\bibfield  {journal} {\bibinfo  {journal} {JCAP}\ }\textbf {\bibinfo {volume} {11}},\ \bibinfo {pages} {014} (\bibinfo {year} {2003})},\ \Eprint {https://arxiv.org/abs/astro-ph/0202346} {astro-ph/0202346} \BibitemShut {NoStop}%
\bibitem [{\citenamefont {Sahni}\ \emph {et~al.}(2014)\citenamefont {Sahni}, \citenamefont {Shafieloo},\ and\ \citenamefont {Starobinsky}}]{Sahni:2014ooa}%
  \BibitemOpen
  \bibfield  {author} {\bibinfo {author} {\bibfnamefont {V.}~\bibnamefont {Sahni}}, \bibinfo {author} {\bibfnamefont {A.}~\bibnamefont {Shafieloo}},\ and\ \bibinfo {author} {\bibfnamefont {A.~A.}\ \bibnamefont {Starobinsky}},\ }\bibfield  {title} {\bibinfo {title} {{Model independent evidence for dark energy evolution from Baryon Acoustic Oscillations}},\ }\href {https://doi.org/10.1088/2041-8205/793/2/L40} {\bibfield  {journal} {\bibinfo  {journal} {Astrophys. J. Lett.}\ }\textbf {\bibinfo {volume} {793}},\ \bibinfo {pages} {L40} (\bibinfo {year} {2014})},\ \Eprint {https://arxiv.org/abs/1406.2209} {1406.2209} \BibitemShut {NoStop}%
\bibitem [{\citenamefont {Dutta}\ \emph {et~al.}(2020)\citenamefont {Dutta}, \citenamefont {Ruchika}, \citenamefont {Roy}, \citenamefont {Sen},\ and\ \citenamefont {Sheikh-Jabbari}}]{Dutta:2018vmq}%
  \BibitemOpen
  \bibfield  {author} {\bibinfo {author} {\bibfnamefont {K.}~\bibnamefont {Dutta}}, \bibinfo {author} {\bibnamefont {Ruchika}}, \bibinfo {author} {\bibfnamefont {A.}~\bibnamefont {Roy}}, \bibinfo {author} {\bibfnamefont {A.~A.}\ \bibnamefont {Sen}},\ and\ \bibinfo {author} {\bibfnamefont {M.~M.}\ \bibnamefont {Sheikh-Jabbari}},\ }\bibfield  {title} {\bibinfo {title} {{Beyond $\Lambda$CDM with low and high redshift data: implications for dark energy}},\ }\href {https://doi.org/10.1007/s10714-020-2665-4} {\bibfield  {journal} {\bibinfo  {journal} {Gen. Rel. Grav.}\ }\textbf {\bibinfo {volume} {52}},\ \bibinfo {pages} {15} (\bibinfo {year} {2020})},\ \Eprint {https://arxiv.org/abs/1808.06623} {1808.06623} \BibitemShut {NoStop}%
\bibitem [{\citenamefont {Visinelli}\ \emph {et~al.}(2019)\citenamefont {Visinelli}, \citenamefont {Vagnozzi},\ and\ \citenamefont {Danielsson}}]{Visinelli:2019qqu}%
  \BibitemOpen
  \bibfield  {author} {\bibinfo {author} {\bibfnamefont {L.}~\bibnamefont {Visinelli}}, \bibinfo {author} {\bibfnamefont {S.}~\bibnamefont {Vagnozzi}},\ and\ \bibinfo {author} {\bibfnamefont {U.}~\bibnamefont {Danielsson}},\ }\bibfield  {title} {\bibinfo {title} {{Revisiting a negative cosmological constant from low-redshift data}},\ }\href {https://doi.org/10.3390/sym11081035} {\bibfield  {journal} {\bibinfo  {journal} {Symmetry}\ }\textbf {\bibinfo {volume} {11}},\ \bibinfo {pages} {1035} (\bibinfo {year} {2019})},\ \Eprint {https://arxiv.org/abs/1907.07953} {1907.07953} \BibitemShut {NoStop}%
\bibitem [{\citenamefont {Akarsu}\ \emph {et~al.}(2020)\citenamefont {Akarsu}, \citenamefont {Barrow}, \citenamefont {Escamilla},\ and\ \citenamefont {Vazquez}}]{Akarsu:2019hmw}%
  \BibitemOpen
  \bibfield  {author} {\bibinfo {author} {\bibfnamefont {{\"O}.}~\bibnamefont {Akarsu}}, \bibinfo {author} {\bibfnamefont {J.~D.}\ \bibnamefont {Barrow}}, \bibinfo {author} {\bibfnamefont {L.~A.}\ \bibnamefont {Escamilla}},\ and\ \bibinfo {author} {\bibfnamefont {J.~A.}\ \bibnamefont {Vazquez}},\ }\bibfield  {title} {\bibinfo {title} {{Graduated dark energy: Observational hints of a spontaneous sign switch in the cosmological constant}},\ }\href {https://doi.org/10.1103/PhysRevD.101.063528} {\bibfield  {journal} {\bibinfo  {journal} {Phys. Rev. D}\ }\textbf {\bibinfo {volume} {101}},\ \bibinfo {pages} {063528} (\bibinfo {year} {2020})},\ \Eprint {https://arxiv.org/abs/1912.08751} {1912.08751} \BibitemShut {NoStop}%
\bibitem [{\citenamefont {Calder{\'o}n}\ \emph {et~al.}(2021)\citenamefont {Calder{\'o}n}, \citenamefont {Gannouji}, \citenamefont {L'Huillier},\ and\ \citenamefont {Polarski}}]{Calderon:2020hoc}%
  \BibitemOpen
  \bibfield  {author} {\bibinfo {author} {\bibfnamefont {R.}~\bibnamefont {Calder{\'o}n}}, \bibinfo {author} {\bibfnamefont {R.}~\bibnamefont {Gannouji}}, \bibinfo {author} {\bibfnamefont {B.}~\bibnamefont {L'Huillier}},\ and\ \bibinfo {author} {\bibfnamefont {D.}~\bibnamefont {Polarski}},\ }\bibfield  {title} {\bibinfo {title} {{Negative cosmological constant in the dark sector?}},\ }\href {https://doi.org/10.1103/PhysRevD.103.023526} {\bibfield  {journal} {\bibinfo  {journal} {Phys. Rev. D}\ }\textbf {\bibinfo {volume} {103}},\ \bibinfo {pages} {023526} (\bibinfo {year} {2021})},\ \Eprint {https://arxiv.org/abs/2008.10237} {2008.10237} \BibitemShut {NoStop}%
\bibitem [{\citenamefont {Acquaviva}\ \emph {et~al.}(2021)\citenamefont {Acquaviva}, \citenamefont {Akarsu}, \citenamefont {Katirci},\ and\ \citenamefont {Vazquez}}]{Acquaviva:2021jov}%
  \BibitemOpen
  \bibfield  {author} {\bibinfo {author} {\bibfnamefont {G.}~\bibnamefont {Acquaviva}}, \bibinfo {author} {\bibfnamefont {{\"O}.}~\bibnamefont {Akarsu}}, \bibinfo {author} {\bibfnamefont {N.}~\bibnamefont {Katirci}},\ and\ \bibinfo {author} {\bibfnamefont {J.~A.}\ \bibnamefont {Vazquez}},\ }\bibfield  {title} {\bibinfo {title} {{Simple-graduated dark energy and spatial curvature}},\ }\href {https://doi.org/10.1103/PhysRevD.104.023505} {\bibfield  {journal} {\bibinfo  {journal} {Phys. Rev. D}\ }\textbf {\bibinfo {volume} {104}},\ \bibinfo {pages} {023505} (\bibinfo {year} {2021})},\ \Eprint {https://arxiv.org/abs/2104.02623} {2104.02623} \BibitemShut {NoStop}%
\bibitem [{\citenamefont {Sen}\ \emph {et~al.}(2023)\citenamefont {Sen}, \citenamefont {Adil},\ and\ \citenamefont {Sen}}]{Sen:2021wld}%
  \BibitemOpen
  \bibfield  {author} {\bibinfo {author} {\bibfnamefont {A.~A.}\ \bibnamefont {Sen}}, \bibinfo {author} {\bibfnamefont {S.~A.}\ \bibnamefont {Adil}},\ and\ \bibinfo {author} {\bibfnamefont {S.}~\bibnamefont {Sen}},\ }\bibfield  {title} {\bibinfo {title} {{Do cosmological observations allow a negative $\Lambda$?}},\ }\href {https://doi.org/10.1093/mnras/stac2796} {\bibfield  {journal} {\bibinfo  {journal} {Mon. Not. Roy. Astron. Soc.}\ }\textbf {\bibinfo {volume} {518}},\ \bibinfo {pages} {1098} (\bibinfo {year} {2023})},\ \Eprint {https://arxiv.org/abs/2112.10641} {2112.10641} \BibitemShut {NoStop}%
\bibitem [{\citenamefont {Adil}\ \emph {et~al.}(2024)\citenamefont {Adil}, \citenamefont {Akarsu}, \citenamefont {Di~Valentino}, \citenamefont {Nunes}, \citenamefont {{\"O}z{\"u}lker}, \citenamefont {Sen},\ and\ \citenamefont {Specogna}}]{Adil:2023exv}%
  \BibitemOpen
  \bibfield  {author} {\bibinfo {author} {\bibfnamefont {S.~A.}\ \bibnamefont {Adil}}, \bibinfo {author} {\bibfnamefont {{\"O}.}~\bibnamefont {Akarsu}}, \bibinfo {author} {\bibfnamefont {E.}~\bibnamefont {Di~Valentino}}, \bibinfo {author} {\bibfnamefont {R.~C.}\ \bibnamefont {Nunes}}, \bibinfo {author} {\bibfnamefont {E.}~\bibnamefont {{\"O}z{\"u}lker}}, \bibinfo {author} {\bibfnamefont {A.~A.}\ \bibnamefont {Sen}},\ and\ \bibinfo {author} {\bibfnamefont {E.}~\bibnamefont {Specogna}},\ }\bibfield  {title} {\bibinfo {title} {{Omnipotent dark energy: A phenomenological answer to the Hubble tension}},\ }\href {https://doi.org/10.1103/PhysRevD.109.023527} {\bibfield  {journal} {\bibinfo  {journal} {Phys. Rev. D}\ }\textbf {\bibinfo {volume} {109}},\ \bibinfo {pages} {023527} (\bibinfo {year} {2024})},\ \Eprint {https://arxiv.org/abs/2306.08046} {2306.08046} \BibitemShut {NoStop}%
\bibitem [{\citenamefont {Malekjani}\ \emph {et~al.}(2024)\citenamefont {Malekjani}, \citenamefont {Mc~Conville}, \citenamefont {{\'O}~Colg{\'a}in}, \citenamefont {Pourojaghi},\ and\ \citenamefont {Sheikh-Jabbari}}]{Malekjani:2023ple}%
  \BibitemOpen
  \bibfield  {author} {\bibinfo {author} {\bibfnamefont {M.}~\bibnamefont {Malekjani}}, \bibinfo {author} {\bibfnamefont {R.}~\bibnamefont {Mc~Conville}}, \bibinfo {author} {\bibfnamefont {E.}~\bibnamefont {{\'O}~Colg{\'a}in}}, \bibinfo {author} {\bibfnamefont {S.}~\bibnamefont {Pourojaghi}},\ and\ \bibinfo {author} {\bibfnamefont {M.~M.}\ \bibnamefont {Sheikh-Jabbari}},\ }\bibfield  {title} {\bibinfo {title} {{On redshift evolution and negative dark energy density in Pantheon+ Supernovae}},\ }\href {https://doi.org/10.1140/epjc/s10052-024-12667-z} {\bibfield  {journal} {\bibinfo  {journal} {Eur. Phys. J. C}\ }\textbf {\bibinfo {volume} {84}},\ \bibinfo {pages} {317} (\bibinfo {year} {2024})},\ \Eprint {https://arxiv.org/abs/2301.12725} {2301.12725} \BibitemShut {NoStop}%
\bibitem [{\citenamefont {Tiwari}\ \emph {et~al.}(2024)\citenamefont {Tiwari}, \citenamefont {Ghosh},\ and\ \citenamefont {Jain}}]{Tiwari:2023jle}%
  \BibitemOpen
  \bibfield  {author} {\bibinfo {author} {\bibfnamefont {Y.}~\bibnamefont {Tiwari}}, \bibinfo {author} {\bibfnamefont {B.}~\bibnamefont {Ghosh}},\ and\ \bibinfo {author} {\bibfnamefont {R.~K.}\ \bibnamefont {Jain}},\ }\bibfield  {title} {\bibinfo {title} {{Towards a possible solution to the Hubble tension with Horndeski gravity}},\ }\href {https://doi.org/10.1140/epjc/s10052-024-12577-0} {\bibfield  {journal} {\bibinfo  {journal} {Eur. Phys. J. C}\ }\textbf {\bibinfo {volume} {84}},\ \bibinfo {pages} {220} (\bibinfo {year} {2024})},\ \Eprint {https://arxiv.org/abs/2301.09382} {2301.09382} \BibitemShut {NoStop}%
\bibitem [{\citenamefont {Akarsu}\ \emph {et~al.}(2026{\natexlab{c}})\citenamefont {Akarsu}, \citenamefont {De~Felice}, \citenamefont {Di~Valentino}, \citenamefont {Kumar}, \citenamefont {Nunes}, \citenamefont {{\"O}z{\"u}lker}, \citenamefont {Vazquez},\ and\ \citenamefont {Yadav}}]{Akarsu:2024qsi}%
  \BibitemOpen
  \bibfield  {author} {\bibinfo {author} {\bibfnamefont {{\"O}.}~\bibnamefont {Akarsu}}, \bibinfo {author} {\bibfnamefont {A.}~\bibnamefont {De~Felice}}, \bibinfo {author} {\bibfnamefont {E.}~\bibnamefont {Di~Valentino}}, \bibinfo {author} {\bibfnamefont {S.}~\bibnamefont {Kumar}}, \bibinfo {author} {\bibfnamefont {R.~C.}\ \bibnamefont {Nunes}}, \bibinfo {author} {\bibfnamefont {E.}~\bibnamefont {{\"O}z{\"u}lker}}, \bibinfo {author} {\bibfnamefont {J.~A.}\ \bibnamefont {Vazquez}},\ and\ \bibinfo {author} {\bibfnamefont {A.}~\bibnamefont {Yadav}},\ }\bibfield  {title} {\bibinfo {title} {{{\ensuremath{\Lambda}}sCDM cosmology from a type-II minimally modified gravity}},\ }\href {https://doi.org/10.1093/mnras/staf2276} {\bibfield  {journal} {\bibinfo  {journal} {Mon. Not. Roy. Astron. Soc.}\ }\textbf {\bibinfo {volume} {546}},\ \bibinfo {pages} {staf2276} (\bibinfo {year} {2026}{\natexlab{c}})},\ \Eprint {https://arxiv.org/abs/2402.07716} {2402.07716} \BibitemShut {NoStop}%
\bibitem [{\citenamefont {Akarsu}\ \emph {et~al.}(2024{\natexlab{a}})\citenamefont {Akarsu}, \citenamefont {De~Felice}, \citenamefont {Di~Valentino}, \citenamefont {Kumar}, \citenamefont {Nunes}, \citenamefont {{\"O}z{\"u}lker}, \citenamefont {Vazquez},\ and\ \citenamefont {Yadav}}]{Akarsu:2024eoo}%
  \BibitemOpen
  \bibfield  {author} {\bibinfo {author} {\bibfnamefont {{\"O}.}~\bibnamefont {Akarsu}}, \bibinfo {author} {\bibfnamefont {A.}~\bibnamefont {De~Felice}}, \bibinfo {author} {\bibfnamefont {E.}~\bibnamefont {Di~Valentino}}, \bibinfo {author} {\bibfnamefont {S.}~\bibnamefont {Kumar}}, \bibinfo {author} {\bibfnamefont {R.~C.}\ \bibnamefont {Nunes}}, \bibinfo {author} {\bibfnamefont {E.}~\bibnamefont {{\"O}z{\"u}lker}}, \bibinfo {author} {\bibfnamefont {J.~A.}\ \bibnamefont {Vazquez}},\ and\ \bibinfo {author} {\bibfnamefont {A.}~\bibnamefont {Yadav}},\ }\bibfield  {title} {\bibinfo {title} {{Cosmological constraints on {\ensuremath{\Lambda}}sCDM scenario in a type II minimally modified gravity}},\ }\href {https://doi.org/10.1103/PhysRevD.110.103527} {\bibfield  {journal} {\bibinfo  {journal} {Phys. Rev. D}\ }\textbf {\bibinfo {volume} {110}},\ \bibinfo {pages} {103527} (\bibinfo {year} {2024}{\natexlab{a}})},\ \Eprint {https://arxiv.org/abs/2406.07526} {2406.07526} \BibitemShut {NoStop}%
\bibitem [{\citenamefont {Dwivedi}\ and\ \citenamefont {H{\"o}g{\r{a}}s}(2024)}]{Dwivedi:2024okk}%
  \BibitemOpen
  \bibfield  {author} {\bibinfo {author} {\bibfnamefont {S.}~\bibnamefont {Dwivedi}}\ and\ \bibinfo {author} {\bibfnamefont {M.}~\bibnamefont {H{\"o}g{\r{a}}s}},\ }\bibfield  {title} {\bibinfo {title} {{2D BAO vs. 3D BAO: Solving the Hubble Tension with Bimetric Cosmology}},\ }\href {https://doi.org/10.3390/universe10110406} {\bibfield  {journal} {\bibinfo  {journal} {Universe}\ }\textbf {\bibinfo {volume} {10}},\ \bibinfo {pages} {406} (\bibinfo {year} {2024})},\ \Eprint {https://arxiv.org/abs/2407.04322} {2407.04322} \BibitemShut {NoStop}%
\bibitem [{\citenamefont {Akarsu}\ \emph {et~al.}(2025{\natexlab{a}})\citenamefont {Akarsu}, \citenamefont {Bulduk}, \citenamefont {De~Felice}, \citenamefont {Kat{\i}rc{\i}},\ and\ \citenamefont {Uzun}}]{Akarsu:2024nas}%
  \BibitemOpen
  \bibfield  {author} {\bibinfo {author} {\bibfnamefont {O.}~\bibnamefont {Akarsu}}, \bibinfo {author} {\bibfnamefont {B.}~\bibnamefont {Bulduk}}, \bibinfo {author} {\bibfnamefont {A.}~\bibnamefont {De~Felice}}, \bibinfo {author} {\bibfnamefont {N.}~\bibnamefont {Kat{\i}rc{\i}}},\ and\ \bibinfo {author} {\bibfnamefont {N.~M.}\ \bibnamefont {Uzun}},\ }\bibfield  {title} {\bibinfo {title} {{Unexplored regions in teleparallel f(T) gravity: Sign-changing dark energy density}},\ }\href {https://doi.org/10.1103/1xd4-k91h} {\bibfield  {journal} {\bibinfo  {journal} {Phys. Rev. D}\ }\textbf {\bibinfo {volume} {112}},\ \bibinfo {pages} {083532} (\bibinfo {year} {2025}{\natexlab{a}})},\ \Eprint {https://arxiv.org/abs/2410.23068} {2410.23068} \BibitemShut {NoStop}%
\bibitem [{\citenamefont {Akarsu}\ \emph {et~al.}(2025{\natexlab{b}})\citenamefont {Akarsu}, \citenamefont {Perivolaropoulos}, \citenamefont {Tsikoundoura}, \citenamefont {Y{\"u}kselci},\ and\ \citenamefont {Zhuk}}]{Akarsu:2025gwi}%
  \BibitemOpen
  \bibfield  {author} {\bibinfo {author} {\bibfnamefont {{\"O}.}~\bibnamefont {Akarsu}}, \bibinfo {author} {\bibfnamefont {L.}~\bibnamefont {Perivolaropoulos}}, \bibinfo {author} {\bibfnamefont {A.}~\bibnamefont {Tsikoundoura}}, \bibinfo {author} {\bibfnamefont {A.~E.}\ \bibnamefont {Y{\"u}kselci}},\ and\ \bibinfo {author} {\bibfnamefont {A.}~\bibnamefont {Zhuk}},\ }\href@noop {} {\bibinfo {title} {{Dynamical dark energy with AdS-to-dS and dS-to-dS transitions: Implications for the $H_0$ tension}}} (\bibinfo {year} {2025}{\natexlab{b}}),\ \Eprint {https://arxiv.org/abs/2502.14667} {2502.14667} \BibitemShut {NoStop}%
\bibitem [{\citenamefont {Akarsu}\ \emph {et~al.}(2026{\natexlab{d}})\citenamefont {Akarsu}, \citenamefont {Perivolaropoulos}, \citenamefont {Y{\"u}kselci},\ and\ \citenamefont {Zhuk}}]{Akarsu:2026lva}%
  \BibitemOpen
  \bibfield  {author} {\bibinfo {author} {\bibfnamefont {{\"O}.}~\bibnamefont {Akarsu}}, \bibinfo {author} {\bibfnamefont {L.}~\bibnamefont {Perivolaropoulos}}, \bibinfo {author} {\bibfnamefont {A.~E.}\ \bibnamefont {Y{\"u}kselci}},\ and\ \bibinfo {author} {\bibfnamefont {A.}~\bibnamefont {Zhuk}},\ }\href@noop {} {\bibinfo {title} {{A Friendly Phantom: Late-time AdS-to-dS transition and cosmological tensions}}} (\bibinfo {year} {2026}{\natexlab{d}}),\ \Eprint {https://arxiv.org/abs/2606.11062} {2606.11062} \BibitemShut {NoStop}%
\bibitem [{\citenamefont {Bouhmadi-L{\'o}pez}\ and\ \citenamefont {Ibarra-Uriondo}(2025{\natexlab{a}})}]{Bouhmadi-Lopez:2025ggl}%
  \BibitemOpen
  \bibfield  {author} {\bibinfo {author} {\bibfnamefont {M.}~\bibnamefont {Bouhmadi-L{\'o}pez}}\ and\ \bibinfo {author} {\bibfnamefont {B.}~\bibnamefont {Ibarra-Uriondo}},\ }\bibfield  {title} {\bibinfo {title} {{Cosmographic analysis of sign-switching dark energy}},\ }\href {https://doi.org/10.1103/v1cl-pr54} {\bibfield  {journal} {\bibinfo  {journal} {Phys. Rev. D}\ }\textbf {\bibinfo {volume} {112}},\ \bibinfo {pages} {063559} (\bibinfo {year} {2025}{\natexlab{a}})},\ \Eprint {https://arxiv.org/abs/2506.12139} {2506.12139} \BibitemShut {NoStop}%
\bibitem [{\citenamefont {Bouhmadi-L{\'o}pez}\ and\ \citenamefont {Ibarra-Uriondo}(2025{\natexlab{b}})}]{Bouhmadi-Lopez:2025spo}%
  \BibitemOpen
  \bibfield  {author} {\bibinfo {author} {\bibfnamefont {M.}~\bibnamefont {Bouhmadi-L{\'o}pez}}\ and\ \bibinfo {author} {\bibfnamefont {B.}~\bibnamefont {Ibarra-Uriondo}},\ }\bibfield  {title} {\bibinfo {title} {{Cosmological perturbations for smooth sign-switching dark energy models}},\ }\href {https://doi.org/10.1016/j.dark.2025.102129} {\bibfield  {journal} {\bibinfo  {journal} {Phys. Dark Univ.}\ }\textbf {\bibinfo {volume} {50}},\ \bibinfo {pages} {102129} (\bibinfo {year} {2025}{\natexlab{b}})},\ \Eprint {https://arxiv.org/abs/2506.18992} {2506.18992} \BibitemShut {NoStop}%
\bibitem [{\citenamefont {Adil}\ \emph {et~al.}(2026)\citenamefont {Adil}, \citenamefont {Akarsu}, \citenamefont {Bouhmadi-L{\'o}pez}, \citenamefont {Ibarra-Uriondo}, \citenamefont {Kat{\i}rc{\i}},\ and\ \citenamefont {V{\'a}zquez}}]{Adil:2026gjl}%
  \BibitemOpen
  \bibfield  {author} {\bibinfo {author} {\bibfnamefont {S.~A.}\ \bibnamefont {Adil}}, \bibinfo {author} {\bibfnamefont {{\"O}.}~\bibnamefont {Akarsu}}, \bibinfo {author} {\bibfnamefont {M.}~\bibnamefont {Bouhmadi-L{\'o}pez}}, \bibinfo {author} {\bibfnamefont {B.}~\bibnamefont {Ibarra-Uriondo}}, \bibinfo {author} {\bibfnamefont {N.}~\bibnamefont {Kat{\i}rc{\i}}},\ and\ \bibinfo {author} {\bibfnamefont {J.~A.}\ \bibnamefont {V{\'a}zquez}},\ }\href@noop {} {\bibinfo {title} {{Reconstructing sign-switching dark energy histories: Scalar-field regularity, conditional potential comparison, and representative dynamics}}} (\bibinfo {year} {2026}),\ \Eprint {https://arxiv.org/abs/2609.09261} {2609.09261} \BibitemShut {NoStop}%
\bibitem [{\citenamefont {Ozulker}(2022)}]{Ozulker:2022slu}%
  \BibitemOpen
  \bibfield  {author} {\bibinfo {author} {\bibfnamefont {E.}~\bibnamefont {Ozulker}},\ }\bibfield  {title} {\bibinfo {title} {{Is the dark energy equation of state parameter singular?}},\ }\href {https://doi.org/10.1103/PhysRevD.106.063509} {\bibfield  {journal} {\bibinfo  {journal} {Phys. Rev. D}\ }\textbf {\bibinfo {volume} {106}},\ \bibinfo {pages} {063509} (\bibinfo {year} {2022})},\ \Eprint {https://arxiv.org/abs/2203.04167} {2203.04167} \BibitemShut {NoStop}%
\bibitem [{\citenamefont {Abdul-Karim}\ \emph {et~al.}(2025)\citenamefont {Abdul-Karim} \emph {et~al.}}]{DESI:2025DR2}%
  \BibitemOpen
  \bibfield  {author} {\bibinfo {author} {\bibfnamefont {M.}~\bibnamefont {Abdul-Karim}} \emph {et~al.} (\bibinfo {collaboration} {DESI}),\ }\bibfield  {title} {\bibinfo {title} {{DESI DR2 Results II: Measurements of Baryon Acoustic Oscillations and Cosmological Constraints}},\ }\href {https://doi.org/10.1103/tr6y-kpc6} {\bibfield  {journal} {\bibinfo  {journal} {Phys. Rev. D}\ }\textbf {\bibinfo {volume} {112}},\ \bibinfo {pages} {083515} (\bibinfo {year} {2025})},\ \Eprint {https://arxiv.org/abs/2503.14738} {2503.14738} \BibitemShut {NoStop}%
\bibitem [{\citenamefont {Adame}\ \emph {et~al.}(2026)\citenamefont {Adame} \emph {et~al.}}]{DESI:2026Lya}%
  \BibitemOpen
  \bibfield  {author} {\bibinfo {author} {\bibfnamefont {A.~G.}\ \bibnamefont {Adame}} \emph {et~al.} (\bibinfo {collaboration} {DESI}),\ }\href@noop {} {\bibinfo {title} {{DESI DR2 Results IV: Alcock-Paczynski Measurements from the Lyman Alpha Forest and Cosmological Constraints}}} (\bibinfo {year} {2026}),\ \Eprint {https://arxiv.org/abs/2607.27410} {2607.27410} \BibitemShut {NoStop}%
\bibitem [{\citenamefont {Perivolaropoulos}\ and\ \citenamefont {Skara}(2022)}]{Perivolaropoulos:2021jda}%
  \BibitemOpen
  \bibfield  {author} {\bibinfo {author} {\bibfnamefont {L.}~\bibnamefont {Perivolaropoulos}}\ and\ \bibinfo {author} {\bibfnamefont {F.}~\bibnamefont {Skara}},\ }\bibfield  {title} {\bibinfo {title} {{Challenges for {\ensuremath{\Lambda}}CDM: An update}},\ }\href {https://doi.org/10.1016/j.newar.2022.101659} {\bibfield  {journal} {\bibinfo  {journal} {New Astron. Rev.}\ }\textbf {\bibinfo {volume} {95}},\ \bibinfo {pages} {101659} (\bibinfo {year} {2022})},\ \Eprint {https://arxiv.org/abs/2105.05208} {2105.05208} \BibitemShut {NoStop}%
\bibitem [{\citenamefont {Abdalla}\ \emph {et~al.}(2022)\citenamefont {Abdalla} \emph {et~al.}}]{Abdalla:2022yfr}%
  \BibitemOpen
  \bibfield  {author} {\bibinfo {author} {\bibfnamefont {E.}~\bibnamefont {Abdalla}} \emph {et~al.},\ }\bibfield  {title} {\bibinfo {title} {{Cosmology intertwined: A review of the particle physics, astrophysics, and cosmology associated with the cosmological tensions and anomalies}},\ }\href {https://doi.org/10.1016/j.jheap.2022.04.002} {\bibfield  {journal} {\bibinfo  {journal} {JHEAp}\ }\textbf {\bibinfo {volume} {34}},\ \bibinfo {pages} {49} (\bibinfo {year} {2022})},\ \Eprint {https://arxiv.org/abs/2203.06142} {2203.06142} \BibitemShut {NoStop}%
\bibitem [{\citenamefont {Vagnozzi}(2023)}]{Vagnozzi:2023nrq}%
  \BibitemOpen
  \bibfield  {author} {\bibinfo {author} {\bibfnamefont {S.}~\bibnamefont {Vagnozzi}},\ }\bibfield  {title} {\bibinfo {title} {{Seven Hints That Early-Time New Physics Alone Is Not Sufficient to Solve the Hubble Tension}},\ }\href {https://doi.org/10.3390/universe9090393} {\bibfield  {journal} {\bibinfo  {journal} {Universe}\ }\textbf {\bibinfo {volume} {9}},\ \bibinfo {pages} {393} (\bibinfo {year} {2023})},\ \Eprint {https://arxiv.org/abs/2308.16628} {2308.16628} \BibitemShut {NoStop}%
\bibitem [{\citenamefont {Akarsu}\ \emph {et~al.}(2024{\natexlab{b}})\citenamefont {Akarsu}, \citenamefont {Colg{\'a}in}, \citenamefont {Sen},\ and\ \citenamefont {Sheikh-Jabbari}}]{Akarsu:2024qiq}%
  \BibitemOpen
  \bibfield  {author} {\bibinfo {author} {\bibfnamefont {{\"O}.}~\bibnamefont {Akarsu}}, \bibinfo {author} {\bibfnamefont {E.~{\'O}.}\ \bibnamefont {Colg{\'a}in}}, \bibinfo {author} {\bibfnamefont {A.~A.}\ \bibnamefont {Sen}},\ and\ \bibinfo {author} {\bibfnamefont {M.~M.}\ \bibnamefont {Sheikh-Jabbari}},\ }\bibfield  {title} {\bibinfo {title} {{{\ensuremath{\Lambda}}CDM Tensions: Localising Missing Physics through Consistency Checks}},\ }\href {https://doi.org/10.3390/universe10080305} {\bibfield  {journal} {\bibinfo  {journal} {Universe}\ }\textbf {\bibinfo {volume} {10}},\ \bibinfo {pages} {305} (\bibinfo {year} {2024}{\natexlab{b}})},\ \Eprint {https://arxiv.org/abs/2402.04767} {2402.04767} \BibitemShut {NoStop}%
\bibitem [{\citenamefont {Di~Valentino}\ \emph {et~al.}(2025)\citenamefont {Di~Valentino} \emph {et~al.}}]{CosmoVerseNetwork:2025alb}%
  \BibitemOpen
  \bibfield  {author} {\bibinfo {author} {\bibfnamefont {E.}~\bibnamefont {Di~Valentino}} \emph {et~al.} (\bibinfo {collaboration} {CosmoVerse Network}),\ }\bibfield  {title} {\bibinfo {title} {{The CosmoVerse White Paper: Addressing observational tensions in cosmology with systematics and fundamental physics}},\ }\href {https://doi.org/10.1016/j.dark.2025.101965} {\bibfield  {journal} {\bibinfo  {journal} {Phys. Dark Univ.}\ }\textbf {\bibinfo {volume} {49}},\ \bibinfo {pages} {101965} (\bibinfo {year} {2025})},\ \Eprint {https://arxiv.org/abs/2504.01669} {2504.01669} \BibitemShut {NoStop}%
\bibitem [{\citenamefont {Compere}\ \emph {et~al.}(2011)\citenamefont {Compere}, \citenamefont {McFadden}, \citenamefont {Skenderis},\ and\ \citenamefont {Taylor}}]{Compere:2011dx}%
  \BibitemOpen
  \bibfield  {author} {\bibinfo {author} {\bibfnamefont {G.}~\bibnamefont {Compere}}, \bibinfo {author} {\bibfnamefont {P.}~\bibnamefont {McFadden}}, \bibinfo {author} {\bibfnamefont {K.}~\bibnamefont {Skenderis}},\ and\ \bibinfo {author} {\bibfnamefont {M.}~\bibnamefont {Taylor}},\ }\bibfield  {title} {\bibinfo {title} {{The Holographic fluid dual to vacuum Einstein gravity}},\ }\href {https://doi.org/10.1007/JHEP07(2011)050} {\bibfield  {journal} {\bibinfo  {journal} {JHEP}\ }\textbf {\bibinfo {volume} {07}},\ \bibinfo {pages} {050} (\bibinfo {year} {2011})},\ \Eprint {https://arxiv.org/abs/1103.3022} {1103.3022} \BibitemShut {NoStop}%
\bibitem [{\citenamefont {Compere}\ \emph {et~al.}(2012)\citenamefont {Compere}, \citenamefont {McFadden}, \citenamefont {Skenderis},\ and\ \citenamefont {Taylor}}]{Compere:2012mt}%
  \BibitemOpen
  \bibfield  {author} {\bibinfo {author} {\bibfnamefont {G.}~\bibnamefont {Compere}}, \bibinfo {author} {\bibfnamefont {P.}~\bibnamefont {McFadden}}, \bibinfo {author} {\bibfnamefont {K.}~\bibnamefont {Skenderis}},\ and\ \bibinfo {author} {\bibfnamefont {M.}~\bibnamefont {Taylor}},\ }\bibfield  {title} {\bibinfo {title} {{The relativistic fluid dual to vacuum Einstein gravity}},\ }\href {https://doi.org/10.1007/JHEP03(2012)076} {\bibfield  {journal} {\bibinfo  {journal} {JHEP}\ }\textbf {\bibinfo {volume} {03}},\ \bibinfo {pages} {076} (\bibinfo {year} {2012})},\ \Eprint {https://arxiv.org/abs/1201.2678} {1201.2678} \BibitemShut {NoStop}%
\bibitem [{\citenamefont {Eling}\ \emph {et~al.}(2012)\citenamefont {Eling}, \citenamefont {Meyer},\ and\ \citenamefont {Oz}}]{Eling:2012ni}%
  \BibitemOpen
  \bibfield  {author} {\bibinfo {author} {\bibfnamefont {C.}~\bibnamefont {Eling}}, \bibinfo {author} {\bibfnamefont {A.}~\bibnamefont {Meyer}},\ and\ \bibinfo {author} {\bibfnamefont {Y.}~\bibnamefont {Oz}},\ }\bibfield  {title} {\bibinfo {title} {{The Relativistic Rindler Hydrodynamics}},\ }\href {https://doi.org/10.1007/JHEP05(2012)116} {\bibfield  {journal} {\bibinfo  {journal} {JHEP}\ }\textbf {\bibinfo {volume} {05}},\ \bibinfo {pages} {116} (\bibinfo {year} {2012})},\ \Eprint {https://arxiv.org/abs/1201.2705} {1201.2705} \BibitemShut {NoStop}%
\bibitem [{\citenamefont {Afshordi}(2008)}]{Afshordi:2008xu}%
  \BibitemOpen
  \bibfield  {author} {\bibinfo {author} {\bibfnamefont {N.}~\bibnamefont {Afshordi}},\ }\href@noop {} {\bibinfo {title} {{Gravitational Aether and the thermodynamic solution to the cosmological constant problem}}} (\bibinfo {year} {2008}),\ \Eprint {https://arxiv.org/abs/0807.2639} {0807.2639} \BibitemShut {NoStop}%
\bibitem [{\citenamefont {Aslanbeigi}\ \emph {et~al.}(2011)\citenamefont {Aslanbeigi}, \citenamefont {Robbers}, \citenamefont {Foster}, \citenamefont {Kohri},\ and\ \citenamefont {Afshordi}}]{Aslanbeigi:2011si}%
  \BibitemOpen
  \bibfield  {author} {\bibinfo {author} {\bibfnamefont {S.}~\bibnamefont {Aslanbeigi}}, \bibinfo {author} {\bibfnamefont {G.}~\bibnamefont {Robbers}}, \bibinfo {author} {\bibfnamefont {B.~Z.}\ \bibnamefont {Foster}}, \bibinfo {author} {\bibfnamefont {K.}~\bibnamefont {Kohri}},\ and\ \bibinfo {author} {\bibfnamefont {N.}~\bibnamefont {Afshordi}},\ }\bibfield  {title} {\bibinfo {title} {{Phenomenology of Gravitational Aether as a solution to the Old Cosmological Constant Problem}},\ }\href {https://doi.org/10.1103/PhysRevD.84.103522} {\bibfield  {journal} {\bibinfo  {journal} {Phys. Rev. D}\ }\textbf {\bibinfo {volume} {84}},\ \bibinfo {pages} {103522} (\bibinfo {year} {2011})},\ \Eprint {https://arxiv.org/abs/1106.3955} {1106.3955} \BibitemShut {NoStop}%
\bibitem [{\citenamefont {Kuchowicz}(1968)}]{Kuchowicz1968}%
  \BibitemOpen
  \bibfield  {author} {\bibinfo {author} {\bibfnamefont {B.}~\bibnamefont {Kuchowicz}},\ }\bibfield  {title} {\bibinfo {title} {Extensions of the external {Schwarzschild} solution},\ }\href@noop {} {\bibfield  {journal} {\bibinfo  {journal} {Bulletin de l'Acad{\'e}mie Polonaise des Sciences, S{\'e}rie des Sciences Math{\'e}matiques, Astronomiques et Physiques}\ }\textbf {\bibinfo {volume} {16}},\ \bibinfo {pages} {341} (\bibinfo {year} {1968})}\BibitemShut {NoStop}%
\bibitem [{\citenamefont {Semiz}(2022)}]{Semiz:2022iyh}%
  \BibitemOpen
  \bibfield  {author} {\bibinfo {author} {\bibfnamefont {{\.I}.}~\bibnamefont {Semiz}},\ }\href@noop {} {\bibinfo {title} {{The general static spherical perfect fluid solution with EoS parameter {$w=-1/6$}}}} (\bibinfo {year} {2022}),\ \Eprint {https://arxiv.org/abs/2210.16648} {2210.16648} \BibitemShut {NoStop}%
\bibitem [{\citenamefont {Simmonds}\ and\ \citenamefont {Visser}(2026)}]{Simmonds:2026jln}%
  \BibitemOpen
  \bibfield  {author} {\bibinfo {author} {\bibfnamefont {C.}~\bibnamefont {Simmonds}}\ and\ \bibinfo {author} {\bibfnamefont {M.}~\bibnamefont {Visser}},\ }\href@noop {} {\bibinfo {title} {{Revisiting Schwarzschild's constant density star in isotropic coordinates}}} (\bibinfo {year} {2026}),\ \Eprint {https://arxiv.org/abs/2606.01061} {2606.01061} \BibitemShut {NoStop}%
\bibitem [{\citenamefont {Garriga}\ and\ \citenamefont {Mukhanov}(1999)}]{Garriga:1999vw}%
  \BibitemOpen
  \bibfield  {author} {\bibinfo {author} {\bibfnamefont {J.}~\bibnamefont {Garriga}}\ and\ \bibinfo {author} {\bibfnamefont {V.~F.}\ \bibnamefont {Mukhanov}},\ }\bibfield  {title} {\bibinfo {title} {{Perturbations in k-inflation}},\ }\href {https://doi.org/10.1016/S0370-2693(99)00602-4} {\bibfield  {journal} {\bibinfo  {journal} {Phys. Lett. B}\ }\textbf {\bibinfo {volume} {458}},\ \bibinfo {pages} {219} (\bibinfo {year} {1999})},\ \Eprint {https://arxiv.org/abs/hep-th/9904176} {hep-th/9904176} \BibitemShut {NoStop}%
\bibitem [{\citenamefont {Afshordi}\ \emph {et~al.}(2007{\natexlab{a}})\citenamefont {Afshordi}, \citenamefont {Chung},\ and\ \citenamefont {Geshnizjani}}]{Afshordi:2006ad}%
  \BibitemOpen
  \bibfield  {author} {\bibinfo {author} {\bibfnamefont {N.}~\bibnamefont {Afshordi}}, \bibinfo {author} {\bibfnamefont {D.~J.~H.}\ \bibnamefont {Chung}},\ and\ \bibinfo {author} {\bibfnamefont {G.}~\bibnamefont {Geshnizjani}},\ }\bibfield  {title} {\bibinfo {title} {{Causal field theory with an infinite speed of sound}},\ }\href {https://doi.org/10.1103/PhysRevD.75.083513} {\bibfield  {journal} {\bibinfo  {journal} {Phys. Rev. D}\ }\textbf {\bibinfo {volume} {75}},\ \bibinfo {pages} {083513} (\bibinfo {year} {2007}{\natexlab{a}})},\ \Eprint {https://arxiv.org/abs/hep-th/0609150} {hep-th/0609150} \BibitemShut {NoStop}%
\bibitem [{\citenamefont {Afshordi}\ \emph {et~al.}(2007{\natexlab{b}})\citenamefont {Afshordi}, \citenamefont {Chung}, \citenamefont {Doran},\ and\ \citenamefont {Geshnizjani}}]{Afshordi:2007yx}%
  \BibitemOpen
  \bibfield  {author} {\bibinfo {author} {\bibfnamefont {N.}~\bibnamefont {Afshordi}}, \bibinfo {author} {\bibfnamefont {D.~J.~H.}\ \bibnamefont {Chung}}, \bibinfo {author} {\bibfnamefont {M.}~\bibnamefont {Doran}},\ and\ \bibinfo {author} {\bibfnamefont {G.}~\bibnamefont {Geshnizjani}},\ }\bibfield  {title} {\bibinfo {title} {{Cuscuton Cosmology: Dark Energy meets Modified Gravity}},\ }\href {https://doi.org/10.1103/PhysRevD.75.123509} {\bibfield  {journal} {\bibinfo  {journal} {Phys. Rev. D}\ }\textbf {\bibinfo {volume} {75}},\ \bibinfo {pages} {123509} (\bibinfo {year} {2007}{\natexlab{b}})},\ \Eprint {https://arxiv.org/abs/astro-ph/0702002} {astro-ph/0702002} \BibitemShut {NoStop}%
\bibitem [{\citenamefont {Hoyle}(1948)}]{Hoyle:1948zz}%
  \BibitemOpen
  \bibfield  {author} {\bibinfo {author} {\bibfnamefont {F.}~\bibnamefont {Hoyle}},\ }\bibfield  {title} {\bibinfo {title} {{A New Model for the Expanding Universe}},\ }\href {https://doi.org/10.1093/mnras/108.5.372} {\bibfield  {journal} {\bibinfo  {journal} {Mon. Not. Roy. Astron. Soc.}\ }\textbf {\bibinfo {volume} {108}},\ \bibinfo {pages} {372} (\bibinfo {year} {1948})}\BibitemShut {NoStop}%
\bibitem [{\citenamefont {Prigogine}\ \emph {et~al.}(1988)\citenamefont {Prigogine}, \citenamefont {Geheniau}, \citenamefont {Gunzig},\ and\ \citenamefont {Nardone}}]{Prigogine1988}%
  \BibitemOpen
  \bibfield  {author} {\bibinfo {author} {\bibfnamefont {I.}~\bibnamefont {Prigogine}}, \bibinfo {author} {\bibfnamefont {J.}~\bibnamefont {Geheniau}}, \bibinfo {author} {\bibfnamefont {E.}~\bibnamefont {Gunzig}},\ and\ \bibinfo {author} {\bibfnamefont {P.}~\bibnamefont {Nardone}},\ }\bibfield  {title} {\bibinfo {title} {Thermodynamics of cosmological matter creation},\ }\href {https://doi.org/10.1073/pnas.85.20.7428} {\bibfield  {journal} {\bibinfo  {journal} {Proceedings of the National Academy of Sciences}\ }\textbf {\bibinfo {volume} {85}},\ \bibinfo {pages} {7428} (\bibinfo {year} {1988})}\BibitemShut {NoStop}%
\bibitem [{\citenamefont {Calvao}\ \emph {et~al.}(1992)\citenamefont {Calvao}, \citenamefont {Lima},\ and\ \citenamefont {Waga}}]{Calvao:1991wg}%
  \BibitemOpen
  \bibfield  {author} {\bibinfo {author} {\bibfnamefont {M.~O.}\ \bibnamefont {Calvao}}, \bibinfo {author} {\bibfnamefont {J.~A.~S.}\ \bibnamefont {Lima}},\ and\ \bibinfo {author} {\bibfnamefont {I.}~\bibnamefont {Waga}},\ }\bibfield  {title} {\bibinfo {title} {{On the thermodynamics of matter creation in cosmology}},\ }\href {https://doi.org/10.1016/0375-9601(92)90437-Q} {\bibfield  {journal} {\bibinfo  {journal} {Phys. Lett. A}\ }\textbf {\bibinfo {volume} {162}},\ \bibinfo {pages} {223} (\bibinfo {year} {1992})}\BibitemShut {NoStop}%
\bibitem [{\citenamefont {Zimdahl}(1996)}]{Zimdahl:1996fj}%
  \BibitemOpen
  \bibfield  {author} {\bibinfo {author} {\bibfnamefont {W.}~\bibnamefont {Zimdahl}},\ }\bibfield  {title} {\bibinfo {title} {{`Understanding' cosmological bulk viscosity}},\ }\href {https://doi.org/10.1093/mnras/280.4.1239} {\bibfield  {journal} {\bibinfo  {journal} {Mon. Not. Roy. Astron. Soc.}\ }\textbf {\bibinfo {volume} {280}},\ \bibinfo {pages} {1239} (\bibinfo {year} {1996})},\ \Eprint {https://arxiv.org/abs/astro-ph/9602128} {astro-ph/9602128} \BibitemShut {NoStop}%
\bibitem [{\citenamefont {Lima}\ and\ \citenamefont {Germano}(1992)}]{Lima:1992np}%
  \BibitemOpen
  \bibfield  {author} {\bibinfo {author} {\bibfnamefont {J.~A.~S.}\ \bibnamefont {Lima}}\ and\ \bibinfo {author} {\bibfnamefont {A.~S.~M.}\ \bibnamefont {Germano}},\ }\bibfield  {title} {\bibinfo {title} {{On the equivalence of bulk viscosity and matter creation}},\ }\href {https://doi.org/10.1016/0375-9601(92)90890-X} {\bibfield  {journal} {\bibinfo  {journal} {Phys. Lett. A}\ }\textbf {\bibinfo {volume} {170}},\ \bibinfo {pages} {373} (\bibinfo {year} {1992})}\BibitemShut {NoStop}%
\bibitem [{\citenamefont {Bolotin}\ \emph {et~al.}(2014)\citenamefont {Bolotin}, \citenamefont {Kostenko}, \citenamefont {Lemets},\ and\ \citenamefont {Yerokhin}}]{Bolotin:2013jpa}%
  \BibitemOpen
  \bibfield  {author} {\bibinfo {author} {\bibfnamefont {Y.~L.}\ \bibnamefont {Bolotin}}, \bibinfo {author} {\bibfnamefont {A.}~\bibnamefont {Kostenko}}, \bibinfo {author} {\bibfnamefont {O.~A.}\ \bibnamefont {Lemets}},\ and\ \bibinfo {author} {\bibfnamefont {D.~A.}\ \bibnamefont {Yerokhin}},\ }\bibfield  {title} {\bibinfo {title} {{Cosmological Evolution With Interaction Between Dark Energy And Dark Matter}},\ }\href {https://doi.org/10.1142/S0218271815300074} {\bibfield  {journal} {\bibinfo  {journal} {Int. J. Mod. Phys. D}\ }\textbf {\bibinfo {volume} {24}},\ \bibinfo {pages} {1530007} (\bibinfo {year} {2014})},\ \Eprint {https://arxiv.org/abs/1310.0085} {1310.0085} \BibitemShut {NoStop}%
\bibitem [{\citenamefont {Wang}\ \emph {et~al.}(2016)\citenamefont {Wang}, \citenamefont {Abdalla}, \citenamefont {Atrio-Barandela},\ and\ \citenamefont {Pavon}}]{Wang:2016lxa}%
  \BibitemOpen
  \bibfield  {author} {\bibinfo {author} {\bibfnamefont {B.}~\bibnamefont {Wang}}, \bibinfo {author} {\bibfnamefont {E.}~\bibnamefont {Abdalla}}, \bibinfo {author} {\bibfnamefont {F.}~\bibnamefont {Atrio-Barandela}},\ and\ \bibinfo {author} {\bibfnamefont {D.}~\bibnamefont {Pavon}},\ }\bibfield  {title} {\bibinfo {title} {{Dark Matter and Dark Energy Interactions: Theoretical Challenges, Cosmological Implications and Observational Signatures}},\ }\href {https://doi.org/10.1088/0034-4885/79/9/096901} {\bibfield  {journal} {\bibinfo  {journal} {Rept. Prog. Phys.}\ }\textbf {\bibinfo {volume} {79}},\ \bibinfo {pages} {096901} (\bibinfo {year} {2016})},\ \Eprint {https://arxiv.org/abs/1603.08299} {1603.08299} \BibitemShut {NoStop}%
\bibitem [{\citenamefont {Wang}\ \emph {et~al.}(2024)\citenamefont {Wang}, \citenamefont {Abdalla}, \citenamefont {Atrio-Barandela},\ and\ \citenamefont {Pav\'on}}]{Wang:2024vmw}%
  \BibitemOpen
  \bibfield  {author} {\bibinfo {author} {\bibfnamefont {B.}~\bibnamefont {Wang}}, \bibinfo {author} {\bibfnamefont {E.}~\bibnamefont {Abdalla}}, \bibinfo {author} {\bibfnamefont {F.}~\bibnamefont {Atrio-Barandela}},\ and\ \bibinfo {author} {\bibfnamefont {D.}~\bibnamefont {Pav\'on}},\ }\bibfield  {title} {\bibinfo {title} {{Further understanding the interaction between dark energy and dark matter: current status and future directions}},\ }\href {https://doi.org/10.1088/1361-6633/ad2527} {\bibfield  {journal} {\bibinfo  {journal} {Rept. Prog. Phys.}\ }\textbf {\bibinfo {volume} {87}},\ \bibinfo {pages} {036901} (\bibinfo {year} {2024})},\ \Eprint {https://arxiv.org/abs/2402.00819} {2402.00819} \BibitemShut {NoStop}%
\bibitem [{\citenamefont {Bondi}\ and\ \citenamefont {Gold}(1948)}]{Bondi:1948qk}%
  \BibitemOpen
  \bibfield  {author} {\bibinfo {author} {\bibfnamefont {H.}~\bibnamefont {Bondi}}\ and\ \bibinfo {author} {\bibfnamefont {T.}~\bibnamefont {Gold}},\ }\bibfield  {title} {\bibinfo {title} {{The Steady-State Theory of the Expanding Universe}},\ }\href {https://doi.org/10.1093/mnras/108.3.252} {\bibfield  {journal} {\bibinfo  {journal} {Mon. Not. Roy. Astron. Soc.}\ }\textbf {\bibinfo {volume} {108}},\ \bibinfo {pages} {252} (\bibinfo {year} {1948})}\BibitemShut {NoStop}%
\bibitem [{\citenamefont {Hoyle}\ and\ \citenamefont {Narlikar}(1963)}]{Hoyle:1963}%
  \BibitemOpen
  \bibfield  {author} {\bibinfo {author} {\bibfnamefont {F.}~\bibnamefont {Hoyle}}\ and\ \bibinfo {author} {\bibfnamefont {J.~V.}\ \bibnamefont {Narlikar}},\ }\bibfield  {title} {\bibinfo {title} {{Mach's principle and the creation of matter}},\ }\href {https://doi.org/10.1098/rspa.1963.0072} {\bibfield  {journal} {\bibinfo  {journal} {Proc. Roy. Soc. Lond. A}\ }\textbf {\bibinfo {volume} {273}},\ \bibinfo {pages} {1} (\bibinfo {year} {1963})}\BibitemShut {NoStop}%
\bibitem [{\citenamefont {Paliathanasis}\ \emph {et~al.}(2017)\citenamefont {Paliathanasis}, \citenamefont {Barrow},\ and\ \citenamefont {Pan}}]{Paliathanasis:2016dhu}%
  \BibitemOpen
  \bibfield  {author} {\bibinfo {author} {\bibfnamefont {A.}~\bibnamefont {Paliathanasis}}, \bibinfo {author} {\bibfnamefont {J.~D.}\ \bibnamefont {Barrow}},\ and\ \bibinfo {author} {\bibfnamefont {S.}~\bibnamefont {Pan}},\ }\bibfield  {title} {\bibinfo {title} {{Cosmological solutions with gravitational particle production and nonzero curvature}},\ }\href {https://doi.org/10.1103/PhysRevD.95.103516} {\bibfield  {journal} {\bibinfo  {journal} {Phys. Rev. D}\ }\textbf {\bibinfo {volume} {95}},\ \bibinfo {pages} {103516} (\bibinfo {year} {2017})},\ \Eprint {https://arxiv.org/abs/1610.02893} {1610.02893} \BibitemShut {NoStop}%
\bibitem [{\citenamefont {Schiavone}\ \emph {et~al.}(2026)\citenamefont {Schiavone}, \citenamefont {De~Angelis}, \citenamefont {Escamilla}, \citenamefont {Montani},\ and\ \citenamefont {Di~Valentino}}]{Schiavone:2026agq}%
  \BibitemOpen
  \bibfield  {author} {\bibinfo {author} {\bibfnamefont {T.}~\bibnamefont {Schiavone}}, \bibinfo {author} {\bibfnamefont {M.}~\bibnamefont {De~Angelis}}, \bibinfo {author} {\bibfnamefont {L.~A.}\ \bibnamefont {Escamilla}}, \bibinfo {author} {\bibfnamefont {G.}~\bibnamefont {Montani}},\ and\ \bibinfo {author} {\bibfnamefont {E.}~\bibnamefont {Di~Valentino}},\ }\bibfield  {title} {\bibinfo {title} {{Revisiting the matter creation process: Observational constraints on gravitationally induced dark energy and the Hubble tension}},\ }\href {https://doi.org/10.1103/bqwg-ff57} {\bibfield  {journal} {\bibinfo  {journal} {Phys. Rev. D}\ }\textbf {\bibinfo {volume} {113}},\ \bibinfo {pages} {123552} (\bibinfo {year} {2026})},\ \Eprint {https://arxiv.org/abs/2601.14222} {2601.14222} \BibitemShut {NoStop}%
\bibitem [{\citenamefont {Lima}\ \emph {et~al.}(2010)\citenamefont {Lima}, \citenamefont {Jesus},\ and\ \citenamefont {Oliveira}}]{Lima:2009ic}%
  \BibitemOpen
  \bibfield  {author} {\bibinfo {author} {\bibfnamefont {J.~A.~S.}\ \bibnamefont {Lima}}, \bibinfo {author} {\bibfnamefont {J.~F.}\ \bibnamefont {Jesus}},\ and\ \bibinfo {author} {\bibfnamefont {F.~A.}\ \bibnamefont {Oliveira}},\ }\bibfield  {title} {\bibinfo {title} {{CDM Accelerating Cosmology as an Alternative to $\Lambda$CDM model}},\ }\href {https://doi.org/10.1088/1475-7516/2010/11/027} {\bibfield  {journal} {\bibinfo  {journal} {JCAP}\ }\textbf {\bibinfo {volume} {11}},\ \bibinfo {pages} {027} (\bibinfo {year} {2010})},\ \Eprint {https://arxiv.org/abs/0911.5727} {0911.5727} \BibitemShut {NoStop}%
\bibitem [{\citenamefont {Steigman}\ \emph {et~al.}(2009)\citenamefont {Steigman}, \citenamefont {Santos},\ and\ \citenamefont {Lima}}]{Steigman:2008bc}%
  \BibitemOpen
  \bibfield  {author} {\bibinfo {author} {\bibfnamefont {G.}~\bibnamefont {Steigman}}, \bibinfo {author} {\bibfnamefont {R.~C.}\ \bibnamefont {Santos}},\ and\ \bibinfo {author} {\bibfnamefont {J.~A.~S.}\ \bibnamefont {Lima}},\ }\bibfield  {title} {\bibinfo {title} {{An Accelerating Cosmology Without Dark Energy}},\ }\href {https://doi.org/10.1088/1475-7516/2009/06/033} {\bibfield  {journal} {\bibinfo  {journal} {JCAP}\ }\textbf {\bibinfo {volume} {06}},\ \bibinfo {pages} {033} (\bibinfo {year} {2009})},\ \Eprint {https://arxiv.org/abs/0812.3912} {0812.3912} \BibitemShut {NoStop}%
\bibitem [{\citenamefont {Ramos}\ \emph {et~al.}(2014)\citenamefont {Ramos}, \citenamefont {Vargas~dos Santos},\ and\ \citenamefont {Waga}}]{Ramos:2014dba}%
  \BibitemOpen
  \bibfield  {author} {\bibinfo {author} {\bibfnamefont {R.~O.}\ \bibnamefont {Ramos}}, \bibinfo {author} {\bibfnamefont {M.}~\bibnamefont {Vargas~dos Santos}},\ and\ \bibinfo {author} {\bibfnamefont {I.}~\bibnamefont {Waga}},\ }\bibfield  {title} {\bibinfo {title} {{Matter creation and cosmic acceleration}},\ }\href {https://doi.org/10.1103/PhysRevD.89.083524} {\bibfield  {journal} {\bibinfo  {journal} {Phys. Rev. D}\ }\textbf {\bibinfo {volume} {89}},\ \bibinfo {pages} {083524} (\bibinfo {year} {2014})},\ \Eprint {https://arxiv.org/abs/1404.2604} {1404.2604} \BibitemShut {NoStop}%
\bibitem [{\citenamefont {Li}\ and\ \citenamefont {Barrow}(2009)}]{Li:2009mf}%
  \BibitemOpen
  \bibfield  {author} {\bibinfo {author} {\bibfnamefont {B.}~\bibnamefont {Li}}\ and\ \bibinfo {author} {\bibfnamefont {J.~D.}\ \bibnamefont {Barrow}},\ }\bibfield  {title} {\bibinfo {title} {{Does Bulk Viscosity Create a Viable Unified Dark Matter Model?}},\ }\href {https://doi.org/10.1103/PhysRevD.79.103521} {\bibfield  {journal} {\bibinfo  {journal} {Phys. Rev. D}\ }\textbf {\bibinfo {volume} {79}},\ \bibinfo {pages} {103521} (\bibinfo {year} {2009})},\ \Eprint {https://arxiv.org/abs/0902.3163} {0902.3163} \BibitemShut {NoStop}%
\bibitem [{\citenamefont {Velten}\ and\ \citenamefont {Schwarz}(2011)}]{Velten:2011bg}%
  \BibitemOpen
  \bibfield  {author} {\bibinfo {author} {\bibfnamefont {H.}~\bibnamefont {Velten}}\ and\ \bibinfo {author} {\bibfnamefont {D.~J.}\ \bibnamefont {Schwarz}},\ }\bibfield  {title} {\bibinfo {title} {{Constraints on dissipative unified dark matter}},\ }\href {https://doi.org/10.1088/1475-7516/2011/09/016} {\bibfield  {journal} {\bibinfo  {journal} {JCAP}\ }\textbf {\bibinfo {volume} {09}},\ \bibinfo {pages} {016} (\bibinfo {year} {2011})},\ \Eprint {https://arxiv.org/abs/1107.1143} {1107.1143} \BibitemShut {NoStop}%
\bibitem [{\citenamefont {Carvalho}\ \emph {et~al.}(1992)\citenamefont {Carvalho}, \citenamefont {Lima},\ and\ \citenamefont {Waga}}]{Carvalho:1991ut}%
  \BibitemOpen
  \bibfield  {author} {\bibinfo {author} {\bibfnamefont {J.~C.}\ \bibnamefont {Carvalho}}, \bibinfo {author} {\bibfnamefont {J.~A.~S.}\ \bibnamefont {Lima}},\ and\ \bibinfo {author} {\bibfnamefont {I.}~\bibnamefont {Waga}},\ }\bibfield  {title} {\bibinfo {title} {{Cosmological consequences of a time-dependent $\Lambda$ term}},\ }\href {https://doi.org/10.1103/PhysRevD.46.2404} {\bibfield  {journal} {\bibinfo  {journal} {Phys. Rev. D}\ }\textbf {\bibinfo {volume} {46}},\ \bibinfo {pages} {2404} (\bibinfo {year} {1992})}\BibitemShut {NoStop}%
\bibitem [{\citenamefont {Basilakos}\ \emph {et~al.}(2009)\citenamefont {Basilakos}, \citenamefont {Plionis},\ and\ \citenamefont {Sol\`a}}]{Basilakos:2009wi}%
  \BibitemOpen
  \bibfield  {author} {\bibinfo {author} {\bibfnamefont {S.}~\bibnamefont {Basilakos}}, \bibinfo {author} {\bibfnamefont {M.}~\bibnamefont {Plionis}},\ and\ \bibinfo {author} {\bibfnamefont {J.}~\bibnamefont {Sol\`a}},\ }\bibfield  {title} {\bibinfo {title} {{Hubble expansion and structure formation in time varying vacuum models}},\ }\href {https://doi.org/10.1103/PhysRevD.80.083511} {\bibfield  {journal} {\bibinfo  {journal} {Phys. Rev. D}\ }\textbf {\bibinfo {volume} {80}},\ \bibinfo {pages} {083511} (\bibinfo {year} {2009})},\ \Eprint {https://arxiv.org/abs/0907.4555} {0907.4555} \BibitemShut {NoStop}%
\bibitem [{\citenamefont {Akarsu}\ and\ \citenamefont {Uzun}(2023)}]{Akarsu:2023nyl}%
  \BibitemOpen
  \bibfield  {author} {\bibinfo {author} {\bibfnamefont {O.}~\bibnamefont {Akarsu}}\ and\ \bibinfo {author} {\bibfnamefont {N.~M.}\ \bibnamefont {Uzun}},\ }\bibfield  {title} {\bibinfo {title} {{Cosmological models in scale-independent energy-momentum squared gravity}},\ }\href {https://doi.org/10.1016/j.dark.2023.101194} {\bibfield  {journal} {\bibinfo  {journal} {Phys. Dark Univ.}\ }\textbf {\bibinfo {volume} {40}},\ \bibinfo {pages} {101194} (\bibinfo {year} {2023})},\ \Eprint {https://arxiv.org/abs/2301.11204} {2301.11204} \BibitemShut {NoStop}%
\bibitem [{\citenamefont {Caldwell}(2002)}]{Caldwell:1999ew}%
  \BibitemOpen
  \bibfield  {author} {\bibinfo {author} {\bibfnamefont {R.~R.}\ \bibnamefont {Caldwell}},\ }\bibfield  {title} {\bibinfo {title} {{A Phantom menace? Cosmological consequences of a dark energy component with super-negative equation of state}},\ }\href {https://doi.org/10.1016/S0370-2693(02)02589-3} {\bibfield  {journal} {\bibinfo  {journal} {Phys. Lett. B}\ }\textbf {\bibinfo {volume} {545}},\ \bibinfo {pages} {23} (\bibinfo {year} {2002})},\ \Eprint {https://arxiv.org/abs/astro-ph/9908168} {astro-ph/9908168} \BibitemShut {NoStop}%
\bibitem [{\citenamefont {Caldwell}\ \emph {et~al.}(2003)\citenamefont {Caldwell}, \citenamefont {Kamionkowski},\ and\ \citenamefont {Weinberg}}]{Caldwell:2003vq}%
  \BibitemOpen
  \bibfield  {author} {\bibinfo {author} {\bibfnamefont {R.~R.}\ \bibnamefont {Caldwell}}, \bibinfo {author} {\bibfnamefont {M.}~\bibnamefont {Kamionkowski}},\ and\ \bibinfo {author} {\bibfnamefont {N.~N.}\ \bibnamefont {Weinberg}},\ }\bibfield  {title} {\bibinfo {title} {{Phantom Energy: Dark Energy with {$w<-1$} Causes a Cosmic Doomsday}},\ }\href {https://doi.org/10.1103/PhysRevLett.91.071301} {\bibfield  {journal} {\bibinfo  {journal} {Phys. Rev. Lett.}\ }\textbf {\bibinfo {volume} {91}},\ \bibinfo {pages} {071301} (\bibinfo {year} {2003})},\ \Eprint {https://arxiv.org/abs/astro-ph/0302506} {astro-ph/0302506} \BibitemShut {NoStop}%
\bibitem [{\citenamefont {Bouhmadi-L{\'o}pez}\ \emph {et~al.}(2015)\citenamefont {Bouhmadi-L{\'o}pez}, \citenamefont {Errahmani}, \citenamefont {Mart{\'\i}n-Moruno}, \citenamefont {Ouali},\ and\ \citenamefont {Tavakoli}}]{Bouhmadi-Lopez:2014cca}%
  \BibitemOpen
  \bibfield  {author} {\bibinfo {author} {\bibfnamefont {M.}~\bibnamefont {Bouhmadi-L{\'o}pez}}, \bibinfo {author} {\bibfnamefont {A.}~\bibnamefont {Errahmani}}, \bibinfo {author} {\bibfnamefont {P.}~\bibnamefont {Mart{\'\i}n-Moruno}}, \bibinfo {author} {\bibfnamefont {T.}~\bibnamefont {Ouali}},\ and\ \bibinfo {author} {\bibfnamefont {Y.}~\bibnamefont {Tavakoli}},\ }\bibfield  {title} {\bibinfo {title} {{The little sibling of the big rip singularity}},\ }\href {https://doi.org/10.1142/S0218271815500789} {\bibfield  {journal} {\bibinfo  {journal} {Int. J. Mod. Phys. D}\ }\textbf {\bibinfo {volume} {24}},\ \bibinfo {pages} {1550078} (\bibinfo {year} {2015})},\ \Eprint {https://arxiv.org/abs/1407.2446} {1407.2446} \BibitemShut {NoStop}%
\bibitem [{\citenamefont {Diez-Tejedor}(2013)}]{Diez-Tejedor:2013nwa}%
  \BibitemOpen
  \bibfield  {author} {\bibinfo {author} {\bibfnamefont {A.}~\bibnamefont {Diez-Tejedor}},\ }\bibfield  {title} {\bibinfo {title} {{Note on scalars, perfect fluids, constrained field theories, and all that}},\ }\href {https://doi.org/10.1016/j.physletb.2013.10.030} {\bibfield  {journal} {\bibinfo  {journal} {Phys. Lett. B}\ }\textbf {\bibinfo {volume} {727}},\ \bibinfo {pages} {27} (\bibinfo {year} {2013})},\ \Eprint {https://arxiv.org/abs/1309.4756} {1309.4756} \BibitemShut {NoStop}%
\bibitem [{\citenamefont {Faraoni}(2012)}]{Faraoni:2012hn}%
  \BibitemOpen
  \bibfield  {author} {\bibinfo {author} {\bibfnamefont {V.}~\bibnamefont {Faraoni}},\ }\bibfield  {title} {\bibinfo {title} {{Correspondence between a scalar field and an effective perfect fluid}},\ }\href {https://doi.org/10.1103/PhysRevD.85.024040} {\bibfield  {journal} {\bibinfo  {journal} {Phys. Rev. D}\ }\textbf {\bibinfo {volume} {85}},\ \bibinfo {pages} {024040} (\bibinfo {year} {2012})},\ \Eprint {https://arxiv.org/abs/1201.1448} {1201.1448} \BibitemShut {NoStop}%
\bibitem [{\citenamefont {Semiz}(2012)}]{Semiz:2012zz}%
  \BibitemOpen
  \bibfield  {author} {\bibinfo {author} {\bibfnamefont {I.}~\bibnamefont {Semiz}},\ }\bibfield  {title} {\bibinfo {title} {{Comment on `Correspondence between a scalar field and an effective perfect fluid'}},\ }\href {https://doi.org/10.1103/PhysRevD.85.068501} {\bibfield  {journal} {\bibinfo  {journal} {Phys. Rev. D}\ }\textbf {\bibinfo {volume} {85}},\ \bibinfo {pages} {068501} (\bibinfo {year} {2012})}\BibitemShut {NoStop}%
\bibitem [{\citenamefont {Felder}\ \emph {et~al.}(2002)\citenamefont {Felder}, \citenamefont {Frolov}, \citenamefont {Kofman},\ and\ \citenamefont {Linde}}]{Felder:2002jk}%
  \BibitemOpen
  \bibfield  {author} {\bibinfo {author} {\bibfnamefont {G.~N.}\ \bibnamefont {Felder}}, \bibinfo {author} {\bibfnamefont {A.~V.}\ \bibnamefont {Frolov}}, \bibinfo {author} {\bibfnamefont {L.}~\bibnamefont {Kofman}},\ and\ \bibinfo {author} {\bibfnamefont {A.~D.}\ \bibnamefont {Linde}},\ }\bibfield  {title} {\bibinfo {title} {{Cosmology with negative potentials}},\ }\href {https://doi.org/10.1103/PhysRevD.66.023507} {\bibfield  {journal} {\bibinfo  {journal} {Phys. Rev. D}\ }\textbf {\bibinfo {volume} {66}},\ \bibinfo {pages} {023507} (\bibinfo {year} {2002})},\ \Eprint {https://arxiv.org/abs/hep-th/0202017} {hep-th/0202017} \BibitemShut {NoStop}%
\bibitem [{\citenamefont {Armendariz-Picon}\ \emph {et~al.}(2001)\citenamefont {Armendariz-Picon}, \citenamefont {Mukhanov},\ and\ \citenamefont {Steinhardt}}]{Armendariz-Picon:2000ulo}%
  \BibitemOpen
  \bibfield  {author} {\bibinfo {author} {\bibfnamefont {C.}~\bibnamefont {Armendariz-Picon}}, \bibinfo {author} {\bibfnamefont {V.~F.}\ \bibnamefont {Mukhanov}},\ and\ \bibinfo {author} {\bibfnamefont {P.~J.}\ \bibnamefont {Steinhardt}},\ }\bibfield  {title} {\bibinfo {title} {{Essentials of k essence}},\ }\href {https://doi.org/10.1103/PhysRevD.63.103510} {\bibfield  {journal} {\bibinfo  {journal} {Phys. Rev. D}\ }\textbf {\bibinfo {volume} {63}},\ \bibinfo {pages} {103510} (\bibinfo {year} {2001})},\ \Eprint {https://arxiv.org/abs/astro-ph/0006373} {astro-ph/0006373} \BibitemShut {NoStop}%
\bibitem [{\citenamefont {Serish}\ \emph {et~al.}(2026)\citenamefont {Serish}, \citenamefont {Hosseini~Mansoori}, \citenamefont {Felegary}, \citenamefont {Akarsu},\ and\ \citenamefont {Sami}}]{Serish:2025ian}%
  \BibitemOpen
  \bibfield  {author} {\bibinfo {author} {\bibfnamefont {T.~F.}\ \bibnamefont {Serish}}, \bibinfo {author} {\bibfnamefont {S.~A.}\ \bibnamefont {Hosseini~Mansoori}}, \bibinfo {author} {\bibfnamefont {F.}~\bibnamefont {Felegary}}, \bibinfo {author} {\bibfnamefont {{\"O}.}~\bibnamefont {Akarsu}},\ and\ \bibinfo {author} {\bibfnamefont {M.}~\bibnamefont {Sami}},\ }\bibfield  {title} {\bibinfo {title} {{k-inflation: non-separable case meets ACT measurements}},\ }\href {https://doi.org/10.1088/1475-7516/2026/04/031} {\bibfield  {journal} {\bibinfo  {journal} {JCAP}\ }\textbf {\bibinfo {volume} {04}},\ \bibinfo {pages} {031} (\bibinfo {year} {2026})},\ \Eprint {https://arxiv.org/abs/2511.16621} {2511.16621} \BibitemShut {NoStop}%
\bibitem [{\citenamefont {Ayon-Beato}\ \emph {et~al.}(2006)\citenamefont {Ayon-Beato}, \citenamefont {Martinez},\ and\ \citenamefont {Zanelli}}]{Ayon-Beato:2004nzi}%
  \BibitemOpen
  \bibfield  {author} {\bibinfo {author} {\bibfnamefont {E.}~\bibnamefont {Ayon-Beato}}, \bibinfo {author} {\bibfnamefont {C.}~\bibnamefont {Martinez}},\ and\ \bibinfo {author} {\bibfnamefont {J.}~\bibnamefont {Zanelli}},\ }\bibfield  {title} {\bibinfo {title} {{Stealth scalar field overflying a $2+1$ black hole}},\ }\href {https://doi.org/10.1007/s10714-005-0213-x} {\bibfield  {journal} {\bibinfo  {journal} {Gen. Rel. Grav.}\ }\textbf {\bibinfo {volume} {38}},\ \bibinfo {pages} {145} (\bibinfo {year} {2006})},\ \Eprint {https://arxiv.org/abs/hep-th/0403228} {hep-th/0403228} \BibitemShut {NoStop}%
\bibitem [{\citenamefont {Ay{\'o}n-Beato}\ \emph {et~al.}(2013)\citenamefont {Ay{\'o}n-Beato}, \citenamefont {Garc{\'\i}a}, \citenamefont {Ram{\'\i}rez-Baca},\ and\ \citenamefont {Terrero-Escalante}}]{Ayon-Beato:2013bsa}%
  \BibitemOpen
  \bibfield  {author} {\bibinfo {author} {\bibfnamefont {E.}~\bibnamefont {Ay{\'o}n-Beato}}, \bibinfo {author} {\bibfnamefont {A.~A.}\ \bibnamefont {Garc{\'\i}a}}, \bibinfo {author} {\bibfnamefont {P.~I.}\ \bibnamefont {Ram{\'\i}rez-Baca}},\ and\ \bibinfo {author} {\bibfnamefont {C.~A.}\ \bibnamefont {Terrero-Escalante}},\ }\bibfield  {title} {\bibinfo {title} {{Conformal stealth for any standard cosmology}},\ }\href {https://doi.org/10.1103/PhysRevD.88.063523} {\bibfield  {journal} {\bibinfo  {journal} {Phys. Rev. D}\ }\textbf {\bibinfo {volume} {88}},\ \bibinfo {pages} {063523} (\bibinfo {year} {2013})},\ \Eprint {https://arxiv.org/abs/1307.6534} {1307.6534} \BibitemShut {NoStop}%
\bibitem [{\citenamefont {Blanco}\ \emph {et~al.}(2026)\citenamefont {Blanco}, \citenamefont {Campuzano}, \citenamefont {Corona-Oran},\ and\ \citenamefont {C{\'a}rdenas}}]{Blanco:2025kjb}%
  \BibitemOpen
  \bibfield  {author} {\bibinfo {author} {\bibfnamefont {J.}~\bibnamefont {Blanco}}, \bibinfo {author} {\bibfnamefont {C.}~\bibnamefont {Campuzano}}, \bibinfo {author} {\bibfnamefont {J.~C.}\ \bibnamefont {Corona-Oran}},\ and\ \bibinfo {author} {\bibfnamefont {V.~H.}\ \bibnamefont {C{\'a}rdenas}},\ }\bibfield  {title} {\bibinfo {title} {{Non-geometrical perturbation on homogeneous stealth dust}},\ }\href {https://doi.org/10.31349/RevMexFis.72.020702} {\bibfield  {journal} {\bibinfo  {journal} {Rev. Mex. Fis.}\ }\textbf {\bibinfo {volume} {72}},\ \bibinfo {pages} {020702} (\bibinfo {year} {2026})},\ \Eprint {https://arxiv.org/abs/2504.16313} {2504.16313} \BibitemShut {NoStop}%
\bibitem [{\citenamefont {Aguilar-P{\'e}rez}\ \emph {et~al.}(2026)\citenamefont {Aguilar-P{\'e}rez}, \citenamefont {Campuzano}, \citenamefont {C{\'a}rdenas}, \citenamefont {Cruz},\ and\ \citenamefont {Saavedra}}]{Aguilar-Perez:2026jus}%
  \BibitemOpen
  \bibfield  {author} {\bibinfo {author} {\bibfnamefont {G.}~\bibnamefont {Aguilar-P{\'e}rez}}, \bibinfo {author} {\bibfnamefont {C.}~\bibnamefont {Campuzano}}, \bibinfo {author} {\bibfnamefont {V.~H.}\ \bibnamefont {C{\'a}rdenas}}, \bibinfo {author} {\bibfnamefont {M.}~\bibnamefont {Cruz}},\ and\ \bibinfo {author} {\bibfnamefont {J.}~\bibnamefont {Saavedra}},\ }\href@noop {} {\bibinfo {title} {{Cosmological Stealth fields and Non-Equilibrium thermodynamics}}} (\bibinfo {year} {2026}),\ \Eprint {https://arxiv.org/abs/2606.31180} {2606.31180} \BibitemShut {NoStop}%
\bibitem [{\citenamefont {Fang}\ \emph {et~al.}(2008)\citenamefont {Fang}, \citenamefont {Hu},\ and\ \citenamefont {Lewis}}]{Fang:2008sn}%
  \BibitemOpen
  \bibfield  {author} {\bibinfo {author} {\bibfnamefont {W.}~\bibnamefont {Fang}}, \bibinfo {author} {\bibfnamefont {W.}~\bibnamefont {Hu}},\ and\ \bibinfo {author} {\bibfnamefont {A.}~\bibnamefont {Lewis}},\ }\bibfield  {title} {\bibinfo {title} {{Crossing the Phantom Divide with Parameterized Post-Friedmann Dark Energy}},\ }\href {https://doi.org/10.1103/PhysRevD.78.087303} {\bibfield  {journal} {\bibinfo  {journal} {Phys. Rev. D}\ }\textbf {\bibinfo {volume} {78}},\ \bibinfo {pages} {087303} (\bibinfo {year} {2008})},\ \Eprint {https://arxiv.org/abs/0808.3125} {0808.3125} \BibitemShut {NoStop}%
\bibitem [{\citenamefont {Kodama}\ and\ \citenamefont {Sasaki}(1984)}]{Kodama:1984ziu}%
  \BibitemOpen
  \bibfield  {author} {\bibinfo {author} {\bibfnamefont {H.}~\bibnamefont {Kodama}}\ and\ \bibinfo {author} {\bibfnamefont {M.}~\bibnamefont {Sasaki}},\ }\bibfield  {title} {\bibinfo {title} {{Cosmological Perturbation Theory}},\ }\href {https://doi.org/10.1143/PTPS.78.1} {\bibfield  {journal} {\bibinfo  {journal} {Prog. Theor. Phys. Suppl.}\ }\textbf {\bibinfo {volume} {78}},\ \bibinfo {pages} {1} (\bibinfo {year} {1984})}\BibitemShut {NoStop}%
\bibitem [{\citenamefont {Malik}\ \emph {et~al.}(2003)\citenamefont {Malik}, \citenamefont {Wands},\ and\ \citenamefont {Ungarelli}}]{Malik:2002jb}%
  \BibitemOpen
  \bibfield  {author} {\bibinfo {author} {\bibfnamefont {K.~A.}\ \bibnamefont {Malik}}, \bibinfo {author} {\bibfnamefont {D.}~\bibnamefont {Wands}},\ and\ \bibinfo {author} {\bibfnamefont {C.}~\bibnamefont {Ungarelli}},\ }\bibfield  {title} {\bibinfo {title} {{Large scale curvature and entropy perturbations for multiple interacting fluids}},\ }\href {https://doi.org/10.1103/PhysRevD.67.063516} {\bibfield  {journal} {\bibinfo  {journal} {Phys. Rev. D}\ }\textbf {\bibinfo {volume} {67}},\ \bibinfo {pages} {063516} (\bibinfo {year} {2003})},\ \Eprint {https://arxiv.org/abs/astro-ph/0211602} {astro-ph/0211602} \BibitemShut {NoStop}%
\bibitem [{\citenamefont {Malik}\ and\ \citenamefont {Wands}(2005)}]{Malik:2004tf}%
  \BibitemOpen
  \bibfield  {author} {\bibinfo {author} {\bibfnamefont {K.~A.}\ \bibnamefont {Malik}}\ and\ \bibinfo {author} {\bibfnamefont {D.}~\bibnamefont {Wands}},\ }\bibfield  {title} {\bibinfo {title} {{Adiabatic and entropy perturbations with interacting fluids and fields}},\ }\href {https://doi.org/10.1088/1475-7516/2005/02/007} {\bibfield  {journal} {\bibinfo  {journal} {JCAP}\ }\textbf {\bibinfo {volume} {02}},\ \bibinfo {pages} {007} (\bibinfo {year} {2005})},\ \Eprint {https://arxiv.org/abs/astro-ph/0411703} {astro-ph/0411703} \BibitemShut {NoStop}%
\bibitem [{\citenamefont {Valiviita}\ \emph {et~al.}(2008)\citenamefont {Valiviita}, \citenamefont {Majerotto},\ and\ \citenamefont {Maartens}}]{Valiviita:2008iv}%
  \BibitemOpen
  \bibfield  {author} {\bibinfo {author} {\bibfnamefont {J.}~\bibnamefont {Valiviita}}, \bibinfo {author} {\bibfnamefont {E.}~\bibnamefont {Majerotto}},\ and\ \bibinfo {author} {\bibfnamefont {R.}~\bibnamefont {Maartens}},\ }\bibfield  {title} {\bibinfo {title} {{Large-scale instability in interacting dark energy and dark matter fluids}},\ }\href {https://doi.org/10.1088/1475-7516/2008/07/020} {\bibfield  {journal} {\bibinfo  {journal} {JCAP}\ }\textbf {\bibinfo {volume} {07}},\ \bibinfo {pages} {020} (\bibinfo {year} {2008})},\ \Eprint {https://arxiv.org/abs/0804.0232} {0804.0232} \BibitemShut {NoStop}%
\bibitem [{\citenamefont {Ma}\ and\ \citenamefont {Bertschinger}(1995)}]{Ma:1995ey}%
  \BibitemOpen
  \bibfield  {author} {\bibinfo {author} {\bibfnamefont {C.-P.}\ \bibnamefont {Ma}}\ and\ \bibinfo {author} {\bibfnamefont {E.}~\bibnamefont {Bertschinger}},\ }\bibfield  {title} {\bibinfo {title} {Cosmological perturbation theory in the synchronous and conformal newtonian gauges},\ }\href {https://doi.org/10.1086/176550} {\bibfield  {journal} {\bibinfo  {journal} {Astrophys. J.}\ }\textbf {\bibinfo {volume} {455}},\ \bibinfo {pages} {7} (\bibinfo {year} {1995})},\ \Eprint {https://arxiv.org/abs/astro-ph/9506072} {astro-ph/9506072} \BibitemShut {NoStop}%
\bibitem [{\citenamefont {He}\ \emph {et~al.}(2009)\citenamefont {He}, \citenamefont {Wang},\ and\ \citenamefont {Abdalla}}]{He:2008si}%
  \BibitemOpen
  \bibfield  {author} {\bibinfo {author} {\bibfnamefont {J.-H.}\ \bibnamefont {He}}, \bibinfo {author} {\bibfnamefont {B.}~\bibnamefont {Wang}},\ and\ \bibinfo {author} {\bibfnamefont {E.}~\bibnamefont {Abdalla}},\ }\bibfield  {title} {\bibinfo {title} {{Stability of the curvature perturbation in dark sectors' mutual interacting models}},\ }\href {https://doi.org/10.1016/j.physletb.2008.11.062} {\bibfield  {journal} {\bibinfo  {journal} {Phys. Lett. B}\ }\textbf {\bibinfo {volume} {671}},\ \bibinfo {pages} {139} (\bibinfo {year} {2009})},\ \Eprint {https://arxiv.org/abs/0807.3471} {0807.3471} \BibitemShut {NoStop}%
\bibitem [{\citenamefont {Gavela}\ \emph {et~al.}(2010)\citenamefont {Gavela}, \citenamefont {Lopez~Honorez}, \citenamefont {Mena},\ and\ \citenamefont {Rigolin}}]{Gavela:2010tm}%
  \BibitemOpen
  \bibfield  {author} {\bibinfo {author} {\bibfnamefont {M.~B.}\ \bibnamefont {Gavela}}, \bibinfo {author} {\bibfnamefont {L.}~\bibnamefont {Lopez~Honorez}}, \bibinfo {author} {\bibfnamefont {O.}~\bibnamefont {Mena}},\ and\ \bibinfo {author} {\bibfnamefont {S.}~\bibnamefont {Rigolin}},\ }\bibfield  {title} {\bibinfo {title} {{Dark Coupling and Gauge Invariance}},\ }\href {https://doi.org/10.1088/1475-7516/2010/11/044} {\bibfield  {journal} {\bibinfo  {journal} {JCAP}\ }\textbf {\bibinfo {volume} {11}},\ \bibinfo {pages} {044} (\bibinfo {year} {2010})},\ \Eprint {https://arxiv.org/abs/1005.0295} {1005.0295} \BibitemShut {NoStop}%
\bibitem [{\citenamefont {Lopez~Honorez}\ \emph {et~al.}(2010)\citenamefont {Lopez~Honorez}, \citenamefont {Reid}, \citenamefont {Mena}, \citenamefont {Verde},\ and\ \citenamefont {Jimenez}}]{LopezHonorez:2010esq}%
  \BibitemOpen
  \bibfield  {author} {\bibinfo {author} {\bibfnamefont {L.}~\bibnamefont {Lopez~Honorez}}, \bibinfo {author} {\bibfnamefont {B.~A.}\ \bibnamefont {Reid}}, \bibinfo {author} {\bibfnamefont {O.}~\bibnamefont {Mena}}, \bibinfo {author} {\bibfnamefont {L.}~\bibnamefont {Verde}},\ and\ \bibinfo {author} {\bibfnamefont {R.}~\bibnamefont {Jimenez}},\ }\bibfield  {title} {\bibinfo {title} {{Coupled dark matter-dark energy in light of near Universe observations}},\ }\href {https://doi.org/10.1088/1475-7516/2010/09/029} {\bibfield  {journal} {\bibinfo  {journal} {JCAP}\ }\textbf {\bibinfo {volume} {09}},\ \bibinfo {pages} {029} (\bibinfo {year} {2010})},\ \Eprint {https://arxiv.org/abs/1006.0877} {1006.0877} \BibitemShut {NoStop}%
\bibitem [{\citenamefont {Clemson}\ \emph {et~al.}(2012)\citenamefont {Clemson}, \citenamefont {Koyama}, \citenamefont {Zhao}, \citenamefont {Maartens},\ and\ \citenamefont {Valiviita}}]{Clemson:2011an}%
  \BibitemOpen
  \bibfield  {author} {\bibinfo {author} {\bibfnamefont {T.}~\bibnamefont {Clemson}}, \bibinfo {author} {\bibfnamefont {K.}~\bibnamefont {Koyama}}, \bibinfo {author} {\bibfnamefont {G.-B.}\ \bibnamefont {Zhao}}, \bibinfo {author} {\bibfnamefont {R.}~\bibnamefont {Maartens}},\ and\ \bibinfo {author} {\bibfnamefont {J.}~\bibnamefont {Valiviita}},\ }\bibfield  {title} {\bibinfo {title} {{Interacting Dark Energy -- constraints and degeneracies}},\ }\href {https://doi.org/10.1103/PhysRevD.85.043007} {\bibfield  {journal} {\bibinfo  {journal} {Phys. Rev. D}\ }\textbf {\bibinfo {volume} {85}},\ \bibinfo {pages} {043007} (\bibinfo {year} {2012})},\ \Eprint {https://arxiv.org/abs/1109.6234} {1109.6234} \BibitemShut {NoStop}%
\bibitem [{\citenamefont {Bouhmadi-L{\'o}pez}\ \emph {et~al.}(2026)\citenamefont {Bouhmadi-L{\'o}pez}, \citenamefont {Chiang},\ and\ \citenamefont {Ibarra-Uriondo}}]{Bouhmadi-Lopez:2026vyc}%
  \BibitemOpen
  \bibfield  {author} {\bibinfo {author} {\bibfnamefont {M.}~\bibnamefont {Bouhmadi-L{\'o}pez}}, \bibinfo {author} {\bibfnamefont {H.-W.}\ \bibnamefont {Chiang}},\ and\ \bibinfo {author} {\bibfnamefont {B.}~\bibnamefont {Ibarra-Uriondo}},\ }\href@noop {} {\bibinfo {title} {{Alleviating the Hubble Tension with Smooth Sign-Switching Dark Energy: Full CMB Constraints with DESI and PantheonPlus}}} (\bibinfo {year} {2026}),\ \Eprint {https://arxiv.org/abs/2607.05044} {2607.05044} \BibitemShut {NoStop}%
\bibitem [{\citenamefont {Vikman}(2005)}]{Vikman:2004dc}%
  \BibitemOpen
  \bibfield  {author} {\bibinfo {author} {\bibfnamefont {A.}~\bibnamefont {Vikman}},\ }\bibfield  {title} {\bibinfo {title} {{Can dark energy evolve to the phantom?}},\ }\href {https://doi.org/10.1103/PhysRevD.71.023515} {\bibfield  {journal} {\bibinfo  {journal} {Phys. Rev. D}\ }\textbf {\bibinfo {volume} {71}},\ \bibinfo {pages} {023515} (\bibinfo {year} {2005})},\ \Eprint {https://arxiv.org/abs/astro-ph/0407107} {astro-ph/0407107} \BibitemShut {NoStop}%
\bibitem [{\citenamefont {Hu}(2005)}]{Hu:2004kh}%
  \BibitemOpen
  \bibfield  {author} {\bibinfo {author} {\bibfnamefont {W.}~\bibnamefont {Hu}},\ }\bibfield  {title} {\bibinfo {title} {{Crossing the phantom divide: Dark energy internal degrees of freedom}},\ }\href {https://doi.org/10.1103/PhysRevD.71.047301} {\bibfield  {journal} {\bibinfo  {journal} {Phys. Rev. D}\ }\textbf {\bibinfo {volume} {71}},\ \bibinfo {pages} {047301} (\bibinfo {year} {2005})},\ \Eprint {https://arxiv.org/abs/astro-ph/0410680} {astro-ph/0410680} \BibitemShut {NoStop}%
\bibitem [{\citenamefont {Kunz}\ and\ \citenamefont {Sapone}(2006)}]{Kunz:2006wc}%
  \BibitemOpen
  \bibfield  {author} {\bibinfo {author} {\bibfnamefont {M.}~\bibnamefont {Kunz}}\ and\ \bibinfo {author} {\bibfnamefont {D.}~\bibnamefont {Sapone}},\ }\bibfield  {title} {\bibinfo {title} {{Crossing the Phantom Divide}},\ }\href {https://doi.org/10.1103/PhysRevD.74.123503} {\bibfield  {journal} {\bibinfo  {journal} {Phys. Rev. D}\ }\textbf {\bibinfo {volume} {74}},\ \bibinfo {pages} {123503} (\bibinfo {year} {2006})},\ \Eprint {https://arxiv.org/abs/astro-ph/0609040} {astro-ph/0609040} \BibitemShut {NoStop}%
\bibitem [{\citenamefont {Babichev}\ \emph {et~al.}(2008)\citenamefont {Babichev}, \citenamefont {Mukhanov},\ and\ \citenamefont {Vikman}}]{Babichev:2007dw}%
  \BibitemOpen
  \bibfield  {author} {\bibinfo {author} {\bibfnamefont {E.}~\bibnamefont {Babichev}}, \bibinfo {author} {\bibfnamefont {V.}~\bibnamefont {Mukhanov}},\ and\ \bibinfo {author} {\bibfnamefont {A.}~\bibnamefont {Vikman}},\ }\bibfield  {title} {\bibinfo {title} {{k-Essence, superluminal propagation, causality and emergent geometry}},\ }\href {https://doi.org/10.1088/1126-6708/2008/02/101} {\bibfield  {journal} {\bibinfo  {journal} {JHEP}\ }\textbf {\bibinfo {volume} {02}},\ \bibinfo {pages} {101} (\bibinfo {year} {2008})},\ \Eprint {https://arxiv.org/abs/0708.0561} {0708.0561} \BibitemShut {NoStop}%
\bibitem [{\citenamefont {Jacobs}(1968)}]{Jacobs:1968}%
  \BibitemOpen
  \bibfield  {author} {\bibinfo {author} {\bibfnamefont {K.~C.}\ \bibnamefont {Jacobs}},\ }\bibfield  {title} {\bibinfo {title} {{Spatially Homogeneous and Euclidean Cosmological Models with Shear}},\ }\href {https://doi.org/10.1086/149694} {\bibfield  {journal} {\bibinfo  {journal} {Astrophys. J.}\ }\textbf {\bibinfo {volume} {153}},\ \bibinfo {pages} {661} (\bibinfo {year} {1968})}\BibitemShut {NoStop}%
\bibitem [{\citenamefont {Koivisto}\ and\ \citenamefont {Mota}(2008)}]{Koivisto:2008ig}%
  \BibitemOpen
  \bibfield  {author} {\bibinfo {author} {\bibfnamefont {T.}~\bibnamefont {Koivisto}}\ and\ \bibinfo {author} {\bibfnamefont {D.~F.}\ \bibnamefont {Mota}},\ }\bibfield  {title} {\bibinfo {title} {{Anisotropic dark energy: dynamics of the background and perturbations}},\ }\href {https://doi.org/10.1088/1475-7516/2008/06/018} {\bibfield  {journal} {\bibinfo  {journal} {JCAP}\ }\textbf {\bibinfo {volume} {06}},\ \bibinfo {pages} {018} (\bibinfo {year} {2008})},\ \Eprint {https://arxiv.org/abs/0801.3676} {0801.3676} \BibitemShut {NoStop}%
\bibitem [{\citenamefont {Ellis}\ and\ \citenamefont {van Elst}(1999)}]{Ellis:1998ct}%
  \BibitemOpen
  \bibfield  {author} {\bibinfo {author} {\bibfnamefont {G.~F.~R.}\ \bibnamefont {Ellis}}\ and\ \bibinfo {author} {\bibfnamefont {H.}~\bibnamefont {van Elst}},\ }\bibfield  {title} {\bibinfo {title} {{Cosmological models: Cargese lectures 1998}},\ }\href {https://doi.org/10.1007/978-94-011-4455-1_1} {\bibfield  {journal} {\bibinfo  {journal} {NATO Sci. Ser. C}\ }\textbf {\bibinfo {volume} {541}},\ \bibinfo {pages} {1} (\bibinfo {year} {1999})},\ \Eprint {https://arxiv.org/abs/gr-qc/9812046} {gr-qc/9812046} \BibitemShut {NoStop}%
\bibitem [{\citenamefont {Tsagas}\ \emph {et~al.}(2008)\citenamefont {Tsagas}, \citenamefont {Challinor},\ and\ \citenamefont {Maartens}}]{Tsagas:2007yx}%
  \BibitemOpen
  \bibfield  {author} {\bibinfo {author} {\bibfnamefont {C.~G.}\ \bibnamefont {Tsagas}}, \bibinfo {author} {\bibfnamefont {A.}~\bibnamefont {Challinor}},\ and\ \bibinfo {author} {\bibfnamefont {R.}~\bibnamefont {Maartens}},\ }\bibfield  {title} {\bibinfo {title} {{Relativistic cosmology and large-scale structure}},\ }\href {https://doi.org/10.1016/j.physrep.2008.03.003} {\bibfield  {journal} {\bibinfo  {journal} {Phys. Rept.}\ }\textbf {\bibinfo {volume} {465}},\ \bibinfo {pages} {61} (\bibinfo {year} {2008})},\ \Eprint {https://arxiv.org/abs/0705.4397} {0705.4397} \BibitemShut {NoStop}%
\bibitem [{\citenamefont {Ellis}\ \emph {et~al.}(2012)\citenamefont {Ellis}, \citenamefont {Maartens},\ and\ \citenamefont {MacCallum}}]{Ellis2012}%
  \BibitemOpen
  \bibfield  {author} {\bibinfo {author} {\bibfnamefont {G.}~\bibnamefont {Ellis}}, \bibinfo {author} {\bibfnamefont {R.}~\bibnamefont {Maartens}},\ and\ \bibinfo {author} {\bibfnamefont {M.}~\bibnamefont {MacCallum}},\ }\href@noop {} {\emph {\bibinfo {title} {{Relativistic Cosmology}}}}\ (\bibinfo  {publisher} {Cambridge University Press},\ \bibinfo {year} {2012})\BibitemShut {NoStop}%
\bibitem [{\citenamefont {Collins}\ and\ \citenamefont {Hawking}(1973{\natexlab{a}})}]{Collins:1973lda}%
  \BibitemOpen
  \bibfield  {author} {\bibinfo {author} {\bibfnamefont {C.~B.}\ \bibnamefont {Collins}}\ and\ \bibinfo {author} {\bibfnamefont {S.~W.}\ \bibnamefont {Hawking}},\ }\bibfield  {title} {\bibinfo {title} {{The rotation and distortion of the Universe}},\ }\href@noop {} {\bibfield  {journal} {\bibinfo  {journal} {Mon. Not. Roy. Astron. Soc.}\ }\textbf {\bibinfo {volume} {162}},\ \bibinfo {pages} {307} (\bibinfo {year} {1973}{\natexlab{a}})}\BibitemShut {NoStop}%
\bibitem [{\citenamefont {Wainwright}\ and\ \citenamefont {Ellis}(1997)}]{Wainwright:1997}%
  \BibitemOpen
  \bibinfo {editor} {\bibfnamefont {J.}~\bibnamefont {Wainwright}}\ and\ \bibinfo {editor} {\bibfnamefont {G.~F.~R.}\ \bibnamefont {Ellis}},\ eds.,\ \href@noop {} {\emph {\bibinfo {title} {{Dynamical Systems in Cosmology}}}}\ (\bibinfo  {publisher} {Cambridge University Press},\ \bibinfo {address} {Cambridge},\ \bibinfo {year} {1997})\BibitemShut {NoStop}%
\bibitem [{\citenamefont {Collins}\ and\ \citenamefont {Hawking}(1973{\natexlab{b}})}]{Collins:1972tf}%
  \BibitemOpen
  \bibfield  {author} {\bibinfo {author} {\bibfnamefont {C.~B.}\ \bibnamefont {Collins}}\ and\ \bibinfo {author} {\bibfnamefont {S.~W.}\ \bibnamefont {Hawking}},\ }\bibfield  {title} {\bibinfo {title} {{Why is the Universe isotropic?}},\ }\href {https://doi.org/10.1086/151965} {\bibfield  {journal} {\bibinfo  {journal} {Astrophys. J.}\ }\textbf {\bibinfo {volume} {180}},\ \bibinfo {pages} {317} (\bibinfo {year} {1973}{\natexlab{b}})}\BibitemShut {NoStop}%
\bibitem [{\citenamefont {Groen}\ and\ \citenamefont {Hervik}(2007)}]{Groen:2007zz}%
  \BibitemOpen
  \bibfield  {author} {\bibinfo {author} {\bibfnamefont {O.}~\bibnamefont {Groen}}\ and\ \bibinfo {author} {\bibfnamefont {S.}~\bibnamefont {Hervik}},\ }\href@noop {} {\emph {\bibinfo {title} {{Einstein's general theory of relativity: With modern applications in cosmology}}}}\ (\bibinfo  {publisher} {Springer},\ \bibinfo {year} {2007})\BibitemShut {NoStop}%
\bibitem [{\citenamefont {Chevallier}\ and\ \citenamefont {Polarski}(2001)}]{Chevallier:2000qy}%
  \BibitemOpen
  \bibfield  {author} {\bibinfo {author} {\bibfnamefont {M.}~\bibnamefont {Chevallier}}\ and\ \bibinfo {author} {\bibfnamefont {D.}~\bibnamefont {Polarski}},\ }\bibfield  {title} {\bibinfo {title} {Accelerating universes with scaling dark matter},\ }\href {https://doi.org/10.1142/S0218271801000822} {\bibfield  {journal} {\bibinfo  {journal} {Int. J. Mod. Phys. D}\ }\textbf {\bibinfo {volume} {10}},\ \bibinfo {pages} {213} (\bibinfo {year} {2001})},\ \Eprint {https://arxiv.org/abs/gr-qc/0009008} {gr-qc/0009008} \BibitemShut {NoStop}%
\bibitem [{\citenamefont {Linder}(2003)}]{Linder:2002et}%
  \BibitemOpen
  \bibfield  {author} {\bibinfo {author} {\bibfnamefont {E.~V.}\ \bibnamefont {Linder}},\ }\bibfield  {title} {\bibinfo {title} {Exploring the expansion history of the universe},\ }\href {https://doi.org/10.1103/PhysRevLett.90.091301} {\bibfield  {journal} {\bibinfo  {journal} {Phys. Rev. Lett.}\ }\textbf {\bibinfo {volume} {90}},\ \bibinfo {pages} {091301} (\bibinfo {year} {2003})},\ \Eprint {https://arxiv.org/abs/astro-ph/0208512} {astro-ph/0208512} \BibitemShut {NoStop}%
\bibitem [{\citenamefont {Kunz}(2009)}]{Kunz:2007rk}%
  \BibitemOpen
  \bibfield  {author} {\bibinfo {author} {\bibfnamefont {M.}~\bibnamefont {Kunz}},\ }\bibfield  {title} {\bibinfo {title} {Degeneracy between the dark components resulting from the fact that gravity only measures the total energy-momentum tensor},\ }\href {https://doi.org/10.1103/PhysRevD.80.123001} {\bibfield  {journal} {\bibinfo  {journal} {Phys. Rev. D}\ }\textbf {\bibinfo {volume} {80}},\ \bibinfo {pages} {123001} (\bibinfo {year} {2009})},\ \Eprint {https://arxiv.org/abs/astro-ph/0702615} {astro-ph/0702615} \BibitemShut {NoStop}%
\bibitem [{\citenamefont {Pedrotti}\ \emph {et~al.}(2026)\citenamefont {Pedrotti}, \citenamefont {Escamilla}, \citenamefont {Marra}, \citenamefont {Perivolaropoulos},\ and\ \citenamefont {Vagnozzi}}]{Pedrotti:2025ccw}%
  \BibitemOpen
  \bibfield  {author} {\bibinfo {author} {\bibfnamefont {D.}~\bibnamefont {Pedrotti}}, \bibinfo {author} {\bibfnamefont {L.~A.}\ \bibnamefont {Escamilla}}, \bibinfo {author} {\bibfnamefont {V.}~\bibnamefont {Marra}}, \bibinfo {author} {\bibfnamefont {L.}~\bibnamefont {Perivolaropoulos}},\ and\ \bibinfo {author} {\bibfnamefont {S.}~\bibnamefont {Vagnozzi}},\ }\bibfield  {title} {\bibinfo {title} {{BAO} miscalibration cannot rescue late-time solutions to the {H}ubble tension},\ }\href {https://doi.org/10.1103/pn9j-8whx} {\bibfield  {journal} {\bibinfo  {journal} {Phys. Rev. D}\ }\textbf {\bibinfo {volume} {113}},\ \bibinfo {pages} {043507} (\bibinfo {year} {2026})},\ \Eprint {https://arxiv.org/abs/2510.01974} {2510.01974} \BibitemShut {NoStop}%
\bibitem [{\citenamefont {Weinberg}(1989)}]{Weinberg:1988cp}%
  \BibitemOpen
  \bibfield  {author} {\bibinfo {author} {\bibfnamefont {S.}~\bibnamefont {Weinberg}},\ }\bibfield  {title} {\bibinfo {title} {The cosmological constant problem},\ }\href {https://doi.org/10.1103/RevModPhys.61.1} {\bibfield  {journal} {\bibinfo  {journal} {Rev. Mod. Phys.}\ }\textbf {\bibinfo {volume} {61}},\ \bibinfo {pages} {1} (\bibinfo {year} {1989})}\BibitemShut {NoStop}%
\bibitem [{\citenamefont {Amendola}(2000)}]{Amendola:1999er}%
  \BibitemOpen
  \bibfield  {author} {\bibinfo {author} {\bibfnamefont {L.}~\bibnamefont {Amendola}},\ }\bibfield  {title} {\bibinfo {title} {Coupled quintessence},\ }\href {https://doi.org/10.1103/PhysRevD.62.043511} {\bibfield  {journal} {\bibinfo  {journal} {Phys. Rev. D}\ }\textbf {\bibinfo {volume} {62}},\ \bibinfo {pages} {043511} (\bibinfo {year} {2000})},\ \Eprint {https://arxiv.org/abs/astro-ph/9908023} {astro-ph/9908023} \BibitemShut {NoStop}%
\bibitem [{\citenamefont {Copeland}\ \emph {et~al.}(1998)\citenamefont {Copeland}, \citenamefont {Liddle},\ and\ \citenamefont {Wands}}]{Copeland:1997et}%
  \BibitemOpen
  \bibfield  {author} {\bibinfo {author} {\bibfnamefont {E.~J.}\ \bibnamefont {Copeland}}, \bibinfo {author} {\bibfnamefont {A.~R.}\ \bibnamefont {Liddle}},\ and\ \bibinfo {author} {\bibfnamefont {D.}~\bibnamefont {Wands}},\ }\bibfield  {title} {\bibinfo {title} {{Exponential potentials and cosmological scaling solutions}},\ }\href {https://doi.org/10.1103/PhysRevD.57.4686} {\bibfield  {journal} {\bibinfo  {journal} {Phys. Rev. D}\ }\textbf {\bibinfo {volume} {57}},\ \bibinfo {pages} {4686} (\bibinfo {year} {1998})},\ \Eprint {https://arxiv.org/abs/gr-qc/9711068} {gr-qc/9711068} \BibitemShut {NoStop}%
\bibitem [{\citenamefont {Copeland}\ \emph {et~al.}(2006)\citenamefont {Copeland}, \citenamefont {Sami},\ and\ \citenamefont {Tsujikawa}}]{Copeland:2006wr}%
  \BibitemOpen
  \bibfield  {author} {\bibinfo {author} {\bibfnamefont {E.~J.}\ \bibnamefont {Copeland}}, \bibinfo {author} {\bibfnamefont {M.}~\bibnamefont {Sami}},\ and\ \bibinfo {author} {\bibfnamefont {S.}~\bibnamefont {Tsujikawa}},\ }\bibfield  {title} {\bibinfo {title} {{Dynamics of dark energy}},\ }\href {https://doi.org/10.1142/S021827180600942X} {\bibfield  {journal} {\bibinfo  {journal} {Int. J. Mod. Phys. D}\ }\textbf {\bibinfo {volume} {15}},\ \bibinfo {pages} {1753} (\bibinfo {year} {2006})},\ \Eprint {https://arxiv.org/abs/hep-th/0603057} {hep-th/0603057} \BibitemShut {NoStop}%
\bibitem [{\citenamefont {Amendola}(2004)}]{Amendola:2004Repulsive}%
  \BibitemOpen
  \bibfield  {author} {\bibinfo {author} {\bibfnamefont {L.}~\bibnamefont {Amendola}},\ }\bibfield  {title} {\bibinfo {title} {{Phantom energy mediates a long-range repulsive force}},\ }\href {https://doi.org/10.1103/PhysRevLett.93.181102} {\bibfield  {journal} {\bibinfo  {journal} {Phys. Rev. Lett.}\ }\textbf {\bibinfo {volume} {93}},\ \bibinfo {pages} {181102} (\bibinfo {year} {2004})},\ \Eprint {https://arxiv.org/abs/hep-th/0409224} {hep-th/0409224} \BibitemShut {NoStop}%
\end{thebibliography}%

\end{document}